\documentclass[aps,prd,a4paper,showpacs,12pt]{article}
\usepackage[margin=1.0in]{geometry}
\usepackage{multicol}
\usepackage{flushend}

\usepackage{subfigure}
\usepackage {graphicx}
\usepackage{color}
\usepackage{xcolor}
\usepackage{threeparttable}
\usepackage{amsfonts}
\usepackage{times}
\usepackage{setspace}
\usepackage{balance}
\usepackage{lastpage}
\usepackage{amsthm}
\usepackage[tbtags]{amsmath}
\usepackage{nomencl}
\usepackage{pstricks}
\usepackage{pstricks-add}
\usepackage{pst-plot,pstricks-add}

\usepackage{amsmath,amsfonts,amssymb,simplewick}
\usepackage{float}
\usepackage{amsmath,amssymb,amsfonts,epsfig,graphicx,euscript}%
\usepackage{hyperref} 
\hypersetup{bookmarks=true, bookmarksnumbered=true,linktoc=page, pdfstartview={FitH}, 
colorlinks=true, 
citecolor=red, 
filecolor=blue, 
linkcolor=blue, 
urlcolor=blue}
\usepackage{amsmath}% http://ctan.org/pkg/amsmath
\usepackage{amsfonts}% http://ctan.org/pkg/amsfonts
\usepackage{graphicx}
\usepackage{caption}
\usepackage{float}
\usepackage[all]{hypcap}
\numberwithin{equation}{section}
\usepackage{cite}
\usepackage{calc}
\newlength{\spacer}
\newsavebox{\mybox}

\newcommand{\bse}{\begin{subequations}}
	\newcommand{\ese}{\end{subequations}}
\newcommand{\be}{\begin{equation}}
\newcommand{\ee}{\end{equation}}
\newcommand{\bea}{\begin{eqnarray}}
\newcommand{\eea}{\end{eqnarray}}
\newcommand{\ba}{\begin{array}}
	\newcommand{\ea}{\end{array}}

\renewcommand{\thefootnote}{\fnsymbol{footnote}}

\begin{document}
	
	%\hfill%
	%\vbox{
	%    \halign{#\hfil        \cr
	%           IPM/P-2012/010\cr
	%                     }}
	%\vspace{1cm}
	\begin{center}
       { \large{\textbf{ The effects of non-helical component of hypermagnetic field on the evolution of the matter-antimatter asymmetry, vorticity, and hypermagnetic field}}} %\\
%		{ \large{\textbf{ The contribution of the hypermagnetic field to the generation and evolution of the vorticity and the matter-antimatter asymmetry in the early Universe}}} %\\
		\vspace*{1.5cm}
		\begin{center}
			{\bf S. Abbaslu\footnote{s$_{-}$abbasluo@sbu.ac.ir}$^1$, S. Rostam Zadeh\footnote{sh$_{-}$rostamzadeh@ipm.ir}$^2$, A. Rezaei\footnote{amirh.rezaei@mail.sbu.ac.ir}$^1$ and S. S. Gousheh\footnote{ss-gousheh@sbu.ac.ir}$^1$, }\\
			\vspace*{0.5cm}
			{\it{$^1$Department of Physics, Shahid Beheshti University, Tehran, Iran\\$^2$School of Particles and Accelerators, Institute for Research in Fundamental Sciences (IPM), P.O.Box 19395-5531, Tehran, Iran}}\\
		\vspace*{1cm}
		\end{center}
	\end{center}
	\begin{center}
		\today
	\end{center}
	%\vspace{.5cm}
	%\bigskip
	
	\renewcommand*{\thefootnote}{\arabic{footnote}}
	\setcounter{footnote}{0}
	
	%\textcolor{red}{}	
	
	\begin{center}
		\textbf{Abstract}
	\end{center}

We study the evolution of the matter-antimatter asymmetry ($\eta $), the vorticity, and the hypermagnetic field in the symmetric phase of the early Universe, and in the temperature range $100\ \mbox{GeV}\le T\le10\ \mbox{TeV}$. We assume a configuration for the hypermagnetic field which includes both helical and non-helical ($B_z$) components. Consequently, the hypermagnetic field and the fluid vorticity can directly affect each other, the manifestations of which we explore in three scenarios. In the first scenario, we show that in the presence of a small vorticity and a large $\eta_{e_R} $,  helicity can be generated and amplified for an initially strong  $B_z$. The generation of the helical seed is due to the chiral vortical effect (CVE) and/or the advection term, while its growth is mainly due to the chiral magnetic effect (CME) which leads to the production of the baryon asymmetry, as well. The vorticity saturates to a nonzero value which depends on $B_z$, even in the presence of the viscosity, due to the  back-reaction of $B_z$ on the plasma. 
Increasing the initial vorticity,  makes the values of the helicity, $\eta$s, and vorticity reach their saturation curves sooner, but does not change their final values at the onset of the electroweak phase transition. 
The second scenario is similar to the first except we assume that all initial $\eta$s are zero. We find that much higher initial vorticity is required for the generation process and, while the values of $\eta$s do not reach their saturation curves, final $\eta$s of order $10^{-9}$ are possible. In the third scenario, we show that in the presence of only a strong hypermagnetic field, $\eta$s and  vorticity can be generated and amplified. Increasing the initial helicity, increases the final $\eta$s and vorticity.  Although the values of $\eta$s do not reach their saturation curves, final values of order $10^{-7}$ are possible. We find that although the presence of a nonzero initial $B_z$ is necessary in all three scenarios, its increase only increases the final values of vorticity.

\newpage
\section{Introduction}

Observations indicate that the Universe is magnetized on all scales. Magnetic fields exist everywhere in the Universe, from the stars to the galaxies and the intergalactic medium \cite{{1},{2},{3},{4}}. The amplitude of the detected coherent magnetic fields in the Milky Way is in the order of $10^{-6} \mbox{G}$ over the plain of its disc, while that of the magnetic fields existing in the intergalactic medium is in the order of $10^{-15}\mbox{G}$ \cite{magnetic1,magnetic2,magnetic3,magnetic4,magnetic5}.

These fields are very important from  various aspects. They govern the gas-cloud dynamics, influence the formation of the stars, and can be used to determine the energy of the cosmic rays\cite{kandus}.  
 Meanwhile, the origin and the evolution of these fields are under debate. There are two major approaches for studying the evolution of these fields, namely astrophysical and cosmological \cite{{4},{Kulsrud},{kandus},{phd},{phd1}}.
%Astrophysical and cosmological models are the major approach that concern to the evolution of these fields \cite{{4},{Kulsrud},{kandus},{phd},{phd1}}.
On the other hand, historically, the creation and amplification mechanisms of these fields can be divided into three categories depending on the time of their occurrence: before the recombination, during the recombination and after the recombination \cite{{naoz},{4}}. Astrophysical models are considered to be in the category of the processes occurring after and during the recombination \cite{{naoz},{4}}. Recent observations, \cite{{Brandenburg},{T-1},{Taylor},{huan},{Vovk},{Dolag}} as well as the ubiquitous presence of large-scale magnetic fields in the Universe, strengthen the hypothesis of their primordial origin, {\it i.e.,} the cosmological model \cite{kandus}. However, primordial magnetogenesis model has serious problems. For example, the predictions for the seed fields amplified between the inflation and the recombination era, suffer from the smallness of their correlation lengths\footnote{Note that the generated initial correlation length cannot exceed the Hubble horizon, due to the causality.} \cite{cl}, albeit there are some mechanisms that can increase their correlation lengths \cite{{A-B},{P-olesen},{D-t},{A-G},{A-Boyarsky},{T-K},{J-Z},{Wolf}}. The magnetic fields generated during the inflation do not have this problem, but have a weak strength due to the conservation of the flux and expansion of the Universe. In this work, we concentrate on the cosmological origin for the magnetic fields after the inflation, passing over the scale problem.% and do not concern about these problems.
%\textcolor{red}{} \textcolor{green}{} \textcolor{green}{,} \textcolor{orange}{()} \textcolor{purple}{}

Since the non-Abelian gauge fields acquire mass gaps, they have no contribution to the observed long-range magnetic fields, and only the Abelian hypercharge gauge fields contribute to these fields \cite{kajantie}. The evolution of the hypermagnetic fields before the electroweak phase transition is influenced by the non-perturbative anomalous effects.    
%In the evolution of the hypermagnetic fields before the electroweak phase transition, the consideration of the non-perturbative anomalous effects are inevitable. 
The Abelian anomaly equations violate the conservation of the matter currents, and interconnect the evolution of the hypermagnetic fields and the matter-antimatter asymmetries in the symmetric phase of the early Universe \cite{{S-L},{J-S-B},{G-t-H}}. Indeed, the Abelian gauge fields couple to the fermions chirally and this results in two important Abelian anomalous effects. First, the existence of the Abelian anomaly equations as mentioned earlier, and second, the emergence of the Abelian Chern-Simons term in the $\textrm{U}_{\textrm{Y}}(1)$ effective action. This term leads to the chiral magnetic effect (CME),\footnote{The generation of the electric current in the same direction as the magnetic field.} the current of which is $\vec{J}_{\mathrm{cm}}=c_{\mathrm{B}}\vec{B}_{Y}$, where the hypercharge chiral magnetic coefficient $c_{\mathrm{B}}$ depends on the fermionic chemical potentials \cite{31,34,Elahi-2020}.
%$c_{\mathrm{B}}\varpropto\sum_{i=1}^{n_{G}}(-2\mu_{R_{i}}+\mu_{L_{i}}-3\mu_{Q_{i}})$, $n_{G}$ is the number of generations, and $\mu_{L_i}$($\mu_{R_i}$) and $\mu_{Q_i}$ are the common chemical potentials of the left-handed (right-handed) leptons, and the left-handed quarks with different colors, respectively \cite{31,34}. 
Therefore, the matter-antimatter asymmetries are interconnected with the hypermagnetic fields through the CME, as well.
%In fact, the matter-antimatter asymmetries are interconnected with the hypermagnetic field through the chemical potentials appearing in the hypercharge chiral magnetic coefficient $c_{B}$, as well. 
In this context, some people have shown that, in the presence of the matter-antimatter asymmetries, the hypermagnetic field can be amplified from a weak seed field or, in the presence of the strong hypermagnetic field, the matter-antimatter asymmetries can be generated \cite{{26},{27},{28},{30},{31},{32},{33},{34},shiva3}.   

The origin of the matter-antimatter asymmetry in the Universe is another unanswered problem in particle physics and cosmology. The amplitude of the baryon asymmetry of the Universe is measured via different mechanisms, and its accepted current estimate is $\eta_{B}\sim 10^{-10}$ \cite{{baryon1},{baryon2},{baryon3},{baryon4}}. There exist some scenarios that investigate the generation and evolution of the matter-antimatter asymmetry and the hypermagnetic fields, simultaneously \cite{{26},{27},{28},{30},{31},{32},{33},{34},shiva3}. The authors of Ref.\ \cite{26} investigated the production of the matter-antimatter asymmetry in the presence of the primordial hypermagnetic fields. They considered the Abelian anomalous effects and generalized the ordinary magnetohydrodynamic equations to the anomalous magnetohydrodynamics (AMHD). Then, they showed that, depending on the hypermagnetic energy spectrum and particle physics parameters such as the electron Yukawa coupling and the strength of the electroweak phase transition, the matter-antimatter fluctuation can be generated in the plasma. The authors of Ref.\ \cite{long} considered another scenario, in which, first a lepton asymmetry is created and then it is converted to the baryon asymmetry. They showed that, in contrast to the electroweak baryogenesis, leptogenesis yields a right-handed helical magnetic field \cite{vachaspati}. Moreover, the authors of Ref.\ \cite{m.joyce} have presented a model for the generation of hypermagnetic field, assuming a preexisting right-handed electron asymmetry. They took into account the Abelian anomaly and only the first-generation right-handed leptons, then investigated the evolution of the hypermagnetic fields and the right-handed electron asymmetry. The authors of Refs.\ \cite{{28},{30}} also considered the first-generation left-handed leptons and the influence of the weak sphalerons, the effects which were not considered in the earlier works \cite{{26},{27}}.

In all aforementioned studies, the effects of the velocity and the vorticity of the plasma were absent. Recently, it has been shown that the chiral vortical effect (CVE)\footnote{The generation of the electric current in the same direction as the vorticity field.} has an important role in the generation and evolution of the hypermagnetic fields \cite{saeed,saeed2}. This effect was discovered by Vilenkin \cite{35}. He showed that a rotating black hole can produce a chiral neutrino current density as $J(0)=-\Omega T^{2}/12 -\Omega^{3}/48\pi^{2} -\Omega\mu^{2}/4\pi^{2}$ \cite{35}, where $\Omega$ is the angular velocity, $\mu$ is the chiral chemical potential of the neutrino, and $T$ is its temperature. For a single-species plasma, in the broken phase, the vector current which results from the CVE appears as $\vec{J}_{\mathrm{cv}}=\frac{1}{4\pi^2}(\mu_{R}^2-\mu_{L}^2)\vec{\Omega}$, where $\mu_{R}$ and $\mu_{L}$ are the right-handed and the left-handed chemical potentials of the species, respectively \cite{35,35-1,36,son1,a1,a11,a2,a3,a4,a5,a6}. In the symmetric phase, besides the hypercharge chiral magnetic current, the chiral vortical current also appears in the total current, which generates the hypermagnetic fields and affects their evolution.\footnote{The form of the chiral vortical current in the symmetric phase, not including the temperature dependent part,  is given in Ref.\ \cite{saeed}. The complete form is given in Ref.\ \cite{saeed2}, and is restated and used later in this study.} 
 
On the subject of the magnetogenesis, and the chiral magnetic and vortical effects, the authors of Ref.\ \cite{36} have considered an incompressible fluid with a fully non-helical vorticity field. They have assumed that the back-reaction of the magnetic field on the fluid velocity is negligible, and the advection term\footnote{The term $\vec{\nabla}\times(\vec{v}\times\vec{B})$ \cite{36}.} is unimportant in the magnetohydrodynamics equations. The authors of Ref.\ \cite{37} have investigated the chiral anomalous effects on the evolution of the magnetohydrodynamics turbulence, and showed that a maximally helical magnetic field might be generated from an initially non-helical one. They considered an incompressible fluid in the resistive approximation and took into account the chiral magnetic effect, then showed that this chiral effect can support a turbulent inverse cascade.\footnote{The inverse cascade is the transfer of energy from the small scales to the large scales.} In their scenario, only the right-handed electron has been considered in the chiral plasma. The authors of Ref.\ \cite{37.1}, referring to the work done in Ref.\ \cite{38.1}, approximated the evolution of the velocity of the plasma by the Lorentz force. They investigated the evolution of the energy and helicity spectra of the magnetic field in the broken phase, and showed that in a turbulent plasma with a strong seed of the magnetic field, the right-left handed electron asymmetry is enhanced compared to the non-turbulent plasma with zero velocity. Although the effect of the velocity has been considered in Refs.\ \cite{37,37.1}, the effect of the chiral vorticity was not taken into account. 
%on the evolution of the hypermagnetic field and the matter-antimatter asymmetry. 

%we discussed fully about these CVE and CME
In our previous work \cite{saeed}, we investigated the generation and growth of the hypermagnetic field in a chiral vortical plasma, taking into account the CVE and the CME in the symmetric phase of the early Universe, and in the temperature range $100\ \mbox{GeV}\le T\le10\ \mbox{TeV}$. We showed that, in the presence of an initial large right-handed electron asymmetry, the hypermagnetic field can be generated from zero initial value, only if the plasma is also vortical. We also showed that the produced seed of the hypermagnetic field grows due to the CME. Since we had chosen a fully helical configuration for the hypermagnetic field, the plasma was force-free in the absence of the viscosity. Furthermore, the advection term was absent in the AMHD equations, because the chosen configuration for the velocity field was also fully vortical with the same helical configuration as the hypermagnetic field. The main generalization considered in this paper as compared to our previous work is the addition of a non-helical component to the hypermagnetic field, i.e., $B_z$, which, as we shall show, will have important consequences. 

%\textcolor{red}{} \textcolor{green}{} \textcolor{green}{,} \textcolor{orange}{()} \textcolor{purple}{}

The main purpose of this paper is to answer two important questions: First we investigate the possibility to generate and grow matter-antimatter asymmetries along with helical components of hypermagnetic field resulting in a net helicity, starting with a nonzero $B_z$ and a small vorticity, with or without an initial right-handed electron asymmetry $\eta_{e_R} $. Second, we investigate the possibility to generate and grow matter-antimatter asymmetries along with vorticity, starting with a hypermagnetic field that has both helical and non-helical components. Here, we choose the velocity field to be fully helical with the same Chern-Simons configurations as the helical part of the hypermagnetic field. In all cases that we study here, the prominent effects of adding $B_z$ is that a vorticity field can seed helicity through the advection term and helicity in turn back-reacts on the vorticity. Therefore the plasma is no longer force-free even in the absence of viscosity. Moreover, as we shall show, this back-reaction can usually counteract the effects of the immense viscosity.

The organization of the paper is as follows: In Sec.\ \ref{x1}, we briefly review the fermion number violation, due to the Abelian anomaly equations in the symmetric phase of the expanding Universe. In Sec.\ \ref{x2}, we present the anomalous magnetohydrodynamics equations and derive the complete set of evolution equations for the matter-antimatter asymmetries, and the hypermagnetic and velocity fields, taking the CVE and the CME into account, in the FRW metric. In Sec.\ \ref{x3}, we solve the evolution equations obtained in Sec.\ \ref{x2} numerically, show the results, and discuss about them on the basis of the evolution equations. In Sec.\ \ref{x5}, we summarize our results and conclude. 
\section{Fermion Number violation in the Symmetric Phase}\label{x1}
 
Due to the chiral coupling of the hypercharge gauge fields to the fermions in the symmetric phase, the baryon and lepton numbers are violated separately, while their difference $B-L$ remains conserved \cite{{S-L},{J-S-B},{G-t-H},{26},{27},{28},{30},{31},{32},{33},{34}}.
 Global matter current non-conservation occurs for the chiral leptons and quarks and is manifested in the Abelian anomaly equations. In the expanding Universe, these equations for the right-handed and the left-handed electrons, and the baryons are as follows (see Appendix A):\footnote{The covariant derivatives below are to be associated with our choice of the metric $ (1,-R^{2},-R^{2},-R^{2})$. Also, in the following we use the natural units, in which $\hbar=c=1$.} 
 \begin{equation}\label{eq4dgf}
 \begin{split}
 &\nabla_{\mu} j_{{e}_R}^{\mu}=-\frac{1}{4}\left(Y_{R}^{2}\right)\frac{g'^{2}}{16 \pi^2}Y_{\mu\nu}\tilde{Y}^{\mu\nu}=\frac{g'^{2}}{4\pi^{2}}\vec{E}_{Y}.\vec{B}_{Y},\\
 &\nabla_{\mu} j_{{e}_L}^{\mu}=\frac{1}{4}\left(Y_{L}^{2}\right)\frac{g'^{2}}{16 \pi^2}Y_{\mu\nu}\tilde{Y}^{\mu\nu}=-\frac{g'^{2}}{16\pi^{2}}\vec{E}_{Y}.\vec{B}_{Y},
  \end{split}
 \end{equation}
\begin{equation}
 \nabla_{\mu} j_{B}^{\mu}=\frac{1}{N_{c}}\sum_{i=1}^{n_{G}}\left(\nabla_{\mu} j_{Q_{i}}^{\mu}+\nabla_{\mu} j_{u_{R_{i}}}^{\mu}+\nabla_{\mu} j_{u_{d_{i}}}^{\mu}\right)=3\big[\nabla_{\mu} j_{{e}_R}^{\mu}+2\nabla_{\mu}j_{{e}_L}^{\mu}\big],
\end{equation}
where $n_{G}$ is the number of generations, and $N_{c}=3$ is the rank of the SU$(3)$ non-Abelian gauge group.
After taking the spatial average of Eq.\ (\ref{eq4dgf}) we obtain (see Appendix A for details) 
	\begin{equation}\label{eq4aq3w1asd}
	\begin{split}
	&\partial_{t}\left(\frac{n_{e_R}-\bar{n}_{e_R}}{s}\right)=  \frac{g'^{2}}{4\pi^{2}s}\langle\vec{E}_{Y}.\vec{B}_{Y}\rangle,\\&
	\partial_{t}\left(\frac{n_{e_L}-\bar{n}_{e_L}}{s}\right)=  \frac{-g'^{2}}{16\pi^{2}s}\langle\vec{E}_{Y}.\vec{B}_{Y}\rangle,
	\end{split}
	\end{equation}
%
%\textcolor{red}{} \textcolor{green}{} \textcolor{green}{,} \textcolor{orange}{()} \textcolor{purple}{}
%
where $s=2\pi^{2}g^{*}T^{3}/45$ is the entropy density, $g^{*}=106.75$ is the number of relativistic degrees of freedom, and $n_{e_{R,L}}$ and $\bar{n}_{e_{R,L}}$ denote the chiral number densities of the electrons and positrons, respectively.\footnote{We should mention that usually in the literature the difference between the latter two is denoted by $n_{e_{R,L}}$. The distinction we have made here is merely in view of our upcoming work.}
At the temperatures of our interest, the rate of the electron chirality flip processes become larger than the Hubble parameter. Therefore, their effects should also be taken into account in the equations for the violation of the chiral electron numbers. Recalling the relation $n_{f}-\bar{n}_f=\mu_{f}T^{2}/6$ we obtain $\eta_{f}=(n_{f}-\bar{n}_{f})/s=\mu_{f}T^{2}/6s$.\footnote{$\eta_{f}$ with $f=e_{R},e_{L},\nu_{e}^{L}$ is the fermion asymmetry, and $\eta_{B}$ is the baryon asymmetry.} Therefore, the evolution equations for the asymmetries of the chiral electrons and the baryons in terms of  $\eta$ become \cite{{27},{28},{31}} 

\begin{equation}\label{eq4}
\begin{split}
&\frac{d\eta_{e_R}}{dt}=\frac{g'^{2}}{4\pi^{2}s}\langle\vec{E}_{Y}.\vec{B}_{Y}\rangle+\left(\frac{\Gamma_{0}}{t_{EW}}\right)\left(\frac{1-x}{\sqrt{x}}\right)\left(\eta_{e_{L}}-\eta_{e_{R}}\right),\\
&\frac{d\eta_{e_L}}{dt}=\frac{d\eta_{{\nu}_{e}}^{L}}{dt}=-\frac{g'^{2}}{16\pi^{2}s}\langle\vec{E}_{Y}.\vec{B}_{Y}\rangle+\left(\frac{\Gamma_{0}}{2t_{EW}}\right)\left(\frac{1-x}{\sqrt{x}}\right)\left(\eta_{e_{R}}-\eta_{e_{L}}\right),\\
&  \frac{1}{3}\frac{d\eta_{B}}{dt}=\frac{d\eta_{e_{R}}}{dt}+2\frac{d\eta_{e_{L}}}{dt},
\end{split}
\end{equation}
where $\Gamma_{0}=121$, $x=\left(t/t_\mathrm{EW}\right)=\left(T_\mathrm{EW}/T\right)^{2}$ is given by the Friedmann law, $t_\mathrm{EW}=M_{0}/2T_\mathrm{EW}^{2}$, and $M_{0}=M_\mathrm{Pl}/1.66\sqrt{g^{*}}$ is the reduced Planck mass. In the following section, we obtain the magnetohydrodynamic equations.
%The Eqs.\ (\ref{eq4}) are the right-left handed electron and the baryon asymmetry evolution equations  in the symmetric phase. In the next section we return to the magnetohydrodynamic equations together with these Abelian anomaly.

\section{Anomalous Magnetohydrodynamics}\label{x2}
 
 We know that our visible Universe at the present time consists of more than $90\% $  electromagnetic plasma \cite{plasma1,plasma2,plasma3}. The dynamics of the plasma is governed by the laws of Magnetohydrodynamics (MHD) \cite{plasma}. In the presence of the anomaly, the  magnetohydrodynamics is generalized to the Anomalous Magnetohydrodynamic equations (AMHD). In the symmetric phase, the plasma is globally neutral, and we obtain the evolution equations in the Landau-Lifshitz frame as follows: (see
Refs. \cite{{saeed},son1} and also Appendix A for details).\footnote{There are also additional terms of $O(\mu/T)$ in the $c_\mathrm{v}$ and $c_\mathrm{B}$ which are negligible within the confines of our model, {\it i.e.,} our initial conditions and the results of our dynamical equations (see Appendix A).}
\begin{equation}\label{eq5}
\frac{1}{R}\vec{\nabla}.\vec{E}_{Y}=0,\qquad\qquad\qquad\qquad\qquad\qquad\frac{1}{R}\vec{\nabla} .\vec{B}_{Y}=0,	
\end{equation}
\begin{equation}\label{eq6}
\frac{\partial \vec{B}_{Y}}{\partial t}+2H\vec{B}_{Y}=-\frac{1}{R}\vec{\nabla}\times\vec{ E}_{Y},		\qquad\qquad\vec{J}_\mathrm{Ohm}=\sigma\left(\vec{E}_{Y}+\vec{v}\times\vec{B}_{Y}\right),
\end{equation}
\begin{equation}\label{eq7}
\frac{\partial \vec{E}_{Y}}{\partial t}+2H\vec{E}_{Y}=\frac{1}{R}\vec{\nabla}\times\vec{B}_{Y}-\vec{J},\qquad\qquad	\vec{J}=\vec{J}_\mathrm{Ohm}+\vec{J}_\mathrm{cv}+\vec{J}_\mathrm{cm},	
\end{equation}
\begin{equation}\label{eq8}
\vec{J}_\mathrm{cv}=c_\mathrm{v}\vec{w},	\qquad\qquad\qquad\qquad\vec{J}_\mathrm{cm}=c_\mathrm{B}\vec{B}_{Y},	
\end{equation}
\begin{equation}\label{eq9}
\begin{split}
&\left[\frac{\partial}{\partial t}+\frac{1}{R}\left(\vec{v}.\vec{\nabla}\right)+H\right]\vec{v}+\frac{\vec{v}}{\rho+p}\frac{\partial p}{\partial t}=\\&\frac{1}{R}\frac{\vec{\nabla} p}{\rho+p}+\frac{\vec{J}\times\vec{B}_{Y}}{\rho+p}+\frac{\nu}{{R}^{2}}\left[\nabla^{2}\vec{v}+\frac{1}{3}\vec{\nabla}\left(\vec{\nabla}.\vec{v}\right)\right],
\end{split}
\end{equation}
\begin{equation}\label{eq10}
\vec{\omega}=\frac{1}{R}\vec{\nabla}\times\vec{v},
\end{equation}
%
%\textcolor{red}{} \textcolor{green}{} \textcolor{green}{,} \textcolor{orange}{()} \textcolor{purple}{}
%
\begin{equation}\label{eq11}
\frac{\partial \rho}{\partial t}+\frac{1}{R}\vec{\nabla}.\left[(\rho+p)\vec{v}\right]+3H(\rho+p)=0,		
\end{equation}
 \begin{equation}\label{eq12}
c_\mathrm{v}(t)=\frac{g'}{8\pi^{2}}\left(\mu_{e_{R}}^{2}-\mu_{e_{L}}^{2}\right),	
\end{equation}
\begin{equation}\label{eq13}
c_\mathrm{B}(t)=-\frac{g'^{2}}{8\pi^{2}}\left(-2\mu_{e_{R}}+\mu_{e_{L}}-\frac{3}{4}\mu_{B}\right).
\end{equation}
In the above equations, $\vec{\omega}$ and $\vec{v}$ are the vorticity and bulk velocity of the plasma, $g'$ is the coupling constant of the $\textrm{U}_{\textrm{Y}}(1)$, $\rho$ and $p$ are the energy density and pressure of the fluid, $R(t)$ is the scale factor, $H=\dot{R}/R$ is the Hubble parameter, $\sigma=100T$ is the electrical hyperconductivity, and $\nu\simeq1/(5\alpha_{Y}T)$ is the kinematic viscosity, where $\alpha_{Y}$ is the fine structure constant of the $\textrm{U}_{\textrm{Y}}(1)$. In Eqs.\ (\ref{eq7}) and (\ref{eq8}), $\vec{J}_\mathrm{Ohm}, \vec{J}_\mathrm{cv}$, and $\vec{J}_\mathrm{cm}$ are the Ohmic current, the chiral vortical current, and the hypercharge chiral magnetic current, respectively. Moreover, the coefficients $c_\mathrm{v}$ and $c_\mathrm{B}$ are the chiral vortical and the hypercharge chiral magnetic coefficients, which are obtained by considering the quarks and the first-generation leptons and assuming that $\mu_{d_{R}}=\mu_{u_{R}}=\mu_{Q}$ for all generations of the quarks \cite{34,saeed}. Since we consider an incompressible fluid in the comoving frame, {\it i.e.,} $\partial_{t}\rho+3H(\rho+p)=0$ in the Lab frame, the continuity Eq.\ (\ref{eq11}) reduces to $\vec{\nabla} .\vec{v}=0$ \cite{37,saeed}.  

Now we choose the configurations for our hypermagnetic field and the velocity field by using the following orthonormal basis \{$\hat{a}(z,k)=(\cos kz, -\sin kz,0)$, $\hat{b}(z,k)=(\sin kz, \cos kz,0)$, $\hat{z}$\} \cite{2016}. Note that the first two basis elements are Chern-Simons configurations with positive helicity. We can now express these fields as follows,  
\begin{equation}\label{eq15}
\vec{B}_{Y}(t,z) =	B_{z}(t)\hat{z}+B_{a}(t)\hat{a}(z,k)+B_{b}(t)\hat{b}(z,k),
\end{equation}
\begin{equation}\label{eq16}
\vec{v}(t,z) =  v_{a}(t)\hat{a}(z,k)+v_{b}(t)\hat{b}(z,k),
\end{equation}
The vorticities, as given by Eq.\ (\ref{eq10}), reduce to
\begin{equation}\label{}
\vec{w}(t,z) =w_{a}(t)\hat{a}(z,k)+w_{b}(t)\hat{b}(z,k),
\end{equation}
where $w_{i}(t)=(k/R) v_{i}(t)$ for $i=a,b$. Note that the space-dependent part of both the hypermagnetic and velocity fields are encoded in $\hat{a}(z,k)$ and $\hat{b}(z,k)$. These configurations satisfy the divergence-free condition, {\it i.e.,} $\vec{\nabla}.\vec{B}_{Y}=0$ for the hypermagnetic field, and $\vec{\nabla}.\vec{v}=0$ for the incompressible fluid. Therefore, the hypermagnetic and velocity fields can be written in terms of the vector potentials $\vec{A}_{Y}$ and $\vec{S}$, respectively. The vector potential $\vec{A}_{Y}$ can be chosen as
\begin{equation}\label{eq17}
\vec{A}_{Y}(t,x,y,z) =A_{1}(t,x,y,z)R(t)\hat{a}(z,k)+A_{2}(t,x,y,z)R(t)\hat{b}(z,k),
\end{equation}
where 
\begin{equation}\label{eq18}
\begin{split}
&A_{1}(t,x,y,z) =\frac{B_{z}(t)}{2}(x+y)[-\sin kz-\cos kz]+\frac{B_{a}(t)}{k},\\
&A_{2}(t,x,y,z) =\frac{B_{z}(t)}{2}(x+y)[-\sin kz+\cos kz]+\frac{B_{b}(t)}{k}.
\end{split}
\end{equation}
The vector potential $\vec{S}$ can be chosen as,
\begin{equation}\label{eq19}
\vec{S}(t,x,y,z) =\frac{v_{a}(t)}{k}R(t)\hat{a}(z,k)+\frac{v_{b}(t)}{k}R(t)\hat{b}(z,k).
\end{equation}
%where 
%\begin{equation}\label{eq20}
%\begin{split}
%S_{1}(t,x,y,z) =%\frac{v_{z}(t)}{2}(x+y)[-\sin kz-\cos kz]+
%\frac{v_{a}(t)}{k},\\
%S_{2}(t,x,y,z) =%\frac{v_{z}(t)}{2}(x+y)[-\sin kz+\cos kz]+
%\frac{v_{b}(t)}{k}.
%\end{split}
%\end{equation}
%
%\textcolor{red}{} \textcolor{green}{} \textcolor{green}{,} \textcolor{orange}{()} \textcolor{purple}{}
%
 
By using Eqs.\ (\ref{eq15}, \ref{eq17}, \ref{eq18}), we obtain the ensemble averages of the hypermagnetic energy and helicity density as follows 
 \begin{equation}\label{eq22}
 \begin{split}
 E_{B}(t)&=\frac{1}{2}\langle \vec{B_{Y}}(x,t) . \vec{B_{Y}}(x,t) \rangle\\&=\frac{1}{2}B_{Y}^{2}(t)=\frac{1}{2}\left[B_{z}^{2}(t)+B_{a}^{2}(t)+B_{b}^{2}(t)\right],
 \end{split}
 \end{equation}
  \begin{equation}\label{eq23}
 H_{B}(t)=\langle \vec{A_{Y}} . \vec{B_{Y}} \rangle=\frac{R(t)B_{a}^{2}(t)}{k}+\frac{R(t)B_{b}^{2}(t)}{k},
 \end{equation}
where the angle brackets denote the ensemble averaging. It can be seen that the hypermagnetic field becomes fully helical, {\it i.e.}, $E_{B}= (k/2R)H_B$, only in the limit $B_{z}^{2}(t)=0$. Note that, in contrast to the fully helical hypermagnetic field, in the non-helical case, the energy density can be non-zero while the helicity density is zero.

In analogy with the hypermagnetic field, we obtain the fluid kinetic energy and fluid helicity density as \cite{saeed}
 \begin{equation}\label{eq24}
 \begin{split}
 E_{v}(t)&=\frac{\rho}{2}\langle \vec{v} . \vec{v} \rangle \\
 &=\frac{\rho}{2}\left[%v_{z}^{2}(t)+
 v_{a}^{2}(t)+v_{b}^{2}(t)\right],
 \end{split}
 \end{equation}
 and
 \begin{equation}\label{eq25}
 \begin{split}
 H_{\mathrm{v}}(t)&=\sum_{i=1}^{n_{G}}\Big[\frac{1}{24}\left(T_{R_{i}}^{2}+T_{L_{i}}^{2}N_{w}+T_{d_{R_{i}}}^{2}N_{c}+T_{u_{R_{i}}}^{2}N_{c}+T_{Q_{i}}^{2}N_{c}N_{w}\right)\\&+\frac{1}{8\pi^{2}}\left(\mu_{R_{i}}^{2}+\mu_{L_{i}}^{2}N_{w}+\mu_{d_{R_{i}}}^{2}N_{c}+\mu_{u_{R_{i}}}^{2}N_{c}+\mu_{Q_{i}}^{2}N_{c}N_{w}\right)\Big]\langle \vec{v}.\vec{w} \rangle\\&=\sum_{i=1}^{n_{G}}\Big[\frac{15}{24}T^{2}+(\frac{1}{8\pi^{2}})(\mu_{R_{i}}^{2}+2\mu_{L_{i}}^{2}+12\mu_{Q}^{2})\Big]\frac{k}{{R(t)}}\left[v_{a}^{2}(t)+v_{b}^{2}(t)\right],
 \end{split}
 \end{equation}
where we have assumed that all particles, including the quarks and the first-generation leptons, are in thermal equilibrium, and $\mu_{d_{R}}=\mu_{u_{R}}=\mu_{Q}$ for all generations of the quarks \cite{34,saeed}. Our vorticity field is fully helical, since the velocity field contains only Chern-Simons configurations of the same helicity.

Let us now simplify the AMHD equations within the confines of our model. Since we consider a non-relativistic plasma, {\it i.e.,} $v^{2}/c^{2} \ll1$, we can neglect the displacement current in Eq.\ (\ref{eq7}).\footnote{Note that neglecting the displacement current in the comoving frame is equivalent to neglecting the term $\partial_{t} \vec{E}_{Y}+2H\vec{E}_{Y}$ in the Lab frame.} Consequently, we can use Eqs.\ (\ref{eq6}) and (\ref{eq7}), to express the hyperelectric field in terms of the hypermagnetic field as,
\begin{equation}\label{eq14}
\vec{E}_{Y}=-\vec{v}\times\vec{B}_{Y}+\frac{1}{R\sigma}\vec{\nabla}\times\vec{B}_{Y}-\frac{\ c_\mathrm{v}}{\sigma}\vec{\omega}-\frac{c_{B}}{\sigma}\vec{B}_{Y}.
\end{equation} 
By using Eqs.\ (\ref{eq6}) and (\ref{eq14}), we also obtain the evolution equation of the hypermagnetic field as,
\begin{equation}\label{eq26}
\frac{\partial\vec{ B}_{Y}}{\partial t}=\frac{1}{R}\vec{\nabla}\times(\vec{v}\times\vec{B}_{Y})+\frac{1}{R^{2}\sigma}\nabla^{2}\vec{B}_{Y}+\frac{c_\mathrm{v}}{R\sigma}\vec{\nabla}\times\vec{\omega}+\frac{c_{B}}{R\sigma}\vec{\nabla}\times\vec{B}_{Y}-\frac{\vec{ B}_{Y}}{t}.
\end{equation}
The first term on the rhs of Eq.\ (\ref{eq26}) is the advection term. Here, $\vec{v}\times\vec{B}_{Y}$ and its curl are non-zero, in contrast to the case where a fully helical configuration for the hypermagnetic field is taken into account. That is, for our chosen non-helical hypermagnetic field configuration given by Eq.\ (\ref{eq15}) (and helical velocity configuration  given by Eq.\ (\ref{eq16})), we obtain 
\begin{equation}\label{eq27}
\begin{split}
&\frac{1}{R}\vec{\nabla}\times(\vec{v}\times\vec{B}_{Y})=\\&\frac{k}{R} \left[v_{b}(t)B_{z}(t)%-v_{z}(t)B_{b}(t)
\right]\hat{a}(z,k)+\frac{k}{R}\left[%v_{z}(t)B_{a}(t)
-v_{a}(t)B_{z}(t)\right]\hat{b}(z,k).
\end{split}
\end{equation}
In the following, for simplicity, we use the relations $w_{a}(t)=k^{\prime}v_{a}(t)$, and $w_{b}(t)=k^{\prime} v_{b}(t)$, where $k^{\prime}=k/R=kT$. After substituting the chosen configurations for the hypermagnetic and velocity fields in  Eq.\ (\ref{eq26}) and simplifying, we obtain the evolution equation for the hypermagnetic field as follows,
\begin{equation}\label{eq28}
 \begin{split}
 \frac{\partial B_{a}(t)}{\partial t}=&k^{\prime}\left[v_{b}(t)B_{z}(t)%-v_{z}(t)B_{b}(t)
 \right]+\left[-\frac{{k^{\prime}}^{2}}{\sigma}+\frac{{k^{\prime}} c_\mathrm{B}}{\sigma}\right]B_{a}(t)+\frac{{k^{\prime}}^{2}c_\mathrm{v}}{\sigma}v_{a}(t)-\frac{B_{a}(t)}{t},\\
 \frac{\partial B_{b}(t)}{\partial t}=&-k^{\prime}\left[v_{a}(t)B_{z}(t)%-v_{z}(t)B_{a}(t)
 \right]+\left[-\frac{{k^{\prime}}^{2}}{\sigma}+\frac{k^{\prime} c_\mathrm{B}}{\sigma}\right]B_{b}(t)+\frac{{k^{\prime}}^{2} c_\mathrm{v}}{\sigma}v_{b}(t)-\frac{B_{b}(t)}{t},\\
 \frac{\partial B_{z}(t)}{\partial t}=&-\frac{B_{z}(t)}{t}.
  \end{split}
\end{equation}
%
%\textcolor{red}{} \textcolor{green}{} \textcolor{green}{,} \textcolor{orange}{()} \textcolor{purple}{}
%
%Here $c_\mathrm{B}\backsimeq\frac{g'^{2}}{8\pi^{2}}(2\mu_{e_{R}}-\mu_{e_{L}}+\frac{3}{4}\mu_{B})$ and $c_\mathrm{v}\backsimeq\frac{g'}{8\pi^{2}}(\mu_{e_{R}}^{2}-\mu_{e_{L}}^{2} )$ are hypercharge chiral magnetic and chiral vortical coefficients, respectively \cite{34,saeed}.
Let us now consider the evolution of the velocity and the vorticity fields. Due to the homogeneity of the Universe and smallness of the magnetic pressure compared to the fluid radiation pressure, {\it i.e.,}  $B^{2}/(8\pi p)  \ll 1$, we can ignore the gradient of the pressure in the evolution equation of the momentum Eq.\ (\ref{eq9}) \cite{37}, and obtain the evolution of the velocity field as,  
\begin{equation}\label{eq29}
  \begin{split}
 % &\frac{\partial v_{z}(t)}{\partial t}=0\\
  &\frac{\partial v_{a}(t)}{\partial t}=\frac{k^{\prime}}{\rho+p}\left[B_{b}(t)B_{z}(t)\right]-{k^{\prime}}^{2}\nu v_{a}(t),\\%-k^{\prime}v_{z}(t)v_{b}(t),\\
  &\frac{\partial v_{b}(t)}{\partial t}=-\frac{k^{\prime}}{\rho+p}\left[B_{a}(t)B_{z}(t)\right]-{k^{\prime}}^{2}\nu v_{b}(t).%+k^{\prime}v_{z}(t)v_{a}(t).
  \end{split}
\end{equation}
Note that in Eq.\ (\ref{eq9}), unlike the case where a fully helical configuration for the hypermagnetic field is taken into account, the term $\vec{J}\times\vec{B}_{Y}$ is non-zero; and therefore, the hypermagnetic field can affect the evolution of the velocity and the vorticity fields in the plasma. 

After obtaining the evolution equations for the hypermagnetic and the velocity fields, we now focus on the evolution of the hypercharge chiral magnetic coefficient $c_\mathrm{B}$ and the chiral vortical coefficient $ c_\mathrm{v}$ that depend on the matter-antimatter asymmetries. We recall Eqs.\ (\ref{eq4}) and obtain the evolution equations of the matter-antimatter asymmetries using the aforementioned configurations.  To do this, we first use the chosen configurations for the hypermagnetic and velocity fields, given by Eqs.\ (\ref{eq15}) and (\ref{eq16}), in the expression for the hyperelectric field Eq.\ (\ref{eq14}), to obtain 
\begin{equation}\label{eq30}
\begin{split}
\vec{E}_{Y}=&\left[v_{b}(t)B_{a}(t)-v_{a}(t)B_{b}(t)-\frac{c_\mathrm{B}}{\sigma}B_{z}(t)\right]\hat{z}\\&+
\left[%v_{z}(t)B_{b}(t)
-v_{b}(t)B_{z}(t)+\frac{k^{\prime}}{\sigma}B_{a}(t)-\frac{c_\mathrm{v}}{\sigma}k^{\prime} v_{a}(t)-\frac{c_\mathrm{B}}{\sigma}B_{a}(t)\right]\hat{a}(z,k)\\&+
\left[v_{a}(t)B_{z}(t)%-v_{z}(t)B_{a}(t)
+\frac{k^{\prime}}{\sigma}B_{b}(t)-\frac{c_\mathrm{v}}{\sigma}k^{\prime} v_{b}(t)-\frac{c_\mathrm{B}}{\sigma}B_{b}(t)\right]\hat{b}(z,k).
\end{split}
\end{equation}
The Abelian anomaly terms appearing in the evolution equations of the fermion number asymmetries, {\it i.e.,} Eq.\ (\ref{eq4}), are proportional to $\langle\vec{E}_{Y}.\vec{B}_{Y}\rangle$, which we can now calculate using Eq.\ (\ref{eq30}) to obtain,
\begin{equation}\label{eq31}
\begin{split}
\langle\vec{E}_{Y}.\vec{B}_{Y}\rangle=&-\frac{c_\mathrm{B}}{\sigma}\left[B_{z}^{2}(t)+B_{a}^{2}(t)+B_{b}^{2}(t)\right]+\frac{k^{\prime}}{\sigma}\left[B_{a}^{2}(t)+B_{b}^{2}(t)\right]\\&-\frac{c_\mathrm{v}}{\sigma}k^{\prime}\left[v_{a}(t)B_{a}(t)+v_{b}(t)B_{b}(t)\right].
\end{split}
 \end{equation}
By using Eqs.\ (\ref{eq4}) and (\ref{eq31}), and the relations $1\mbox{Gauss}\simeq2\times10^{-20} \mbox{GeV}^{2}$, $x=t/t_{EW}=(T_{EW}/T)^{2}$, $\mu_{f}=(6s/T^2)\eta_f$ for $f=e_{R},e_{L}$ and $B$, we obtain the new forms of the evolution equations of the matter-antimatter asymmetries. The whole set of our evolution equations in terms of $x$ become 

\begin{equation}\label{eq32}	
\begin{split}
\frac{d\eta_{e_R}(x)}{dx}=&F_{0}\left[\left(\frac{B_{a}(x)}{10^{20}G}\right)^{2}+\left(\frac{B_{b}(x)}{10^{20}G}\right)^{2}\right]x^{3/2}\\&-F_{1}\left(\eta_{e_R}(x)-\frac{\eta_{e_L}(x)}{2}+\frac{3}{8}\eta_{B}(x)\right)\left[\left(\frac{B_{z}(x)}{10^{20}G}\right)^{2}+\left(\frac{B_{a}(x)}{10^{20}G}\right)^{2}+\left(\frac{B_{b}(x)}{10^{20}G}\right)^{2}\right]x^{3/2}\\&-F_{2}\left(\eta_{e_R}^{2}(x)-\eta_{e_L}^{2}(x)\right)\left[v_{a}(x)\left(\frac{B_{a}(x)}{10^{20}G}\right)+v_{b}(x)\left(\frac{B_{b}(x)}{10^{20}G}\right)\right]\sqrt{x}\\&-\Gamma_{0}\frac{1-x}{\sqrt{x}}\big(\eta_{e_R}(x)-\eta_{e_L}(x)\big),
\end{split}
\end{equation}
\begin{equation}\label{eq33}	
 \begin{split}
 \frac{d\eta_{e_L}(x)}{dx}=&-\frac{F_{0}}{4}\left[\left(\frac{B_{a}(x)}{10^{20}G}\right)^{2}+\left(\frac{B_{b}(x)}{10^{20}G}\right)^{2}\right]x^{3/2}\\&+\frac{F_{1}}{4}\left(\eta_{e_R}(x)-\frac{\eta_{e_L}(x)}{2}+\frac{3}{8}\eta_{B}(x)\right)\left[\left(\frac{B_{z}(x)}{10^{20}G}\right)^{2}+\left(\frac{B_{a}(x)}{10^{20}G}\right)^{2}+\left(\frac{B_{b}(x)}{10^{20}G}\right)^{2}\right]x^{3/2}\\&+\frac{F_{2}}{4}\left(\eta_{e_R}^{2}(x)-\eta_{e_L}^{2}(x)\right)\left[v_{a}(x)\left(\frac{B_{a}(x)}{10^{20}G}\right)+v_{b}(x)\left(\frac{B_{b}(x)}{10^{20}G}\right)\right]\sqrt{x}\\&+\Gamma_{0}\frac{1-x}{2\sqrt{x}}\big(\eta_{e_R}(x)-\eta_{e_L}(x)\big),
 \end{split}
\end{equation}
\begin{equation}\label{eq34}	
\begin{split}
\frac{d\eta_{B}(x)}{dx}=&\frac{3F_{0}}{2}\left[\left(\frac{B_{a}(x)}{10^{20}G}\right)^{2}+\left(\frac{B_{b}(x)}{10^{20}G}\right)^{2}\right]x^{3/2}\\&-\frac{3F_{1}}{2}\left(\eta_{e_R}(x)-\frac{\eta_{e_L}(x)}{2}+\frac{3}{8}\eta_{B}(x)\right)\left[\left(\frac{B_{z}(x)}{10^{20}G}\right)^{2}+\left(\frac{B_{a}(x)}{10^{20}G}\right)^{2}+\left(\frac{B_{b}(x)}{10^{20}G}\right)^{2}\right]x^{3/2}\\&-\frac{3F_{2}}{2}\left(\eta_{e_R}^{2}(x)-\eta_{e_L}^{2}(x)\right)\left[v_{a}(x)\left(\frac{B_{a}(x)}{10^{20}G}\right)+v_{b}(x)\left(\frac{B_{b}(x)}{10^{20}G}\right)\right]\sqrt{x},
\end{split}
\end{equation}
\begin{equation}\label{eq35}
 \frac{dB_{z}}{dx}=-\frac{B_{z}(x)}{x},
\end{equation}
\begin{equation}\label{eq36}
  \begin{split} 
\frac{dB_{a}(x)}{dx}=&\frac{356k^{\prime\prime}}{\sqrt{x}}\left[F_3\left(\eta_{e_R}(x)-\frac{\eta_{e_L}(x)}{2}+\frac{3}{8}\eta_{B}(x)\right)-\frac{k^{\prime\prime}}{10^{3}}\right]B_{a}(x)\\&+\frac{F_4}{\sqrt{x}}\bigg[v_{b}(x)B_{z}(x)
\bigg]+F_{5}\left(\eta_{e_R}^{2}(x)-\eta_{e_L}^{2}(x)\right)\frac{v_{a}(x)}{x^{3/2}}-\frac{B_{a}(x)}{x},
\end{split}
\end{equation}
\begin{equation}\label{eq37}
 \begin{split} 
 \frac{dB_{b}(x)}{dx}=&\frac{356k^{\prime\prime}}{\sqrt{x}}\left[F_3\left(\eta_{e_R}(x)-\frac{\eta_{e_L}(x)}{2}+\frac{3}{8}\eta_{B}(x)\right)-\frac{k^{\prime\prime}}{10^{3}}\right]B_{b}(x)\\&-\frac{F_{4}}{\sqrt{x}}\bigg[v_{a}(x)B_{z}(x)%-v_{z}B_{a}(x)
 \bigg]+F_5\left(\eta_{e_R}^{2}(x)-\eta_{e_L}^{2}(x)\right)\frac{v_{b}(x)}{x^{3/2}}-\frac{B_{b}(x)}{x},
 \end{split}
\end{equation}

\begin{equation}\label{eq39}
  \begin{split}
  \frac{dv_{a}(x)}{dx}=&F_6\left(\frac{B_{z}(x)}{10^{20}G}\right)\left(\frac{B_{b}(x)}{10^{20}G}\right)x^{3/2}
-\frac{F_{7}}{\alpha_{Y}^{2}\sqrt{x}}v_{a}(x),
  \end{split}
\end{equation}
\begin{equation}\label{eq40}
 \begin{split}
 \frac{dv_{b}(x)}{dx}=& 
 -F_{6}\left(\frac{B_{z}(x)}{10^{20}G}\right)\left(\frac{B_{a}(x)}{10^{20}G}\right)x^{3/2}
-\frac{F_{7}}{\alpha_{Y}^{2}\sqrt{x}}v_{b}(x),
 \end{split}
\end{equation}
where $\alpha_{Y}={g^{\prime}}^{2}/4\pi$, and the coefficients  $F_{i}, i=0,...,7$ are given in the table (\ref{table1}),  and we have used the relation $k^{\prime\prime}=k/10^{-7}$, as well \cite{{37.1},{41},{banerjee}}.
%\begin{table}[H]
% \begin{center}
%\begin{tabular}{|c|c|c|c|c|c|c|}
%	\hline
%	 $F_{0}$& $F_{1}$ & $F_{2}$&$F_3$ \\
%	\hline
%	 $(9.176\times10^{-6}) k^{\prime\prime}$&77.79& $(15.83 )k^{\prime\prime}$&$8477.6$ \\
%	\hline
%	$F_{4}$&$F_{5}$&$F_{6}$&$F_{7}$\\
%	\hline
%	$356\times10^{6}k^{\prime\prime}$&($6.1419\times10^{25}){k^{\prime\prime}}^{2}$&$0.304k^{\prime\prime}$&7.12 ${k^{\prime\prime}}^{2}$\\
%	\hline
%\end{tabular}\label{table1}
%\end{center}
%\end{table}

\begin{table}[h]
	\begin{center}
		\caption{}
		\begin{tabular}{|c|c|}
				\hline
			$F_{0}$&$(9.176\times10^{-6}) k^{\prime\prime}$	\\
				\hline
				 $F_{1}$ &77.79\\
				 \hline
				  $F_{2}$&$(15.83 )k^{\prime\prime}$\\
				  \hline
				  $F_3$&$8477.6$\\
				\hline
				$F_{4}$&$3.56\times10^{8}k^{\prime\prime}$\\
				\hline
			$F_{5}$&($6.1419\times10^{25}){k^{\prime\prime}}^{2}$\\
				\hline
				$F_{6}$&$0.304k^{\prime\prime}$\\
				\hline
		$F_{7}$& $7.12{k^{\prime\prime}}^{2}$	\\
				\hline
			\end{tabular}\label{table1}
	\end{center}
\end{table}

In the next section, we solve this set of coupled differential equations numerically. In particular, we explore how the non-helical hypermagnetic field can affect the evolution of the vorticity and velocity fields, as well as the matter-antimatter asymmetries.  In fact, two of the three scenarios that we explore in the next section are possible only in the presence of non-helical hypermagnetic fields.
%
%\textcolor{red}{} \textcolor{green}{} \textcolor{green}{,} \textcolor{orange}{()} \textcolor{purple}{}
%

\section{Numerical Solution}\label{x3}
%%%%%%%%%%%%%%%%%%%%%%%%%%%%%%%%%%%%%%%%%%%%%%%%%%%
In this section, we solve the evolution equations obtained in Sec.\ \ref{x2} in the temperature range $100 \mbox{GeV}\leq 
T\leq10 \mbox{TeV}$ and in the presence of viscosity. As mentioned earlier, we have chosen monochromatic Chern-Simons configuration for the hypermagnetic and velocity fields with the length scale $2\pi\sqrt{x}/(k T_{EW})$, where $k$ is the comoving wave number.

%\subsection{\small Magnetogenesis due to lepton asymmetry in the presence of the chiral vorticity }\label{x4}
\subsection{ Generation of helicity, vorticity and baryon asymmetry by a large lepton asymmetry and strong non-helical hypermagnetic field}\label{x4}
%Let us investigate the possibility to produce and grow the helicity for a hypermagnetic field which is initially completely non-helical, i.e. $B_{a}^{(0)}=B_{b}^{(0)}=0$ and $B_{z}^{(0)}\neq0$. 
Let us consider a hypermagnetic field which is initially completely non-helical, {\it i.e.,} $B_{a}^{(0)}=B_{b}^{(0)}=0$ and $B_{z}^{(0)}\neq0$, and investigate the possibility to produce its helical components. To accomplish this task, non-zero initial  vorticity is needed which can be produced by non-zero initial $v_{a}$ or $v_{b}$. Given $v_{a}^{(0)}=v_{b}^{(0)}=0$, the vorticity $\omega$ freezes at zero and has no growth, since $v_{a}$ and $v_{b}$ will stay at zero according to Eqs.\ (\ref{eq39},\ref{eq40}). As a result, $B_{a}$ and $B_{b}$ will also remain zero due to Eqs.\ (\ref{eq36},\ref{eq37}). That is, neither vorticity nor helicity can be produced .

\begin{figure}[H]
	\centering
	\subfigure[]{\label{fig:figure:in2} 
	\includegraphics[width=.365\textwidth]{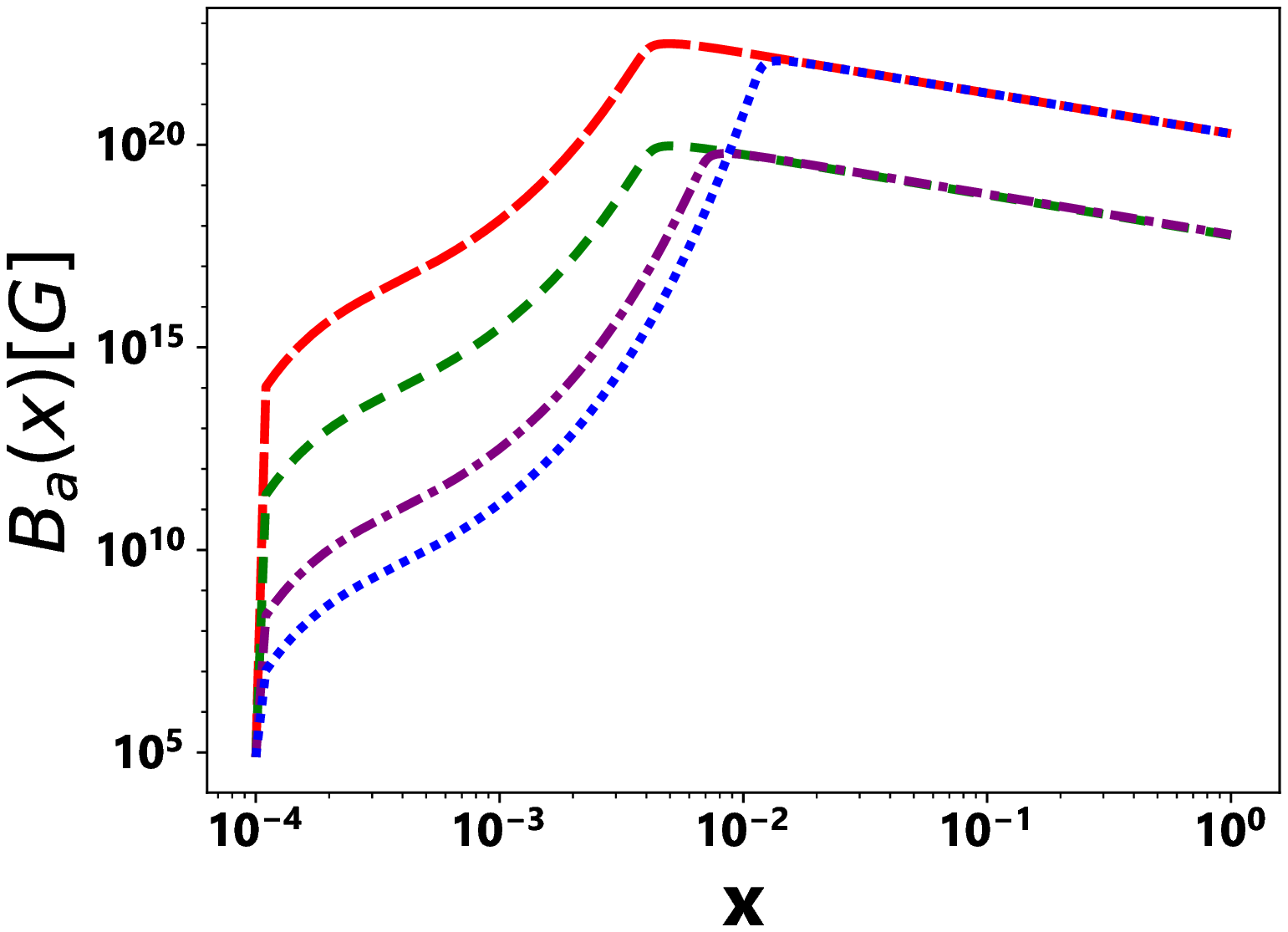}}
	\hspace{10mm}
	\subfigure[]{\label{fig:figure:in211} 
		\includegraphics[width=.36665\textwidth]{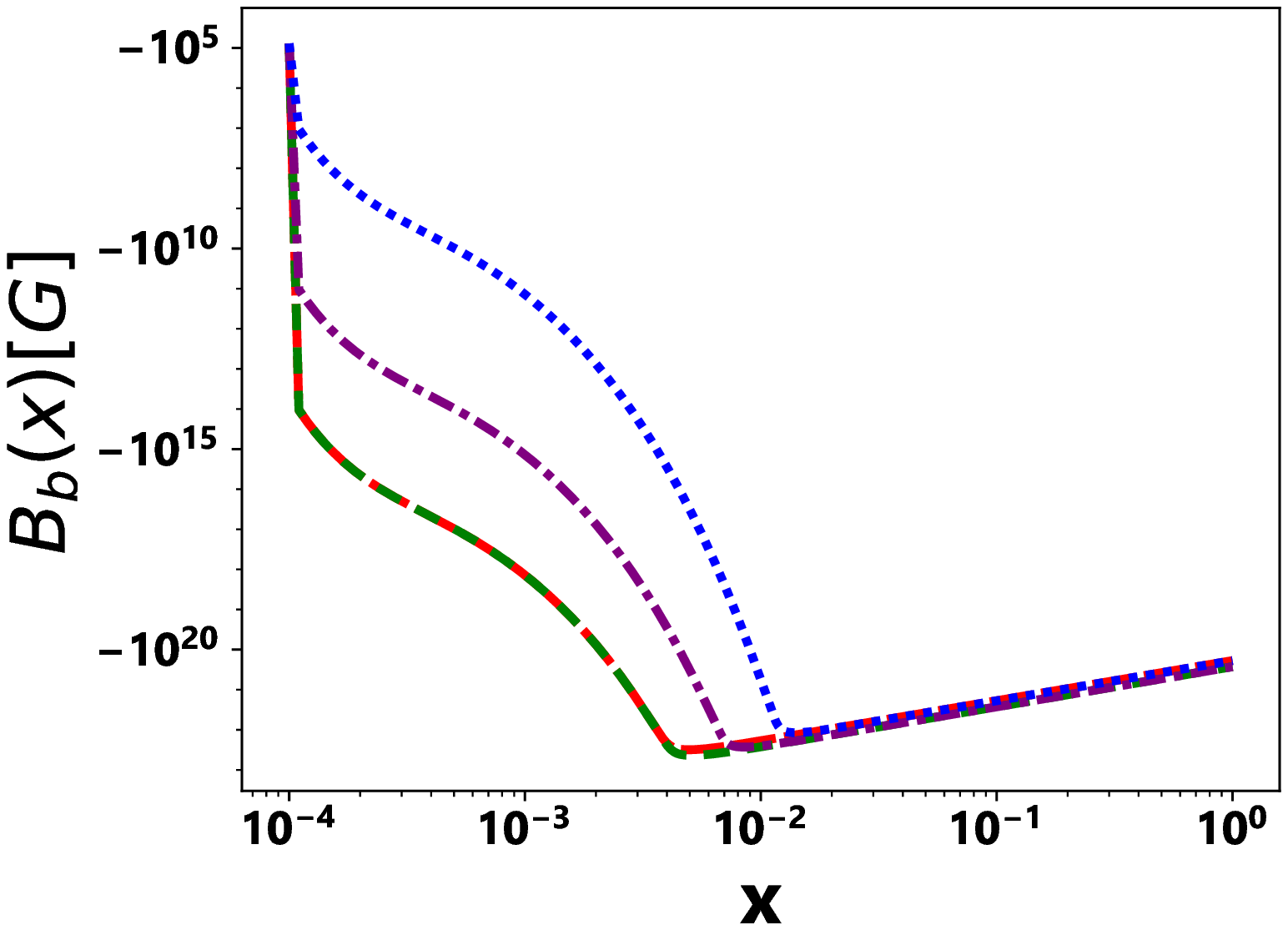}}
  \hspace{8mm}
	\subfigure[]{\label{fig:figure:in222} 
		\includegraphics[width=.365\textwidth]{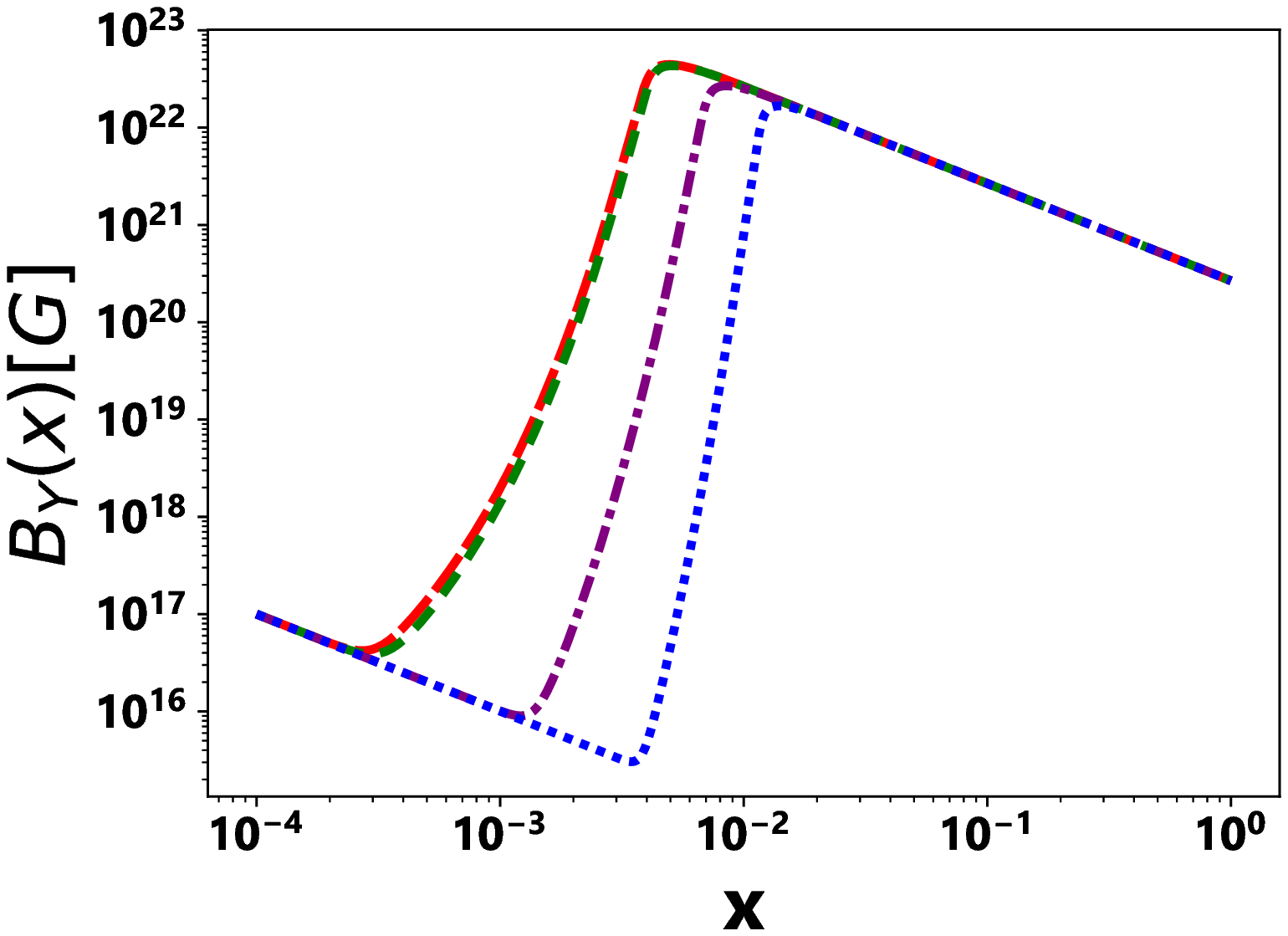}}
	\hspace{8mm}
	\subfigure[]{\label{fig:figure:1.1.31}
	\includegraphics[width=.365\textwidth]{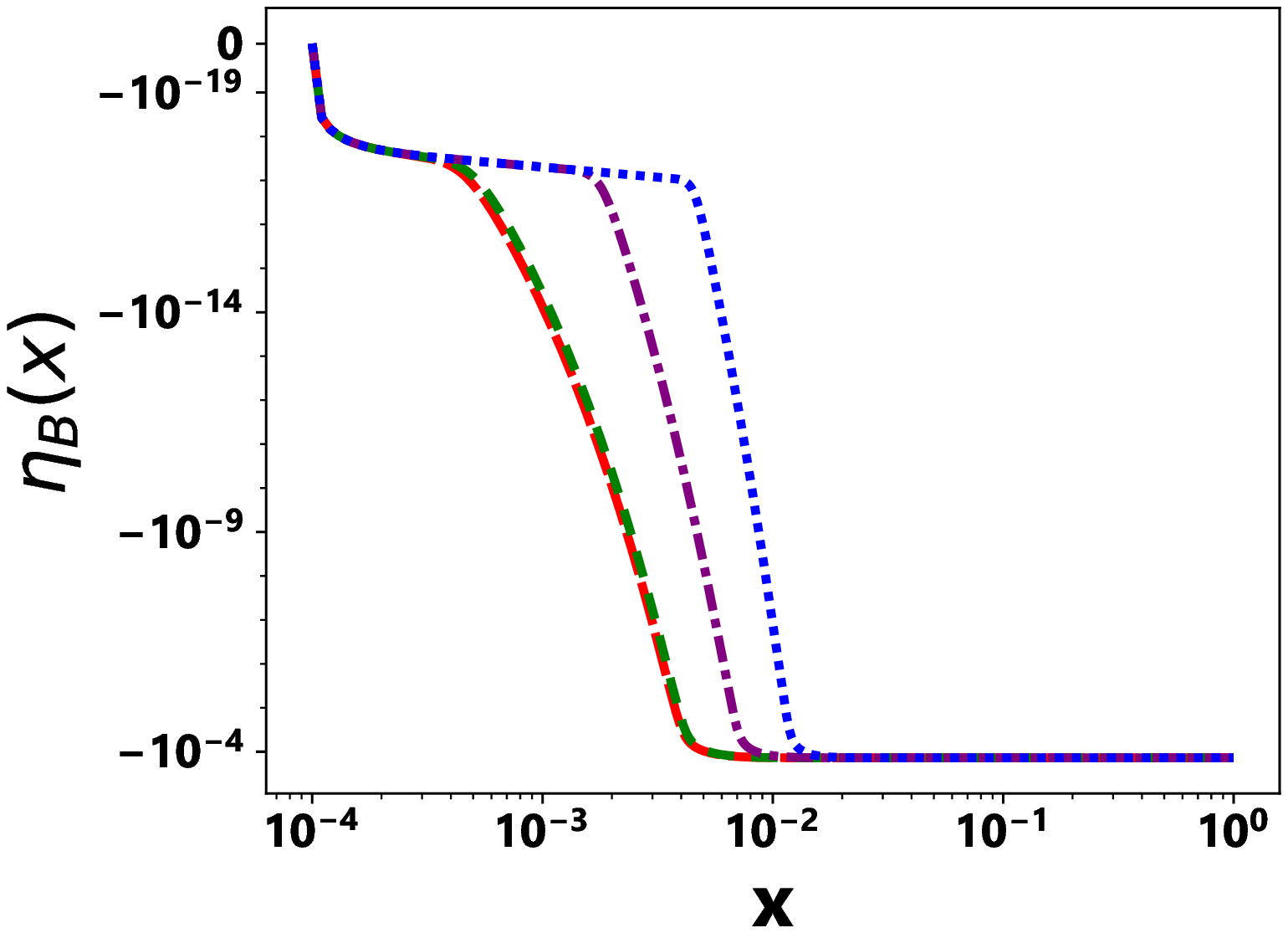}}
	\hspace{8mm}
	\subfigure[]{\label{fig:figure:1.1.21}
		\includegraphics[width=.365\textwidth]{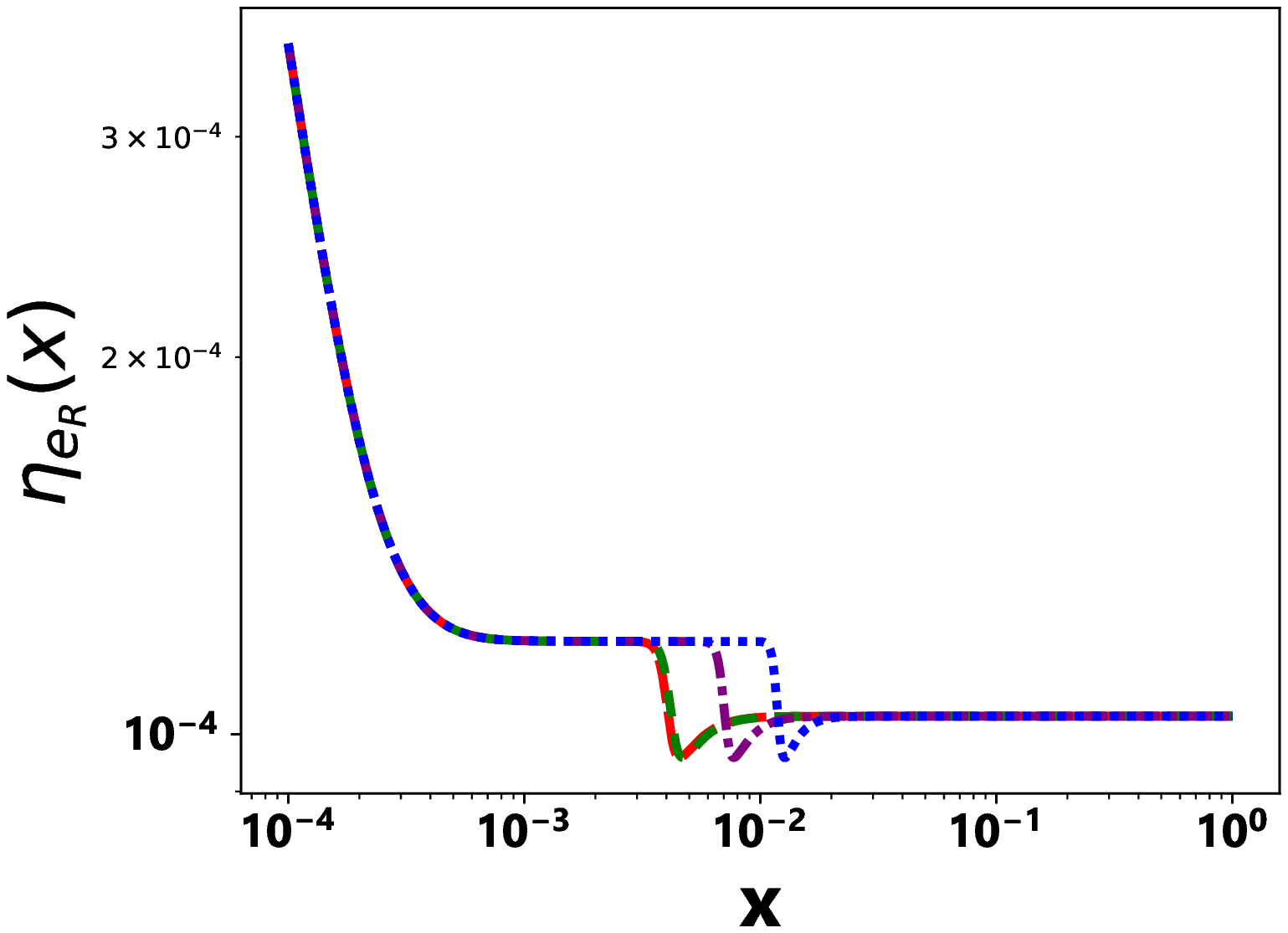}}
	\hspace{8mm}
	\subfigure[]{\label{fig:figure:1.1.31121}
		\includegraphics[width=.365\textwidth]{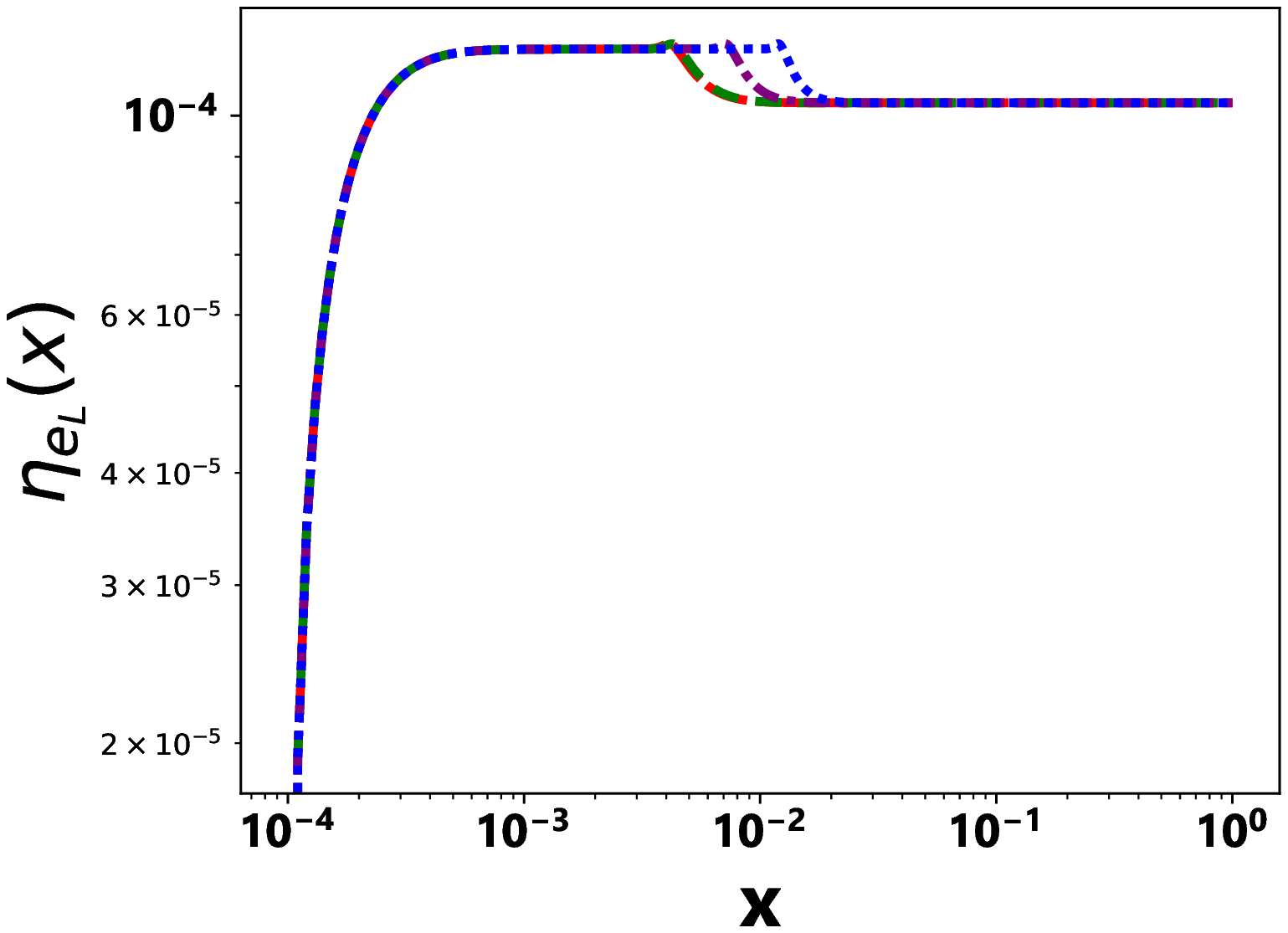}}
	\hspace{8mm}
	\subfigure[]{\label{fig:figure:1va}
		\includegraphics[width=.365\textwidth]{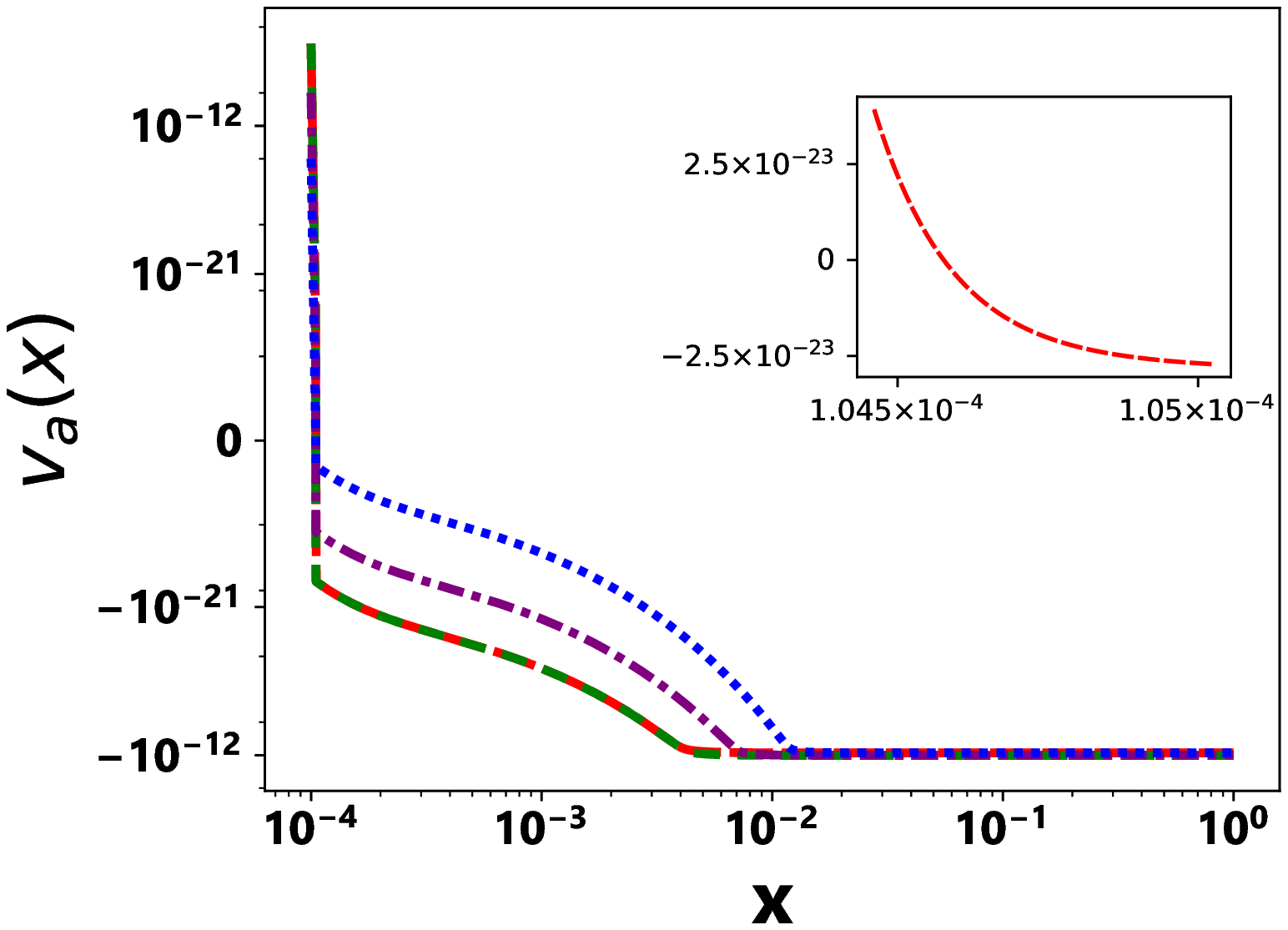}}
	\hspace{8mm}
	\subfigure[]{\label{fig:figure:1vb}
		\includegraphics[width=.365\textwidth]{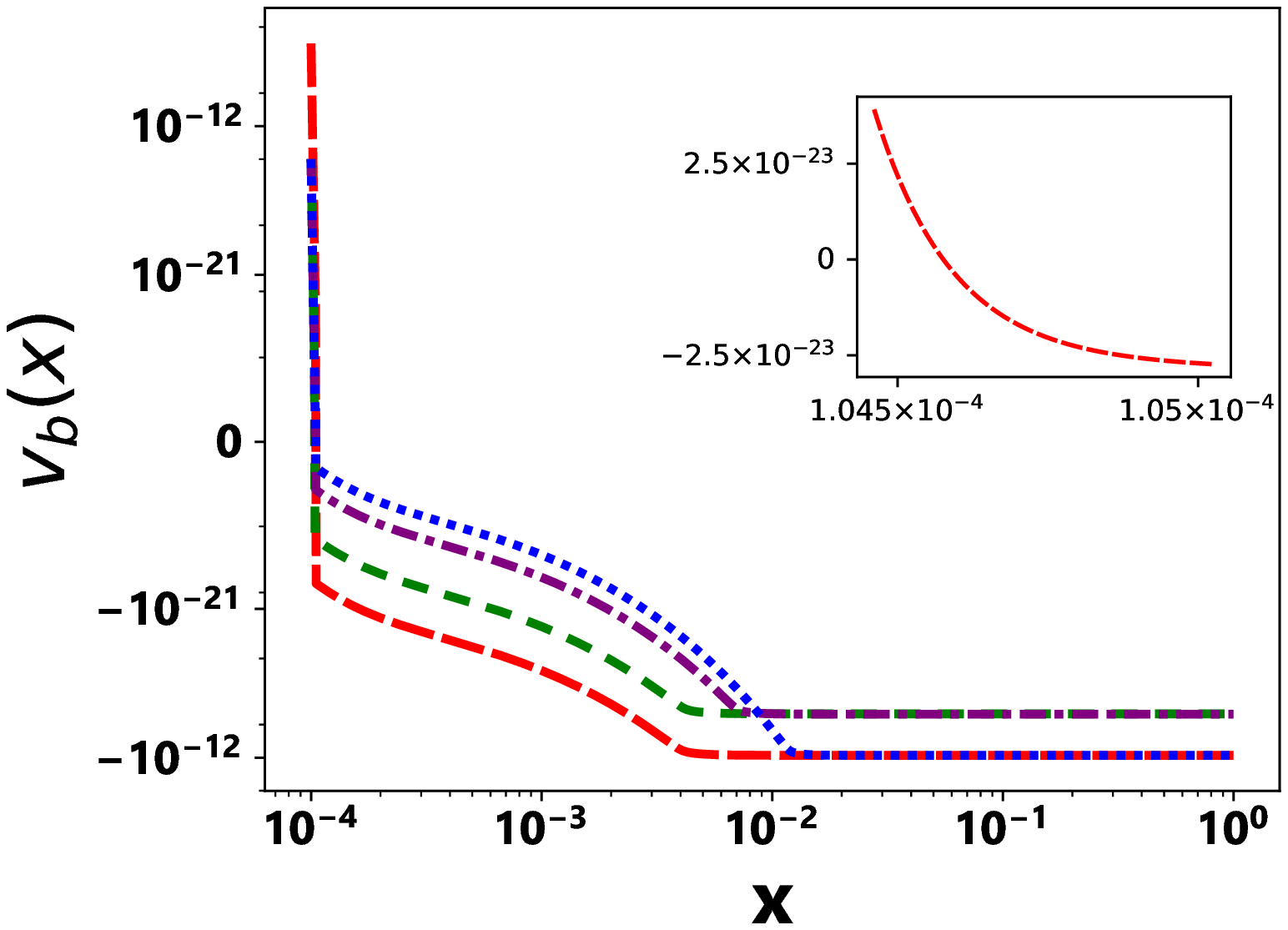}}
	\caption{\footnotesize The time plots of the helical components $B_{a}(x)$ and $ B_{b}(x)$, the hypermagnetic field amplitude $B_{Y}(x)$, the baryon asymmetry $\eta_{B}(x)$, the right-handed electron asymmetry $\eta_{e_R}(x)$, the left-handed electron asymmetry $\eta_{e_L}(x)$, and $v_{a}(x)$ and $v_{b}(x)$ in the presence of the viscosity, with the initial conditions $B_{z}^{(0)}=10^{17} \mbox{G}$, $B_{a}^{(0)}=B_{b}^{(0)}=0$, $\eta_{e_R}^{(0)}=3.56\times10^{-4}$, and $\eta_{e_L}^{(0)}=\eta_{B}^{(0)}=0$. Large dashed (red) line is for, $v_{a}^{(0)}=v_{b}^{(0)}=10^{-7}$, dashed (green) line for $v_{a}^{(0)}=10^{-7}$, $v_{b}^{(0)}=10^{-14}$, dot-dashed (violet) line for $v_{a}^{(0)}=10^{-10}$, $v_{b}^{(0)}=10^{-14}$, and dotted (blue) line for $v_{a}^{(0)}=v_{b}^{(0)}=10^{-14}$.
}\label{fig1.1}
\end{figure}

We first solve the set of coupled differential equations with the initial conditions $k=10^{-7}$, $B_{z}^{(0)}=10^{17}$G, $B_{a}^{(0)}=B_{b}^{(0)}=0$, $\eta_{e_R}^{(0)}=3.56\times10^{-4}$, $\eta_{e_L}^{(0)}=\eta_{B}^{(0)}=0$, and four different sets of values for $v_{a}^{(0)}$ and $v_{b}^{(0)}$.
The results are shown in Fig.\ \ref{fig1.1}. 
It can be seen that the helical components of the hypermagnetic field, $B_{a}$ and $B_{b}$, are generated and amplified from zero initial values. 
The seed for $B_{a}$ ($B_{b}$) can be created due to the second or third terms on the rhs of Eq.\ (\ref{eq36}) (Eq.\ (\ref{eq37})). The former comes from the advection term $\frac{1}{R}\vec{\nabla}\times(\vec{v}\times\vec{B}_{Y})$, and the latter is the chiral vortical term which is responsible for the CVE \cite{saeed}. The effect of vorticity via $v_a$ or $v_b$ appears in both terms; while, the effect of $B_z$ shows up in the former and that of  $\eta_{e_R}^{2}(x)-\eta_{e_L}^{2}(x)$ in the latter. 
Moreover, the first term on the rhs of Eq.\ (\ref{eq36}) (Eq.\ \ref{eq37})) consists of two parts: the CME part and a non-CME part. The CME part is proportional to a signature combination of asymmetries which also appears in the evolution equations for the asymmetries themselves, and we denote by $\Delta \eta := \eta_{e_R}(x)-\eta_{e_L}(x)/2+(3/8)\eta_{B}(x)\sim \sqrt{x}c_\mathrm{B}(x)$, as indicated in Eq.\ (\ref{eq13}). The CME term leads to the growth of the seed of the helical component, while the non-CME one contributes to the saturation of its value \cite{saeed,34}.

Figure \ref{fig1.1} shows that $v_a$, $v_b$, and therefore the vorticity $\omega$, quickly drop due to viscosity, represented by the second terms in  Eqs.\ (\ref{eq39},\ref{eq40}). However, they cross zero and their amplitudes continue to grow due to the first terms in  Eqs.\ (\ref{eq39},\ref{eq40}), which originate from the term $\vec{J}\times\vec{B}_{Y}/(\rho+p)$ in Eq.\ (\ref{eq9}) and signify the back-reaction of the hypermagnetic field on the plasma. As can be seen from Eqs.\ (\ref{eq39},\ref{eq40}), these terms are nonzero only if the hypermagnetic field contains both nonzero helical ($B_a$ and $B_b$) and non-helical ($B_z$) components. We shall henceforth refer to these terms as JB terms. The growth of the amplitudes of $v_a$, $v_b$, and $\omega$ is successful, in spite of the presence of the extremely large viscosity, since not only $B_z$ is large but also $B_a$ and $B_b$ grow due to the CME. Finally, the growth of $B_a$ and $B_b$ stops when $\Delta \eta$, which is their sole growth factor in this case, reaches its minimum and this reduces the first terms in Eqs.\ (\ref{eq36},\ref{eq37}) to zero.
Meanwhile, some of the initial right-handed electron asymmetry $\eta_{e_R}^{(0)}$ is converted into left-handed electron asymmetry $\eta_{e_L}^{(0)}$ by the Higgs chirality flip processes until they equilibrate around $x=10^{-3}$. Then the chirality flip terms, as well as the CVE terms in the evolution equations for $\eta_{e_R}(x)$, $\eta_{e_L}(x)$, and $\eta_{B}(x)$ in Eqs.\ (\ref{eq32},\ref{eq33},\ref{eq34}) go to zero. During this time baryon asymmetry $\eta_B$ continues building up until $\Delta \eta$ reaches its minimum, which makes the sum of the remaining two terms in the evolution equations for $\eta_{e_R}(x)$, $\eta_{e_L}(x)$, and $\eta_{B}(x)$ in Eqs.\ (\ref{eq32},\ref{eq33},\ref{eq34}) go exactly to zero. Then, all three asymmetries reach their constant saturation curves, and $\Delta \eta$ remains constant at its minimum. This makes $B_a$ and $B_b$, and consequently $v_a$, $v_b$, reach their saturation curves.	
We should mention that the saturation curves for the comoving variables are horizontal lines, while those of other variables are inclined lines on the logarithmic scale due to the expansion of the Universe.  

Let us investigate the effect of increasing the initial vorticity on the evolution, when $v_{a}^{(0)}=v_{b}^{(0)}$,\footnote{In this case, the amplitude of $v_{a}$ and $v_{b}$ will remain equal during the evolution due to the symmetric form of the equations with respect to $v_{a}$ and $v_{b}$.} by comparing the two cases, $v_{a}^{(0)}=v_{b}^{(0)}=10^{-14}$ and $v_{a}^{(0)}=v_{b}^{(0)}=10^{-7}$. 
Given the aforementioned values for $B_{z}^{(0)}$ and $\eta_{e_R}^{(0)}$, when $v_{a}^{(0)}=v_{b}^{(0)}$, both of the advection terms are much larger than the chiral vortical terms in Eqs.\ (\ref{eq36},\ref{eq37}). 
%\textcolor{blue}{This is correct, pleas see the Fig \ref{fig1.w}, which show the dominant effect of the advection term in respect to CVE one}. 
Therefore, they are responsible for the production of the seeds for the helical components $B_{a}$ and $B_{b}$, while CME is mainly responsible for their subsequent growth. As can be seen from Eqs.\ (\ref{eq36},\ref{eq37}), $v_{a}^{(0)}$ and $v_{b}^{(0)}$ appear in the advection terms for $B_{b}$ and $B_{a}$, respectively, with opposite signs. Therefore, the seeds that they produce for $B_{b}$ and $B_{a}$, have opposite signs. Incidentally, this is in contrast to the CVE terms for $B_{a}$ and $B_{b}$ in which  $v_{a}^{(0)}$ and $v_{b}^{(0)}$ appear, respectively, and with the same sign. On the other hand, Eqs.\ (\ref{eq39},\ref{eq40}) show that $B_{a}^{(0)}$ and $B_{b}^{(0)}$ appear in the JB terms for $v_{b}$ and $v_{a}$, respectively, with opposite signs. Since the strength of the seeds depend on the values of $v_{a}^{(0)}=v_{b}^{(0)}$ and $B_z^{(0)}$, they become stronger by increasing $v_{a}^{(0)}=v_{b}^{(0)}$, as can be seen in Fig.\ \ref{fig1.1}. The stronger the seeds, the sooner they reach their maxima on the saturation curves, as a result of the growth due to the CME.
%
%\textcolor{red}{} \textcolor{green}{} \textcolor{green}{,} \textcolor{orange}{()} \textcolor{purple}{}
%
%The seeds then grow due to the CME until they reach their maximum values and the stronger ones reach their maxima earlier. 
That is, the maximum values of $B_{a}$, $B_{b}$, and therefore all other variables plotted in Fig.\ \ref{fig1.1}, occur at a higher temperature. However, their final values, at the onset of the electroweak phase transition (EWPT), are almost independent of $v_{a}^{(0)}=v_{b}^{(0)}$. 

Next, we analyze the case where $v_{b}^{(0)} << v_{a}^{(0)}$. As we have indicated before, given our initial conditions for $B_{z}^{(0)}$ and $\eta_{e_R}^{(0)}$, and assuming $v_{a}^{(0)} \approx v_{b}^{(0)}$, the advection terms are dominant over the CVE terms in  Eqs.\ (\ref{eq36},\ref{eq37}) for the evolution of $B_{a}$ and $B_{b}$. For $v_{b}<<v_{a}$, although the advection term  is still dominant for $B_{b}$, the CVE term is dominant for $B_{a}$, both of which contain $v_{a}$. The evolution of $v_{a}$ in turn is controlled by $B_{b}$ through the JB term, as is evident in Eq.\ (\ref{eq39}). As a result, for $v_{b}^{(0)} \leq v_{a}^{(0)}$ the evolution of $v_{a}$ and $B_{b}$ are intertwined with each other and almost independent from that of $v_{b}$, and therefore $B_{a}$. Figure \ref{fig1.1} shows that the evolution of $v_{a}$ and $B_{b}$ for the two cases $v_{a}^{(0)}=10^{-7}$ and $v_{b}^{(0)}=10^{-14}$, and  $v_{a}^{(0)}=v_{b}^{(0)}=10^{-7}$, are almost the same. That is, their evolution is almost completely independent of $v_{b}$. 
Meanwhile, the seed of $B_{a}$, produced by $v_{a}$ via the CVE term, is smaller than the seed of $B_b$, produced by $v_{a}$ via the advection term. As a result, $B_b$ remains much larger than $B_a$ at any instant of time during their growth due to the CME. Therefore, it is the dominant helical component which affects the evolution of all other variables, as can be seen in Fig.\ \ref{fig1.1}. More importantly, as it increases due to CME, it changes $\eta_{e_R}(x)$, $\eta_{e_L}(x)$, and $\eta_{B}(x)$ according to Eqs.\ (\ref{eq32},\ref{eq33},\ref{eq34}) until $\Delta \eta$ goes to its minimum which causes $\eta$s, $B_a$ and $B_b$ to hit their saturation curves. As a result of the latter two, the velocities saturate as well. Therefore, a smaller seed of $B_a$ compared to that of $B_b$, leads to smaller maximum and final values for it.
In the case $v_{a}^{(0)}=10^{-10}$ and $v_{b}^{(0)}=10^{-14}$, the same behavior can be observed, as Fig.\ \ref{fig1.1} shows, except that the temperature at which the concurrent transitions occur becomes smaller, due to the smaller value of the seed of $B_b$. 
As shown in Fig.\ \ref{fig1.1}, the seeds of $B_a$ and $B_b$ are both reduced by the same factor when $v_{a}^{(0)}$ is reduced, in accordance with our argument above.
 
Equation (\ref{eq35}) is the evolution equation of the non-helical part of the hypermagnetic field $B_{z}$. The solution to this equation is $B_{z}(x)=B_{z}^{(0)}(\frac{10^{-4}}{x})$, which shows that $B_{z}$ decreases only due to the expansion of the Universe. As discussed earlier, $B_{z}$ can play important roles in the production of the helical components $B_a$ and $B_b$ through the advection terms, and in the growth of $v_a$, $v_b$, and therefore $\omega$, through the JB terms in their corresponding evolution equations.  
It also affects the evolution of the matter asymmetries both directly and indirectly, through other variables, as indicated in Eqs.\ (\ref{eq32}-\ref{eq34}).
%
%\textcolor{red}{} \textcolor{green}{} \textcolor{green}{,} \textcolor{orange}{()} \textcolor{purple}{}
%
The evolution of the hypermagnetic field amplitude $B_Y$, the baryon asymmetry $\eta_B$, the right-handed electron asymmetry $\eta_{e_R}$ and the left-handed electron asymmetry $\eta_{e_L}$ are shown in Figs.\ \ref{fig:figure:in222}, \ref{fig:figure:1.1.31}, \ref{fig:figure:1.1.21} and \ref{fig:figure:1.1.31121}, respectively. At first, the produced helical components $B_a$ and $B_b$ are small; therefore, $B_Y \simeq B_{z}$ decreases due to the expansion of the Universe. The helical parts then grow due to the CME, leading to the growth of $B_Y$.

\begin{figure}[H]
	\centering
	\subfigure[]{\label{fig:figure:in12s} 
		\includegraphics[width=.365\textwidth]{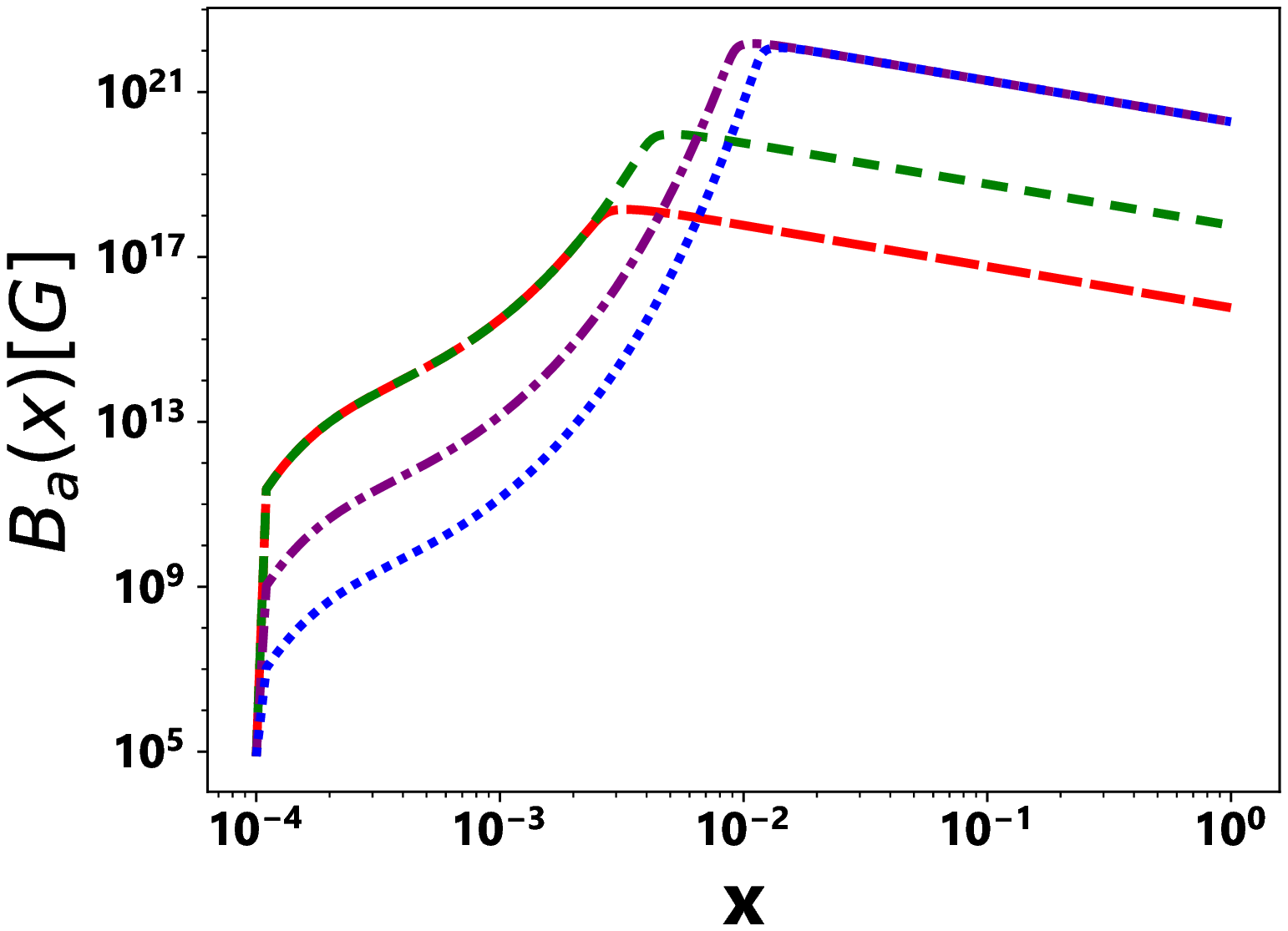}}
	\hspace{10mm}
	\subfigure[]{\label{fig:figure:in2111s} 
		\includegraphics[width=.36665\textwidth]{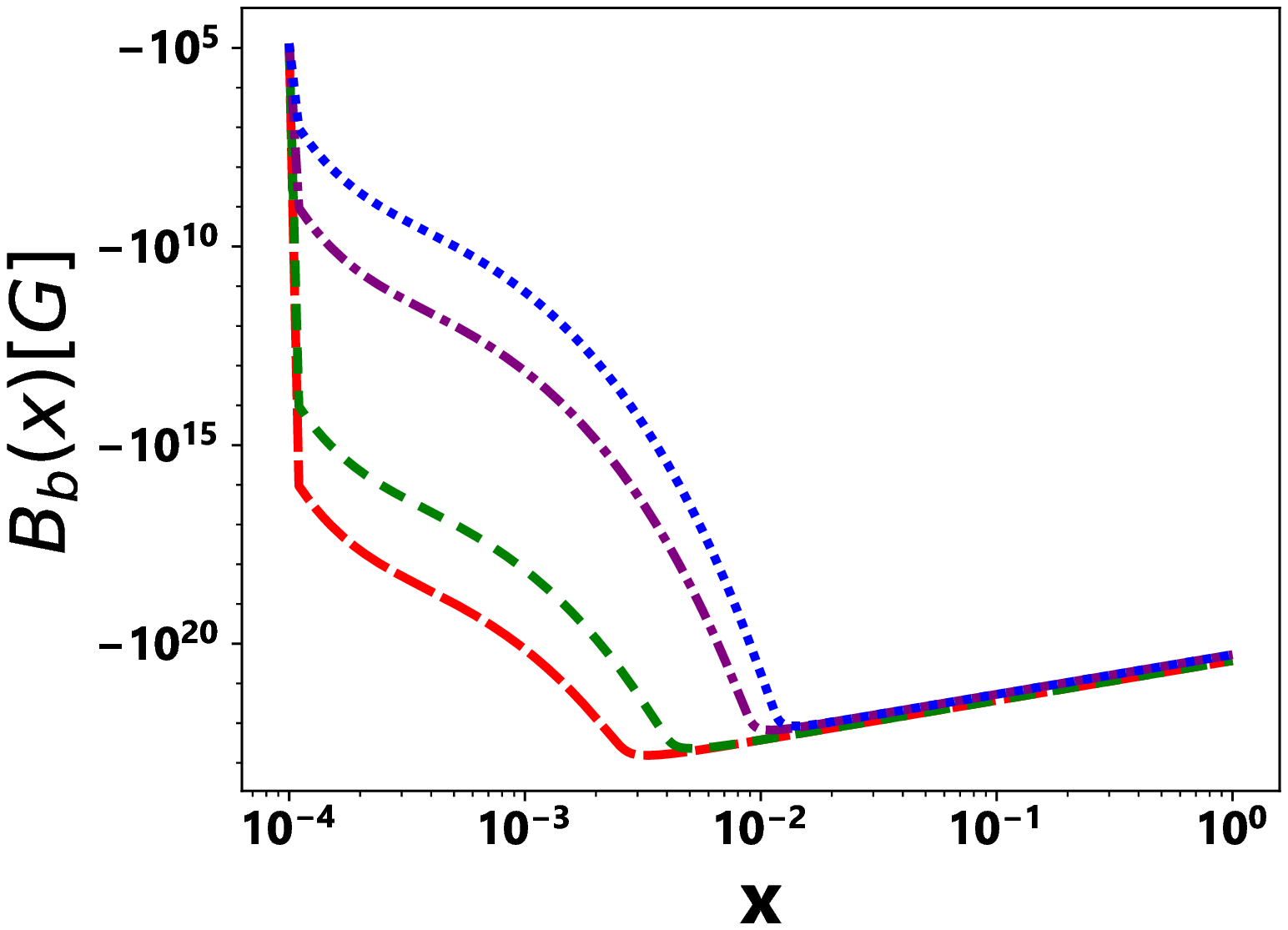}}
	\hspace{8mm}
	\subfigure[]{\label{fig:figure:in2212s} 
		\includegraphics[width=.365\textwidth]{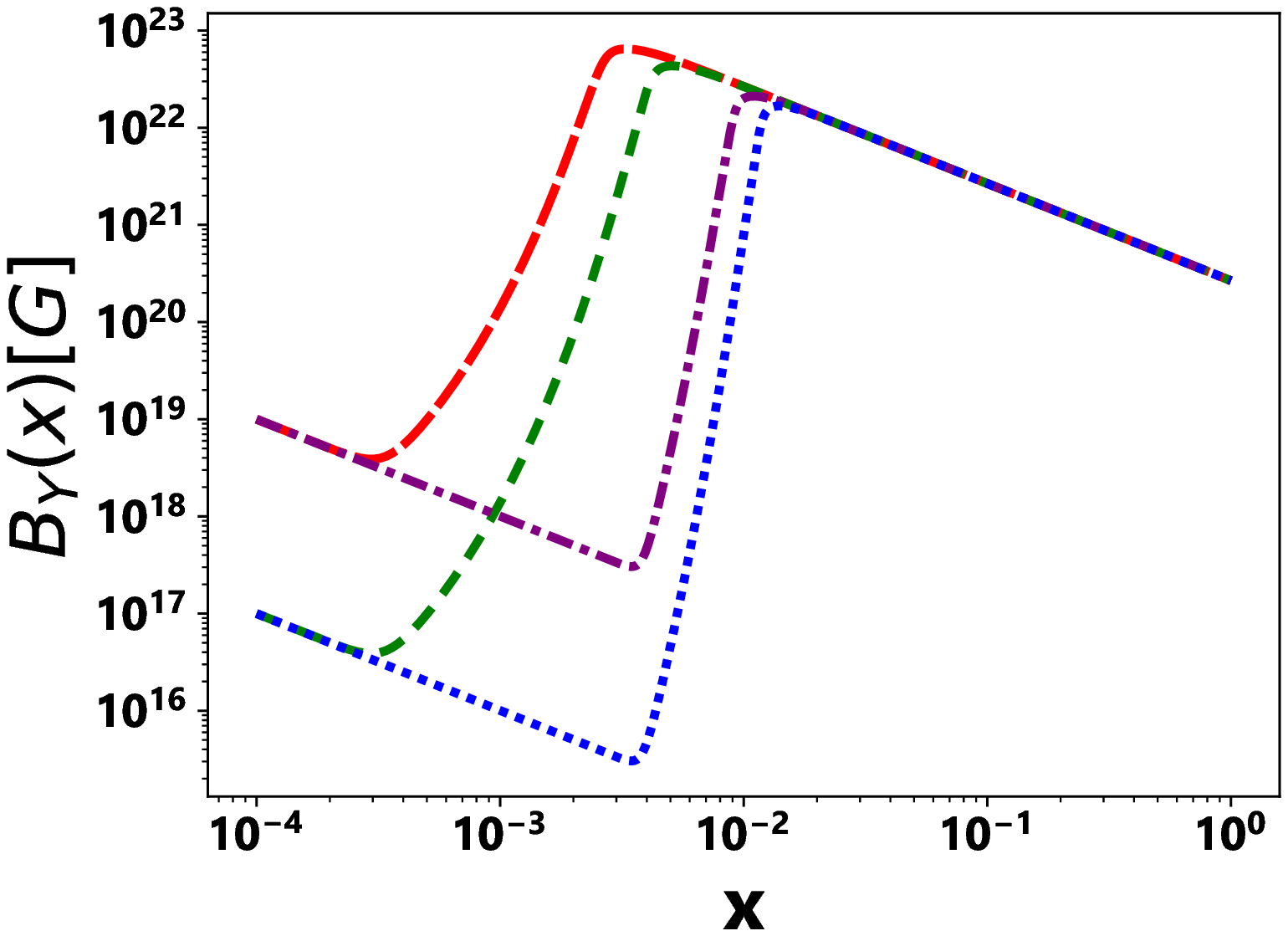}}
	\hspace{8mm}
	\subfigure[]{\label{fig:figure:1.1.311s}
		\includegraphics[width=.365\textwidth]{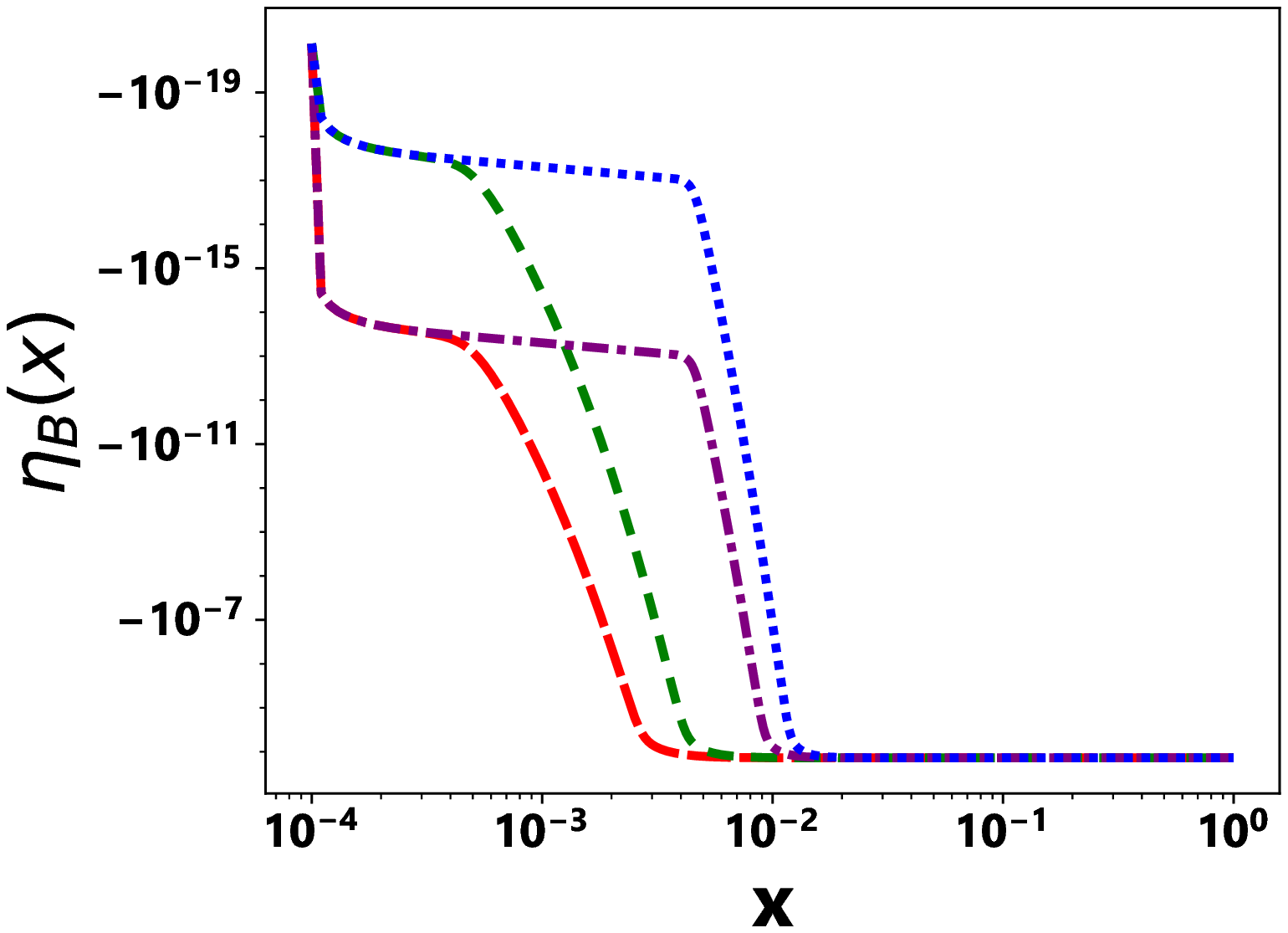}}
	\hspace{8mm}
	\subfigure[]{\label{fig:figure:1.1.211s}
		\includegraphics[width=.365\textwidth]{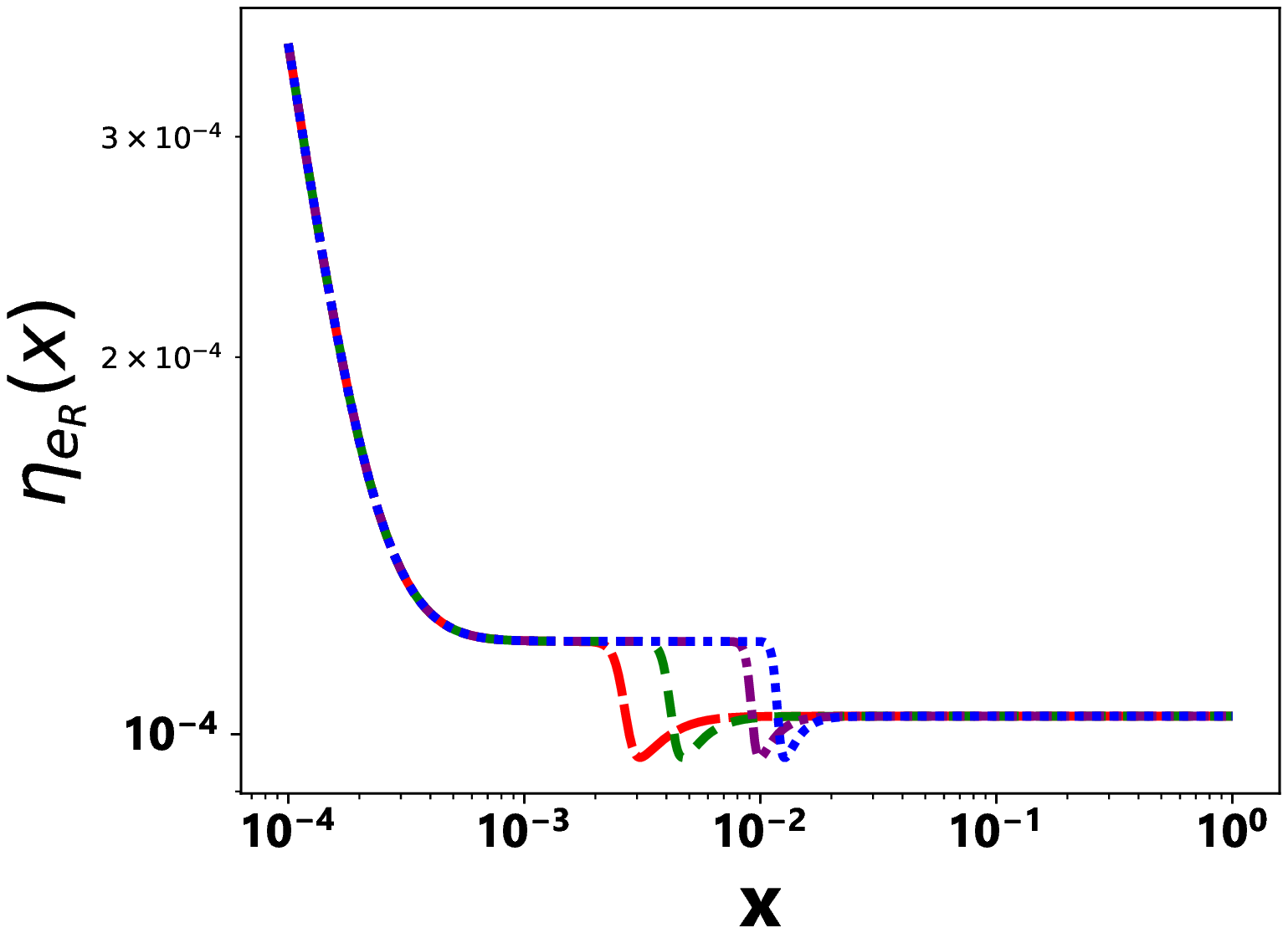}}
	\hspace{8mm}
	\subfigure[]{\label{fig:figure:1.1.311211s}
		\includegraphics[width=.365\textwidth]{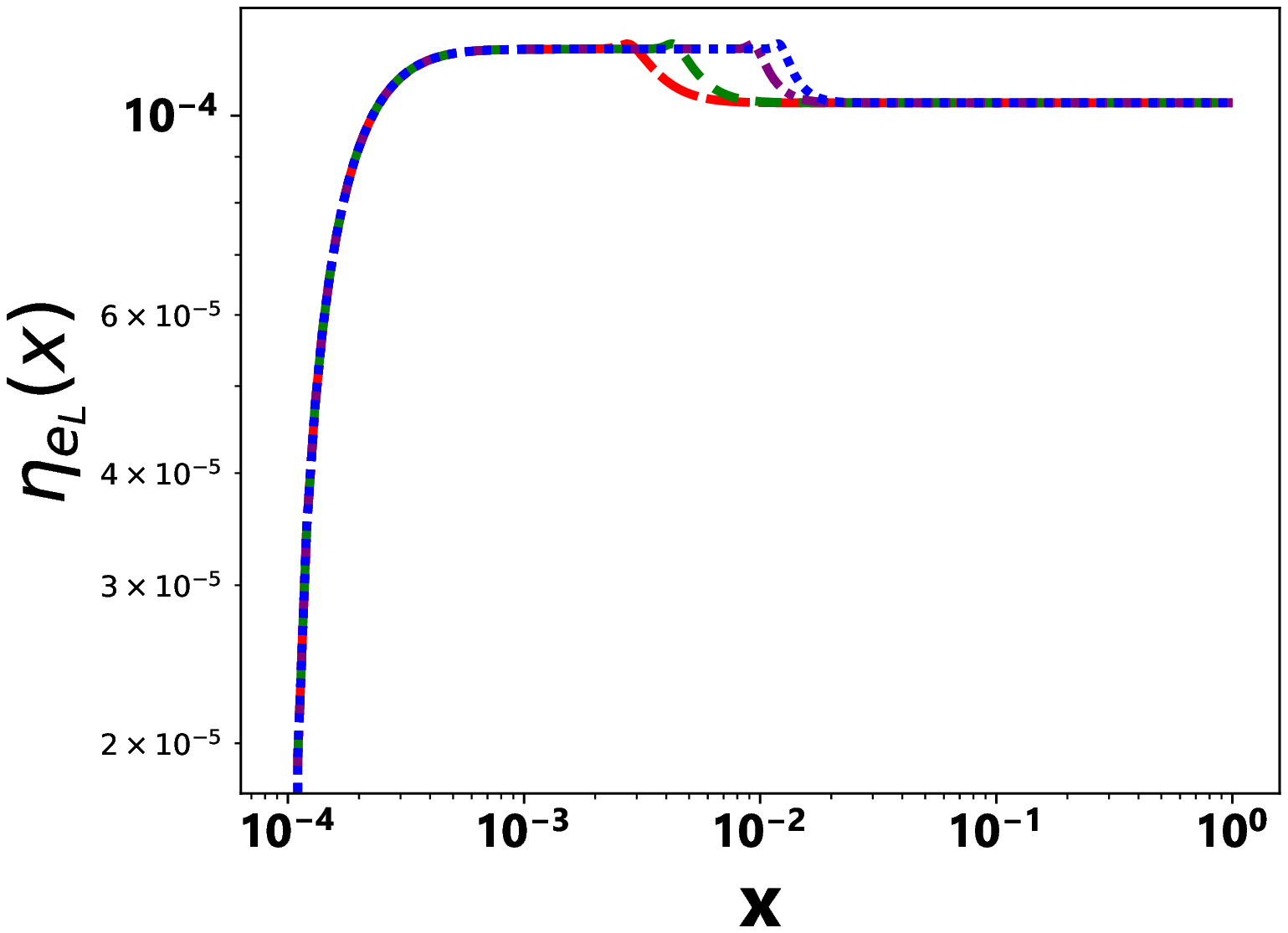}}
	\hspace{8mm}
	\subfigure[]{\label{fig:figure:1va1s}
		\includegraphics[width=.365\textwidth]{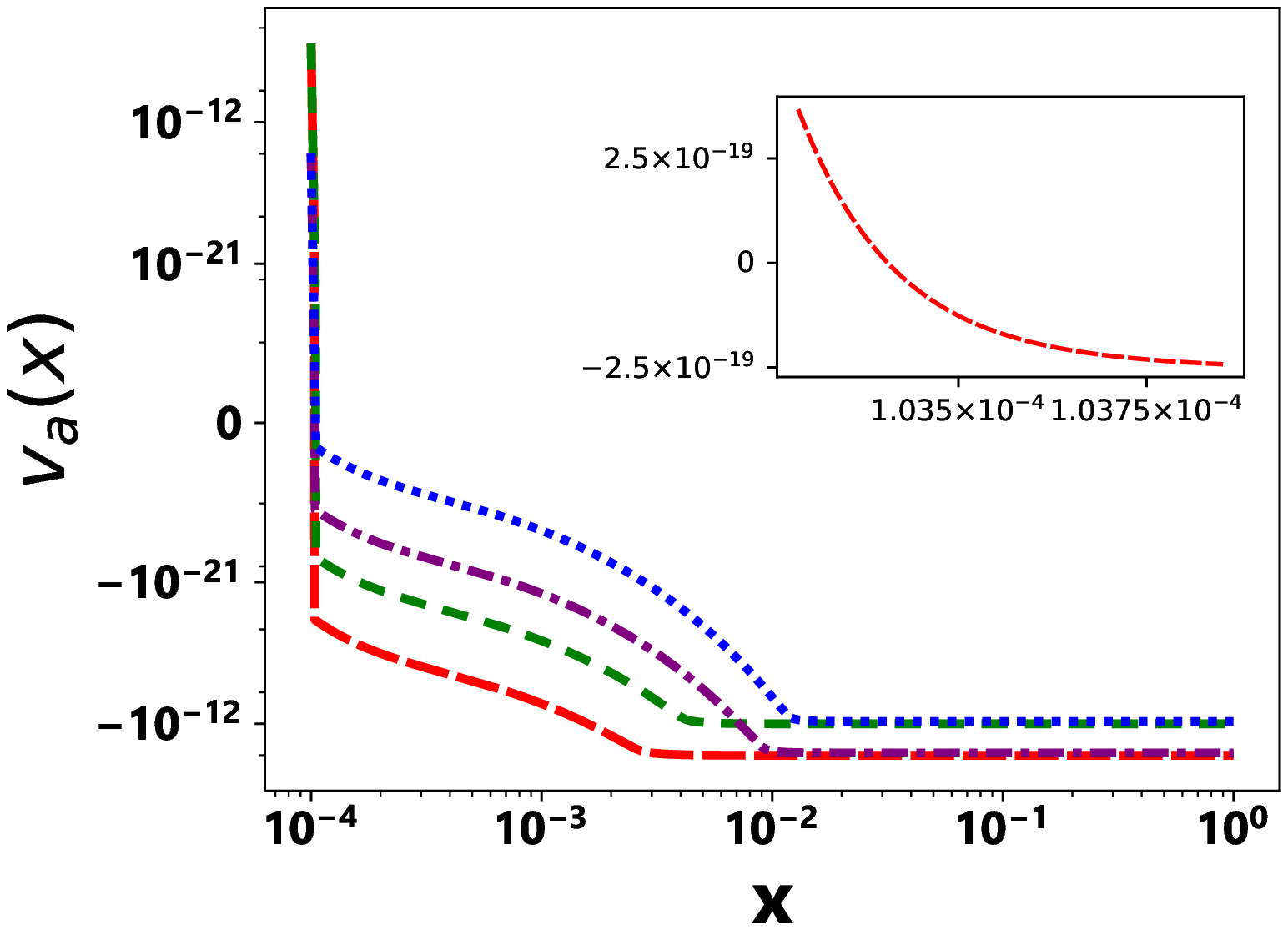}}
	\hspace{8mm}
	\subfigure[]{\label{fig:figure:1vb1s}
		\includegraphics[width=.365\textwidth]{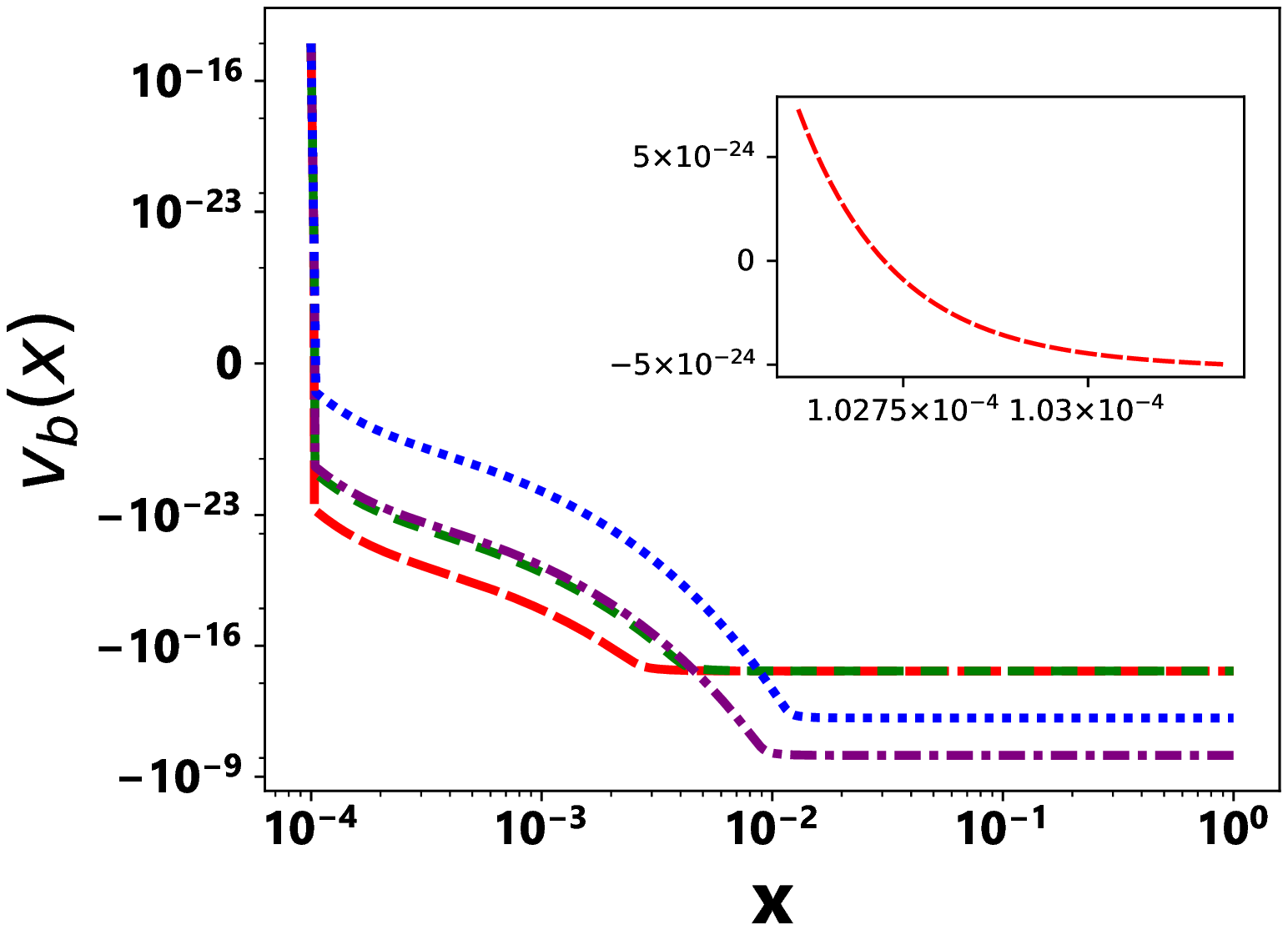}}
	\caption{\footnotesize
		The time plots of the helical components $B_{a}(x)$ and $ B_{b}(x)$, the hypermagnetic field amplitude $B_{Y}(x)$, the baryon asymmetry $\eta_{B}(x)$, the right-handed electron asymmetry $\eta_{e_R}(x)$, the left-handed electron asymmetry $\eta_{e_L}(x)$, and $v_{a}(x)$ and $v_{b}(x)$ with the initial conditions $B_{a}^{(0)}=B_{b}^{(0)}=0$, $\eta_{e_R}^{(0)}=3.56\times10^{-4}$, and $\eta_{e_L}^{(0)}=\eta_{e_B}^{(0)}=0$. Large (red) dashed line is for $v_{a}^{(0)}=10^{-7}$, $v_{b}^{(0)}=10^{-14}$, and $B_{z}^{(0)}=10^{19} \mbox{G}$, dashed (green) line for $v_{a}^{(0)}=10^{-7}$, $v_{b}^{(0)}=10^{-14}$, and $B_{z}^{(0)}=10^{17} \mbox{G}$, dotted-dashed (violet) line for $v_{a}^{(0)}=v_{b}^{(0)}=10^{-14}$, and $B_{z}^{(0)}=10^{19} \mbox{G}$, and dotted (blue) line for $v_{a}^{(0)}=v_{b}^{(0)}=10^{-14}$, and $B_{z}^{(0)}=10^{17} \mbox{G}$. %The time plots of the helical components $B_{a}(x)$ and $\big| B_{b}(x)\big|$, the hypermagnetic field amplitude $B_{Y}(x)$, the baryon asymmetry $\big|\eta_{B}(x)\big|$, the ratio of the helical part to the non-helical part $\sqrt{B_{a}^{2}(x)+B_{b}^{2}(x)}/B_{z}(x)$, the vorticity $\omega(x)$, and $\big|v_{a}(x)\big|$ and $\big|v_{b}(x)\big|$ with the initial conditions $B_{a}^{(0)}=B_{b}^{(0)}=0$, $y_{R}^{(0)}=10^{3}$, and $y_{L}^{(0)}=y_{B}^{(0)}=0$. Large (red) dashed line is for $v_{a}^{(0)}=10^{-7}$, $v_{b}^{(0)}=10^{-14}$, and $B_{z}^{(0)}=10^{19} \mbox{G}$, dashed (green) line for $v_{a}^{(0)}=10^{-7}$, $v_{b}^{(0)}=10^{-14}$, and $B_{z}^{(0)}=10^{17} \mbox{G}$, dotted-dashed (orange) line for $v_{a}^{(0)}=v_{b}^{(0)}=10^{-14}$, and $B_{z}^{(0)}=10^{19} \mbox{G}$, and dotted (blue) line for $v_{a}^{(0)}=v_{b}^{(0)}=10^{-14}$, and $B_{z}^{(0)}=10^{17} \mbox{G}$.
	}\label{fig1.11s}
\end{figure}

Let us now investigate the effect of varying the value of $B_{z}^{(0)}$ on the evolution. We solve the set of coupled differential equations with the initial conditions $k=10^{-7}$, $B_{a}^{(0)}=B_{b}^{(0)}=0$, $\eta_{e_R}^{(0)}=3.56\times10^{-4}$,  $\eta_{e_L}^{(0)}=\eta_{B}^{(0)}=0$, two values for $B_{z}^{(0)}$, {\it i.e.,} $10^{17}$G and $10^{19}$G, and two different sets of values for $v_{a}^{(0)}$ and $v_{b}^{(0)}$, {\it i.e.,} $v_{a}^{(0)}=v_{b}^{(0)}=10^{-14}$, and $v_{a}^{(0)}=10^{-7}$ and $v_{b}^{(0)}=10^{-14}$. The results are shown in Fig.\ \ref{fig1.11s}. 

Let us first investigate the evolution in the two cases $B_{z}^{(0)}=10^{17}$G and $B_{z}^{(0)}=10^{19}$G, when $v_{a}^{(0)}=v_{b}^{(0)}=10^{-14}$. It can be seen that by increasing $B_{z}^{(0)}$, the produced seeds of $B_{a}$ and $B_{b}$ become stronger, as can be seen in Figs.\ \ref{fig:figure:in12s} and \ref{fig:figure:in2111s}, respectively. This is due to the fact that, with the aforementioned initial value for $\eta_{e_R}^{(0)}$ when $v_{a}^{(0)}=v_{b}^{(0)}$, $B_{a}$ and $B_{b}$ are generated via the advection terms in their evolution equations. 
The stronger the seeds, the larger the temperature at which the concurrent transitions to the saturation curves occur.
The final values of  $B_a$, $B_b$, $B_Y$, $\eta_B$, $\eta_{e_R}$ and $\eta_{e_L}$ at the onset of the EWPT are independent  of $B_{z}^{(0)}$ and depend on initial electron asymmetry and wave number. %Fig.\ \ref{fig1.11s} also shows that
Furthermore, by increasing $B_{z}^{(0)}$, the initial hypermagnetic field amplitude $B_{Y}^{(0)}$ increases 
%which leads to the increase of the initial amplitude of $\eta_B$, as well. 
since, initially, the hypermagnetic field has no helical part, {\it i.e.,} $B_{Y}^{(0)}=B_{z}^{(0)}$. A larger $B_Y^{(0)}$ leads to a larger seed for the baryon asymmetry $\eta_B$, as well.
%This leads to the increase of the initial amplitude of $\etaB$, as well. 
The growth of $B_Y$ shows up when the amplitude of the helical part becomes comparable with that of $B_{z}$, as can be seen in Figs.\ \ref{fig:figure:in2212s}. 
This also results in the growth of the amplitude of $\eta_B$, as can be seen in Fig.\ \ref{fig:figure:1.1.311s}. However, this shows up as a drop in the values of $\eta_{e_R}$ and $\eta_{e_L}$ as shown in Figs.\ \ref{fig:figure:1.1.211s} and \ref{fig:figure:1.1.311211s}, respectively. This is in accordance with the conservation of $B-L$. Figures\ \ref{fig:figure:1.1.211s} and \ref{fig:figure:1.1.311211s} also show that, prior to the transition mentioned above, $\eta_{e_R}$ and $\eta_{e_L}$ equilibrate with each other due to chirality-flip processes.

As $B_{z}^{(0)}$ becomes larger, the JB terms become stronger and make the velocities overshoot zero further, when viscosity pushes these to zero, and the final saturated amplitudes of velocities become larger, as can be seen in Figs.\ \ref{fig:figure:1va1s} and \ref{fig:figure:1vb1s}.  Moreover, the JB terms in the evolution equations of $v_a$ and $v_b$, {\it i.e.} Eqs.\ (\ref{eq39}) and (\ref{eq40}), are influenced by $B_{z}$ not only directly, but also indirectly through $B_b$ and $B_a$, respectively, the seeds of which are proportional to $B_{z}$. This has two consequences. First, the amounts of overshoots of $v_a$ and $v_b$  depend approximately on $B_{z}^{(0)}$ squared.  
Second, since the final values of $B_a$ and $B_b$, in case $v_{a}^{(0)}=v_{b}^{(0)}$, are independent  of $B_{z}^{(0)}$, the final values of $v_a$ and $v_b$ depend linearly on $B_{z}^{(0)}$.\footnote{As can be seen in Figs.\ \ref{fig:figure:1va1s} and \ref{fig:figure:1vb1s}, when $B_{z}^{(0)}$ grows by two orders of magnitude, the overshoots of $v_a$ and $v_b$ grow by four orders of magnitude, while their final values grow by two orders of magnitude.}

Let us now investigate the evolution in the two cases $B_{z}^{(0)}=10^{17}$G and $B_{z}^{(0)}=10^{19}$G, when $v_{a}^{(0)}=10^{-7}$ and $v_{b}^{(0)}=10^{-14}$.  As mentioned before, with the initial conditions chosen, the seeds of $B_b$ and $B_a$ 
%are greater as compared to those of the two previous cases, 
are produced via the advection and the chiral vortical terms, respectively. As a result, by increasing $B_{z}^{(0)}$, the seed of $B_b$, which was stronger than that of  $B_a$ to begin with, becomes even stronger while the seed of $B_a$ does not change. Therefore, $B_b$ becomes much larger than before during the evolution and is the dominant helical component. Consequently, the rates of changes of the asymmetries, given in Eqs.\ (\ref{eq32}-\ref{eq34}), are increased and attain the value zero more quickly, when $\Delta \eta$ attains its minimum value. Therefore, the resulting  concurrent transitions of all variables, including $B_a$, occur at a greater temperature. Consequently, the maximum and final values of $B_a$ decrease. That is, by increasing $B_{z}^{(0)}$, the  saturation curve of $B_b$ remains the same, while that of $B_a$ is shifted downward, as Figs.\ \ref{fig:figure:in12s} and \ref{fig:figure:in2111s} show. Indeed, the final values of $B_a$ depend inversely on $B_{z}^{(0)}$. 

%
%\textcolor{red}{} \textcolor{green}{} \textcolor{green}{,} \textcolor{orange}{()} \textcolor{purple}{}
%

The evolution of $v_a$ in these two cases is also similar to the two previous ones. 
That is, as explained above, the amounts of overshoots of $v_a$ depends approximately on $B_{z}^{(0)}$ squared and the final value of $v_a$ depends linearly on $B_{z}^{(0)}$, as can be seen in Fig.\ \ref{fig:figure:1va1s}.
The evolution of $v_b$, which is influenced by that of $B_a$, is  somewhat different from the two previous cases. 
First, since $v_b<<v_a$ in this case, the seeds of $B_a$, which are produced by the CVE and depend on $v_a^{(0)}$, are equal and independent of $B_{z}^{(0)}$. Therefore the amount of overshoot of $v_b$, produced by the JB term in Eq.\ (\ref{eq40}), depends linearly on $B_{z}^{(0)}$. \footnote{The overshoot value of $v_b$ with $B_{z}^{(0)}=10^{19}$G is two orders of magnitude greater than the one with $B_{z}^{(0)}=10^{17}$G.} 
Second, since the final value of $B_a$ depends inversely on $B_{z}^{(0)}$,\footnote{The final value of $B_a$ for $B_{z}^{(0)}=10^{19}$G is two orders of magnitude less than the one with $B_{z}^{(0)}=10^{17}$G.} the saturated values of $v_b$ in these two cases, which can be determined by the product of $B_{z}^{(0)}$ and the saturated value of $B_a$, become equal. 
Incidentally, when we set the viscosity to zero, the overall behavior of the variables remains unchanged. The major differences are that all variables except the velocities reach their saturation curves earlier. As for the velocities, it is the JB terms that make them drop to large negative values rather abruptly. The larger the initial velocity, the later this drop occurs.

\newpage

%
%\textcolor{red}{} \textcolor{green}{} \textcolor{green}{,} \textcolor{orange}{()} \textcolor{purple}{}
%

\subsection{Generation of helicity and matter-antimatter asymmetry by strong non-helical hypermagnetic field and vorticity} 

\begin{figure}[H]
		\centering
		\subfigure[]{\label{fig:figure:in2} 
			\includegraphics[width=.365\textwidth]{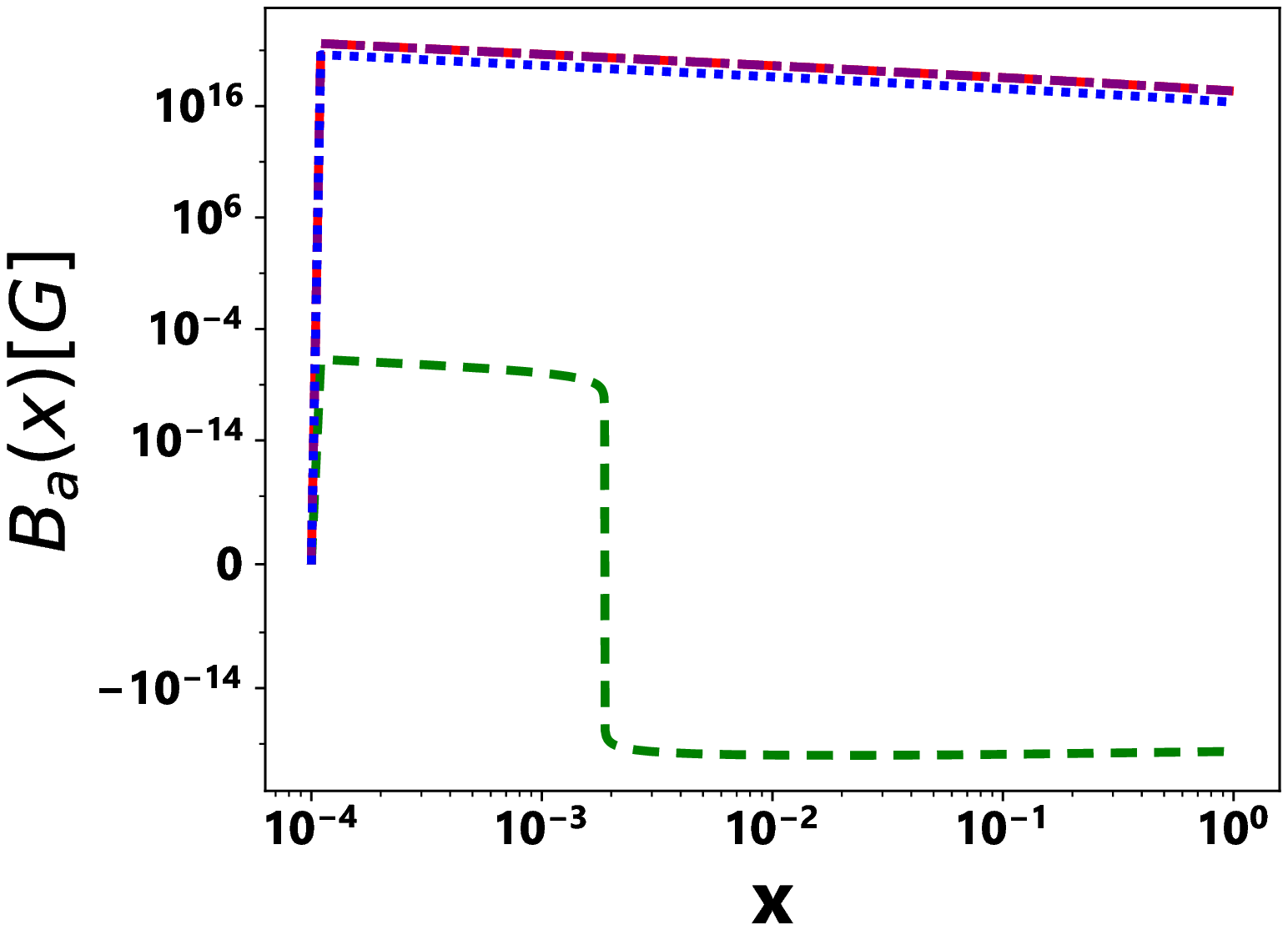}}
		\hspace{10mm}
		\subfigure[]{\label{fig:figure:in211q} 
			\includegraphics[width=.36665\textwidth]{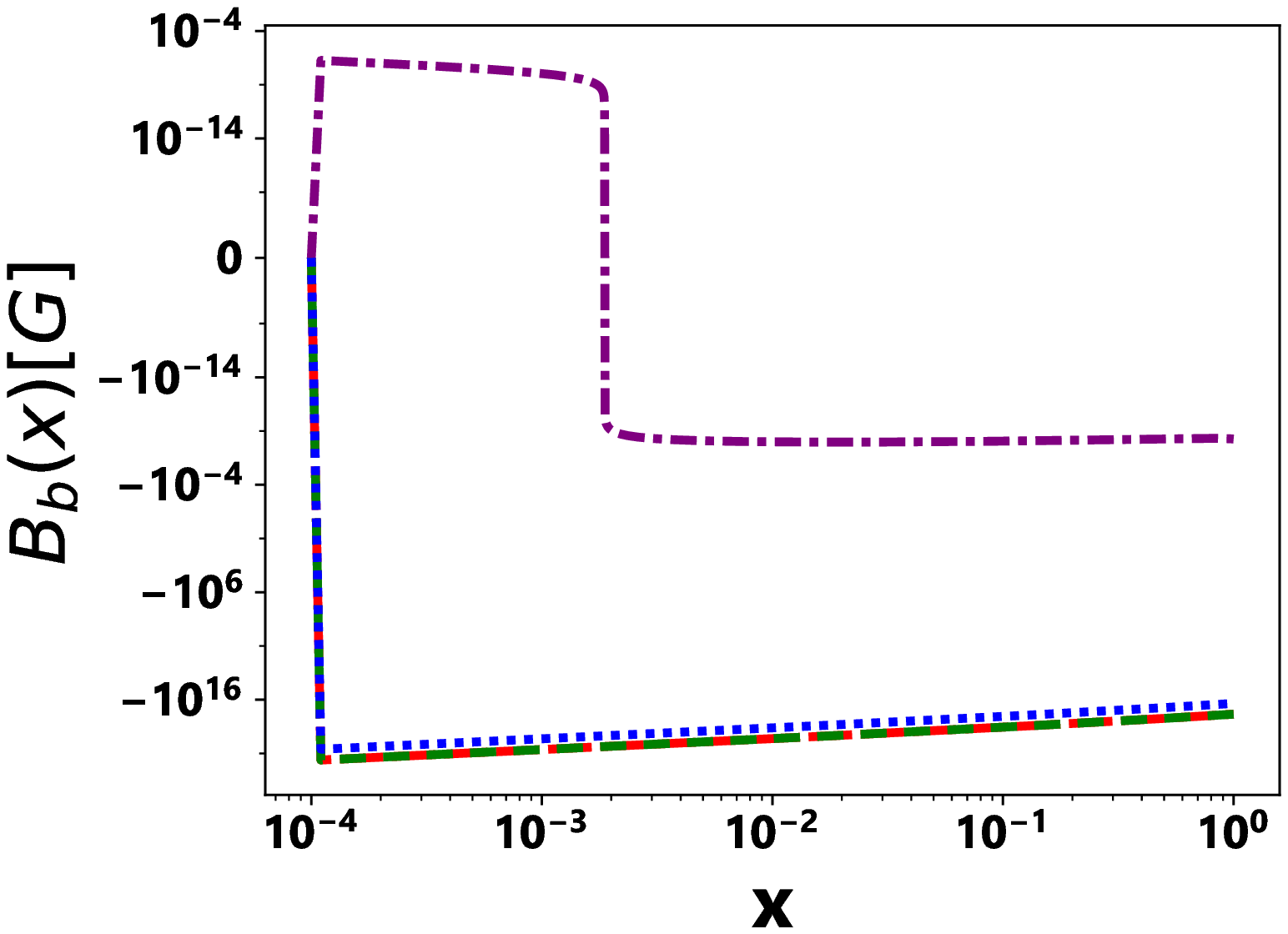}}
		\hspace{8mm}
		\subfigure[]{\label{fig:figure:in222q} 
			\includegraphics[width=.365\textwidth]{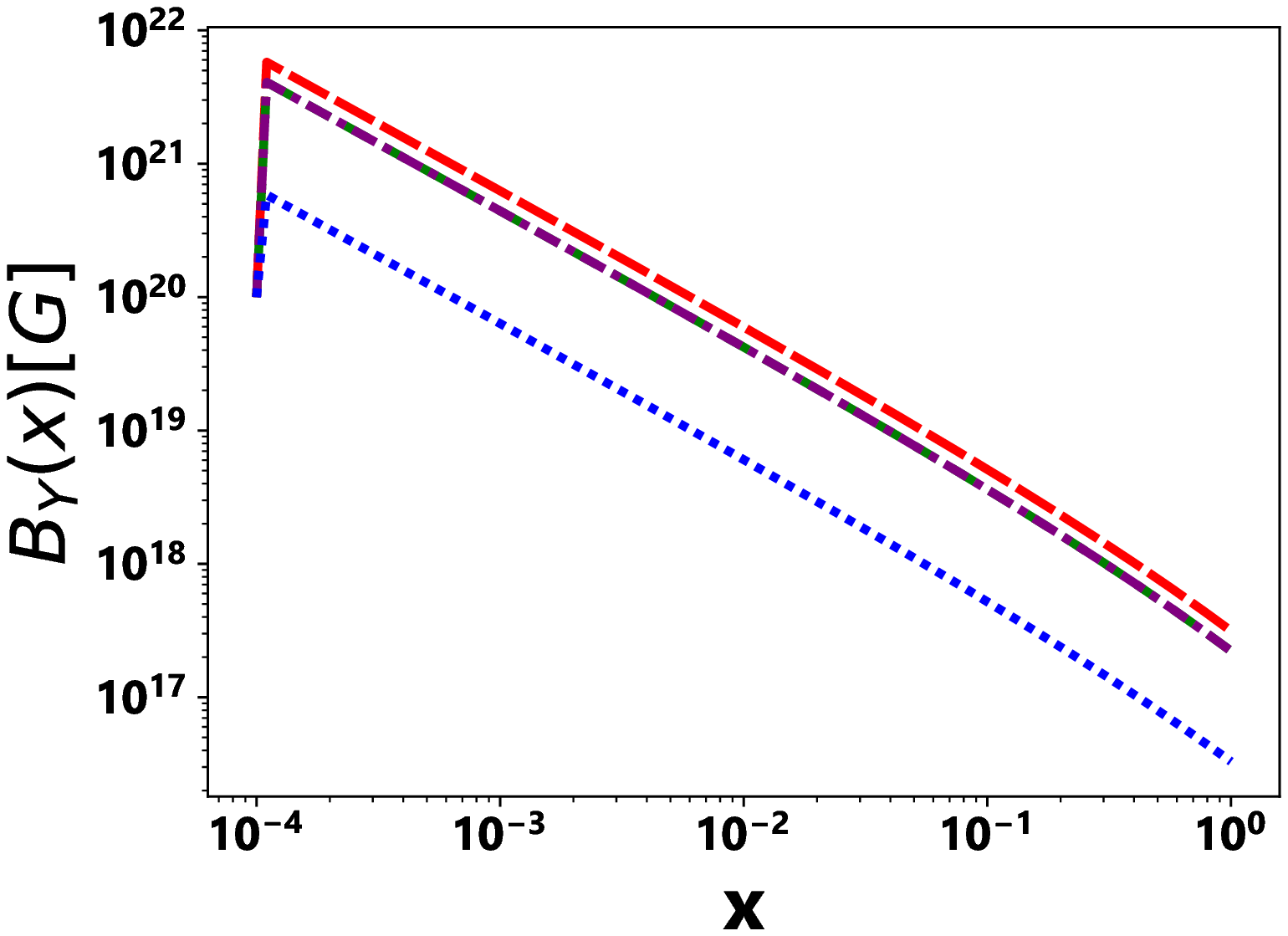}}
		\hspace{8mm}
		\subfigure[]{\label{fig:figure:1.1.31q}
			\includegraphics[width=.365\textwidth]{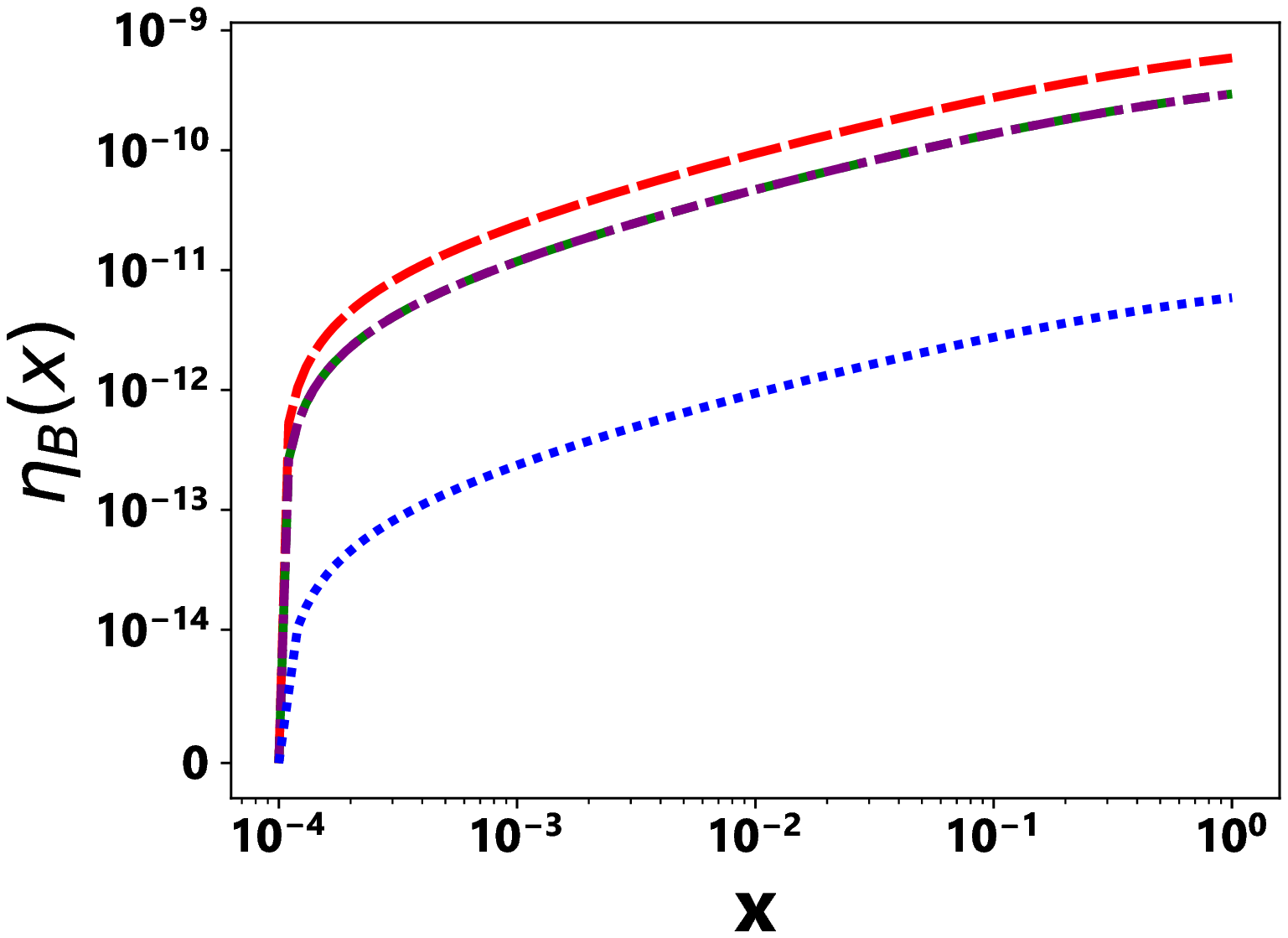}}
		\hspace{8mm}
		\subfigure[]{\label{fig:figure:1.1.21q}
			\includegraphics[width=.365\textwidth]{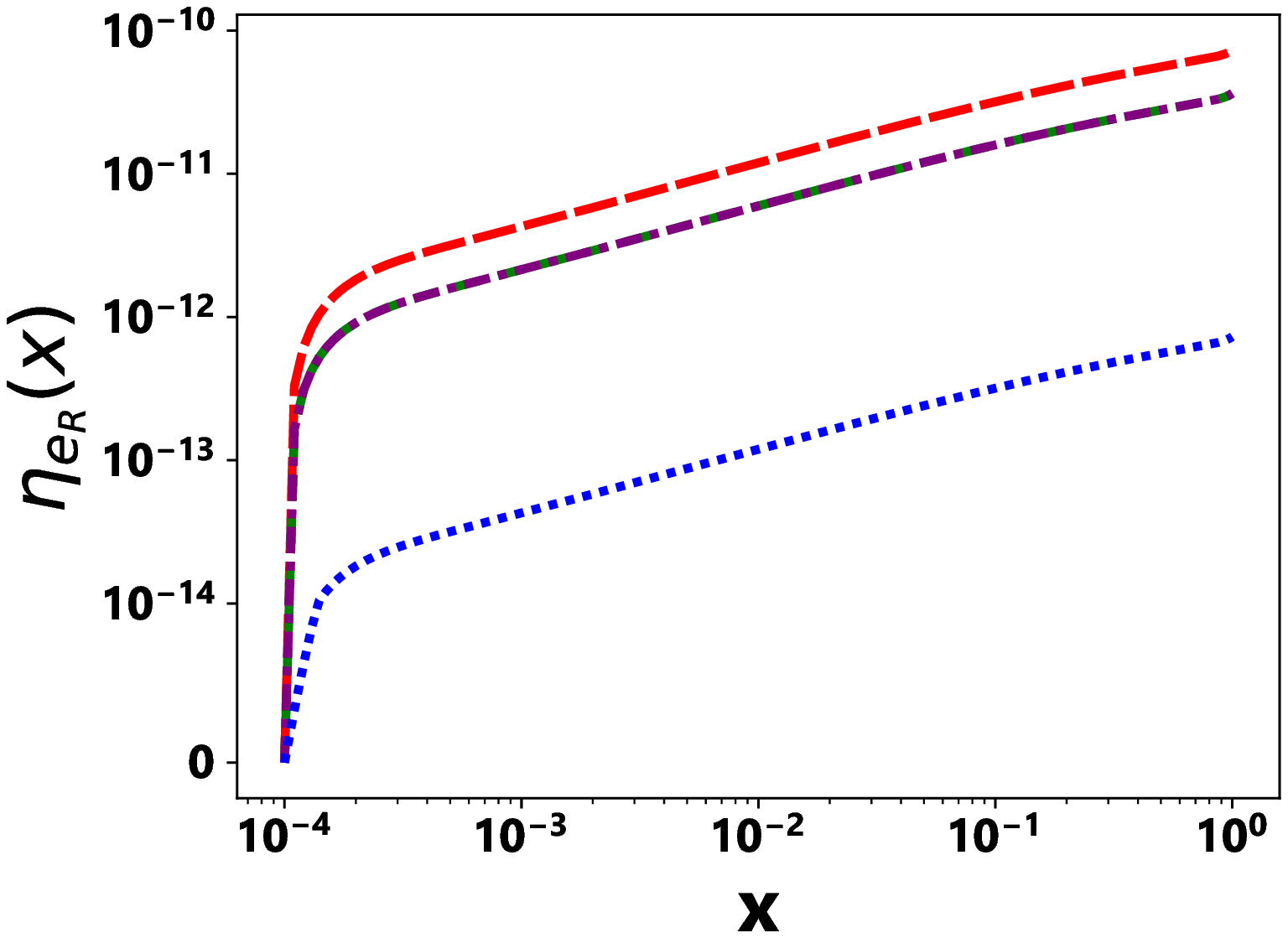}}
		\hspace{8mm}
		\subfigure[]{\label{fig:figure:1.1.31121q}
			\includegraphics[width=.365\textwidth]{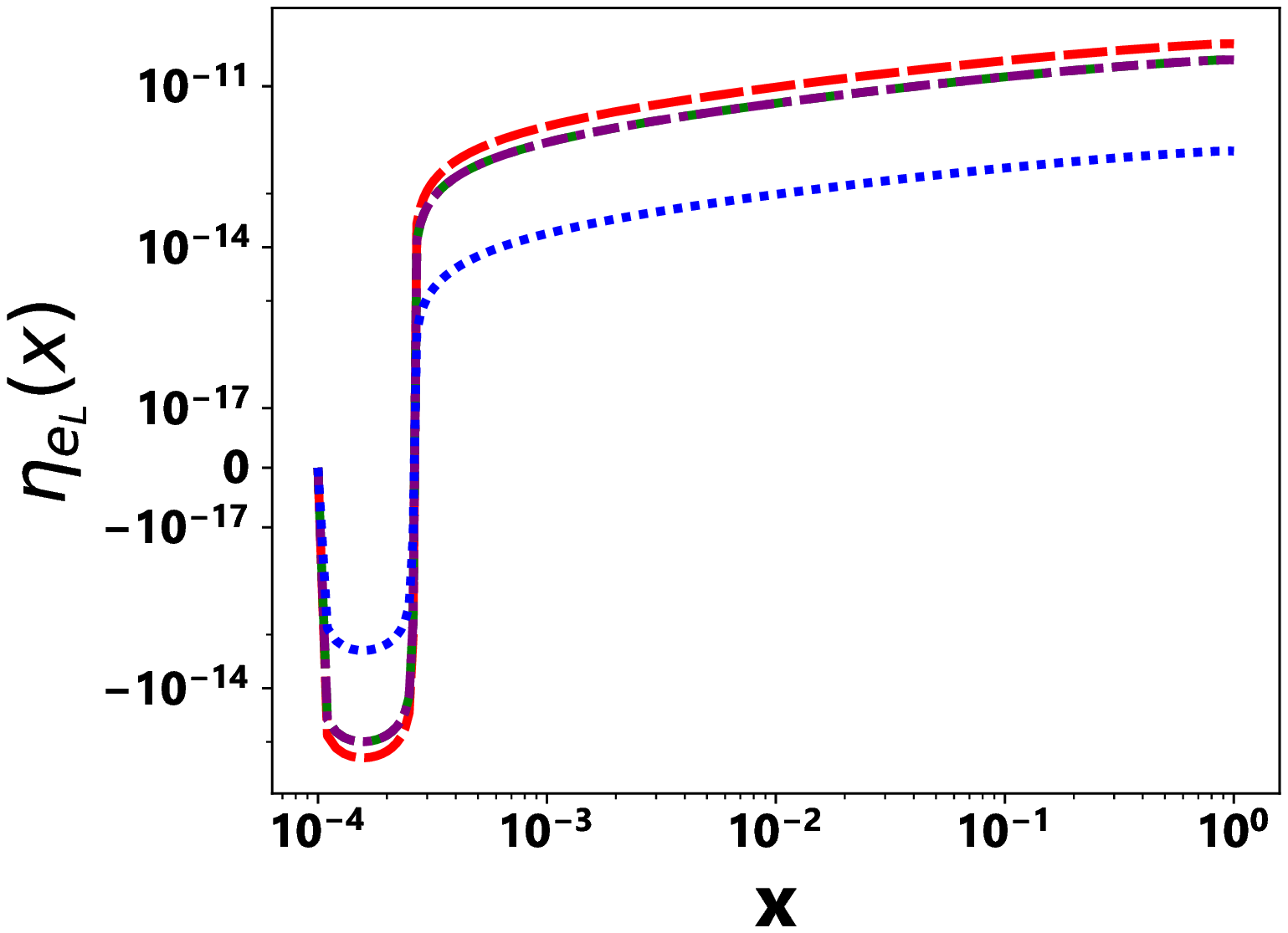}}
		\hspace{8mm}
		\subfigure[]{\label{fig:figure:1vaq}
			\includegraphics[width=.365\textwidth]{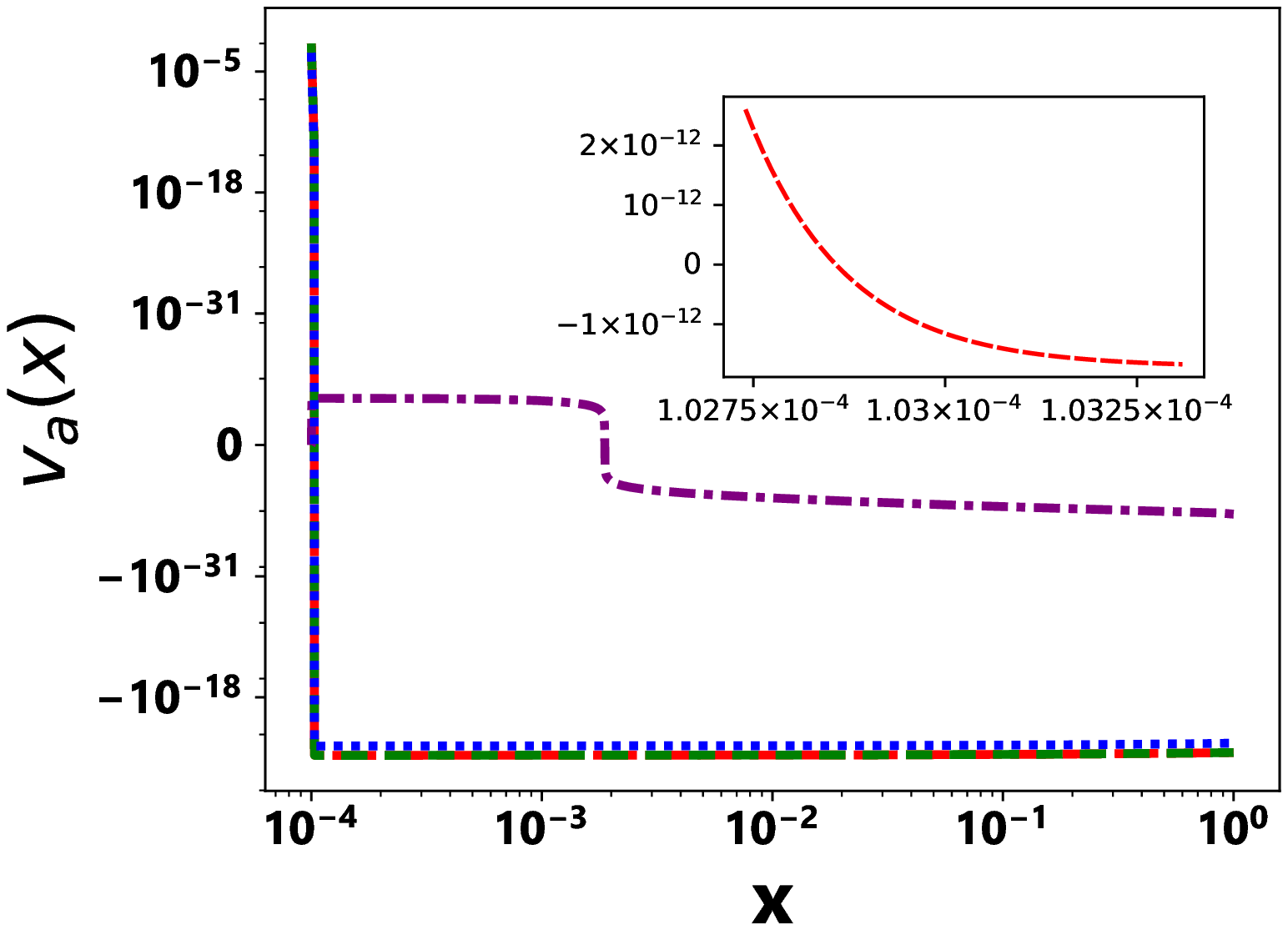}}
		\hspace{8mm}
		\subfigure[]{\label{fig:figure:1vbq}
			\includegraphics[width=.365\textwidth]{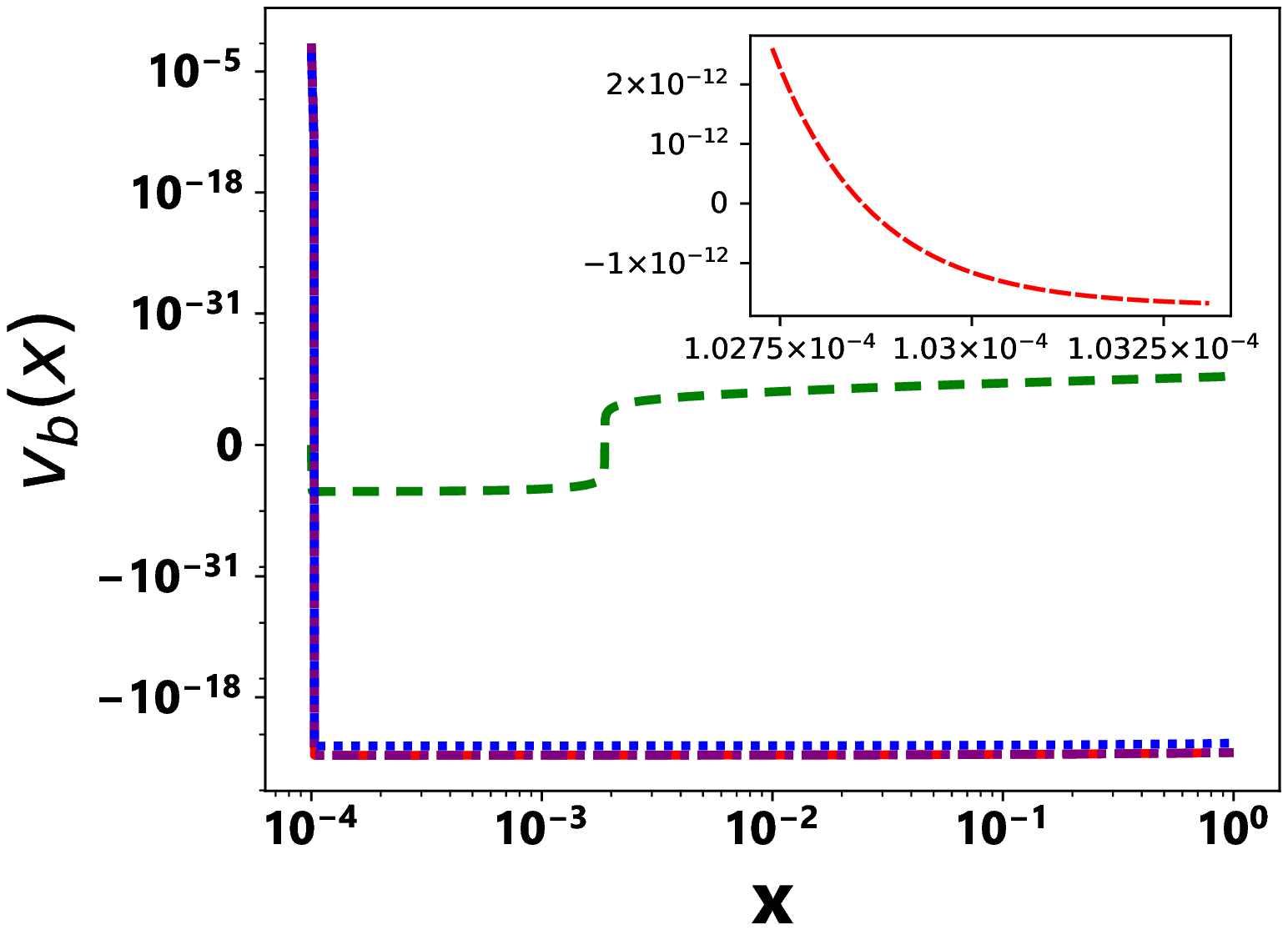}}
		\caption{\footnotesize 
			The time plots of the helical components $B_{a}(x)$ and $ B_{b}(x)$, the hypermagnetic field amplitude $B_{Y}(x)$, the baryon asymmetry $\eta_{B}(x)$, the right-handed electron asymmetry $\eta_{e_R}(x)$, the left-handed electron asymmetry $\eta_{e_L}(x)$, and $v_{a}(x)$ and $v_{b}(x)$ in the presence of the viscosity, with the initial conditions $B_{z}^{(0)}=10^{20} \mbox{G}$, $B_{a}^{(0)}=B_{b}^{(0)}=0$, and $\eta_{e_R}^{(0)}=\eta_{e_L}^{(0)}=\eta_{B}^{(0)}=0$. Large dashed (red) line is for, $v_{a}^{(0)}=v_{b}^{(0)}=10^{-2}$, dashed (green) line for $v_{a}^{(0)}=10^{-2}$, $v_{b}^{(0)}=0$, dot-dashed (violet) line for $v_{a}^{(0)}=0$, $v_{b}^{(0)}=10^{-2}$, and dotted (blue) line for $v_{a}^{(0)}=v_{b}^{(0)}=10^{-3}$.%The time plots of the helical components $B_{a}(x)$ and $ B_{b}(x)$, the hypermagnetic field amplitude $B_{Y}(x)$, the baryon asymmetry $\eta_{B}(x)$, the right-handed electron asymmetry $\eta_{e_R}(x)$, the left-handed electron asymmetry $\eta_{e_L}(x)$, and $v_{a}(x)$ and $v_{b}(x)$ in the presence of the viscosity, with the initial conditions $B_{z}^{(0)}=10^{20} \mbox{G}$, $B_{a}^{(0)}=B_{b}^{(0)}=0$, and $y_{R}^{(0)}=y_{L}^{(0)}=y_{B}^{(0)}=0$. Large dashed (red) line is for, $v_{a}^{(0)}=v_{b}^{(0)}=10^{-2}$, dashed (green) line for $v_{a}^{(0)}=10^{-2}$, $v_{b}^{(0)}=0$, dot-dashed (orange) line for $v_{a}^{(0)}=0$, $v_{b}^{(0)}=10^{-2}$, and dotted (blue) line for $v_{a}^{(0)}=v_{b}^{(0)}=10^{-3}$.
		}\label{fig1.1k}
\end{figure}

Now we investigate the possibility of the helicity and matter-antimatter asymmetry production in the absence of initial matter-antimatter asymmetry. To accomplish this task, non-zero initial values for $B_z$ and $v_{a}$ or $v_{b}$  are needed. We solve the set of coupled differential equations with the initial conditions $k=10^{-7}$, $B_{z}^{(0)}=10^{19}$G, $B_{a}^{(0)}=B_{b}^{(0)}=0$, $\eta_{e_R}^{(0)}=\eta_{e_L}^{(0)}=\eta_{B}^{(0)}=0$, and four different sets of values for $v_{a}^{(0)}$ and $v_{b}^{(0)}$. 
The results are shown in Fig.\ \ref{fig1.1k}. It can be seen that the helical components of the hypermagnetic field, $B_{a}$, $B_{b}$, and matter-antimatter asymmetries are generated and amplified from zero initial values. 
The seeds for $B_{a}$  and $B_{b}$ are produced by the advection terms on the rhs of Eqs.\ (\ref{eq36},\ref{eq37}). Subsequently, these helical components produce the matter-antimatter asymmetries through the first terms in Eqs.\ (\ref{eq32}), (\ref{eq33}) and (\ref{eq34}). 

Let us investigate the effect of initial velocity on the evolution equations for two cases $v_{a}^{(0)}=v_{b}^{(0)}=10^{-3}$ and $v_{a}^{(0)}=v_{b}^{(0)}=10^{-2}$.
As can be seen in Eqs.\ (\ref{eq36},\ref{eq37}) and Figs.\ \ref{fig:figure:in2}, \ref{fig:figure:in211q} and more clearly in Fig.\ \ref{fig:figure:in222q}, the amount of $B_{a}$ and $B_{b}$ produced by the advection term increases linearly with the increase of the initial velocities. Since $v_{a}^{(0)}=v_{b}^{(0)}>0$ the initial hypermagnetic fields produced are  $B_{a}\approx -B_{b}>0$. Therefore, when the viscosity terms in the evolution equations for the velocities, {\it i.e.}, the second terms in Eqs.\ (\ref{eq39}) and (\ref{eq40}), force them to zero, the JB terms, which involve $B_{b}$ and $B_{a}$, respectively, make both velocities overshoot zero and obtain their terminal values, due to counterbalancing effect of viscosity again. Meanwhile, the $F_0$ terms in Eqs.\ (\ref{eq32}), (\ref{eq33}) and (\ref{eq34}) initially produce $\eta_{e_R}>0$, $\eta_{e_L}<0$ and $\eta_{B}>0$. Then the chirality flip processes, represented by the last terms of Eqs.\ (\ref{eq32}), (\ref{eq33}), equilibrate both $\eta_{e_R}$ and $\eta_{e_L}$ to positive values, due to the surplus production of the former. This causes the sudden turnarounds in the graphs for $\eta_{e_L}$. In this case, unlike the case with a large initial value of $\eta_{e_R}$ studied in the last subsection, the $\eta$s and hence $\Delta\eta$ keep increasing, but do not saturate for the initial conditions chosen. Meanwhile, the first terms in Eqs.\ (\ref{eq36},\ref{eq37}), in which the CME terms remain much smaller than the $k''$ terms, together with the last terms, arising from the expansion of the Universe, lead to exponential damping observed in Figs.\ \ref{fig:figure:in2} and \ref{fig:figure:in211q}.

Next, we study two other sets of initial values for velocities which are: \{$v_{a}^{(0)}=10^{-2}$, $v_{b}^{(0)}=0$\} and \{$v_{a}^{(0)}=0$, $v_{b}^{(0)}=10^{-2}$\}. In the first case, $v_{a}^{(0)}$ produces $B_{b}$, by the advection term, which then produces the asymmetries, by the $F_0$ terms in Eqs.\ (\ref{eq32}), (\ref{eq33}) and (\ref{eq34}), with the same signs as before. Subsequently, the net electron chirality produced generates a smaller but positive $B_{a}$ by the CVE term, which appears in Eq.\ (\ref{eq36}) and is proportional to $\eta_{e_R}^2(x)-\eta_{e_L}^2(x)$, which then produces a small negative $v_{b}$ by the JB term in Eq.\ (\ref{eq40}). When both $v_{a}$ or $v_{b}$ become negative, the advection and CVE terms in Eq.\ (\ref{eq36}) together force $B_{a}$ to negative values, while the first and fourth terms have only damping effects in this case. Subsequently, this change of sign of $B_{a}$, forces $v_{b}$ to change sign through the JB term in Eq.\ (\ref{eq40}). The second case, {\it i.e.,} \{$v_{a}^{(0)}=0$, $v_{b}^{(0)}=10^{-2}$\}, can be analyzed similarly.

\newpage
\subsection{Production of matter-antimatter asymmetry and vorticity by strong helical hypermagnetic field}

\begin{figure}[H]
	\centering
	\subfigure[]{\label{fig:figure:in12sa} 
		\includegraphics[width=.365\textwidth]{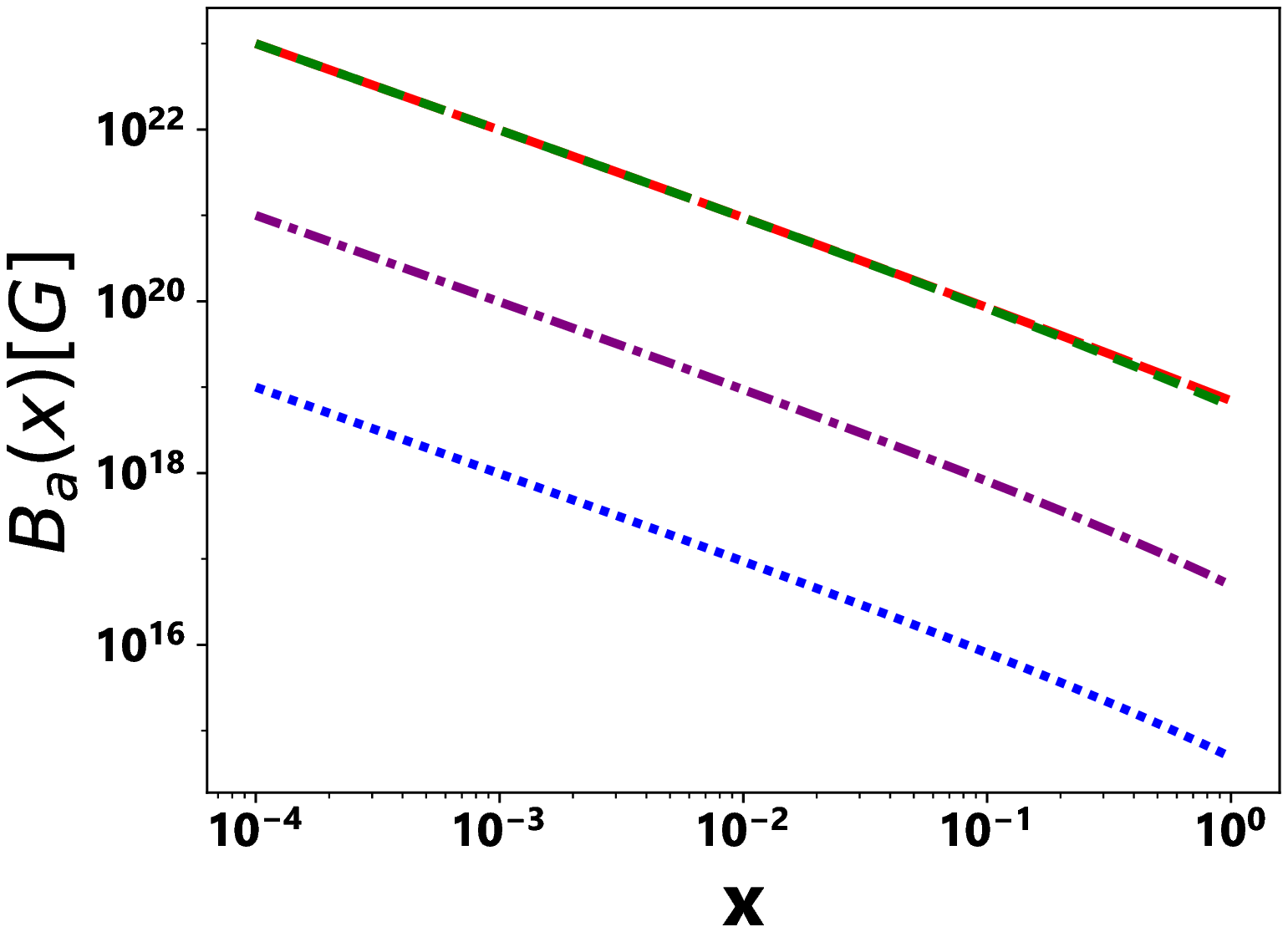}}
	\hspace{10mm}
	\subfigure[]{\label{fig:figure:in2111sa} 
		\includegraphics[width=.36665\textwidth]{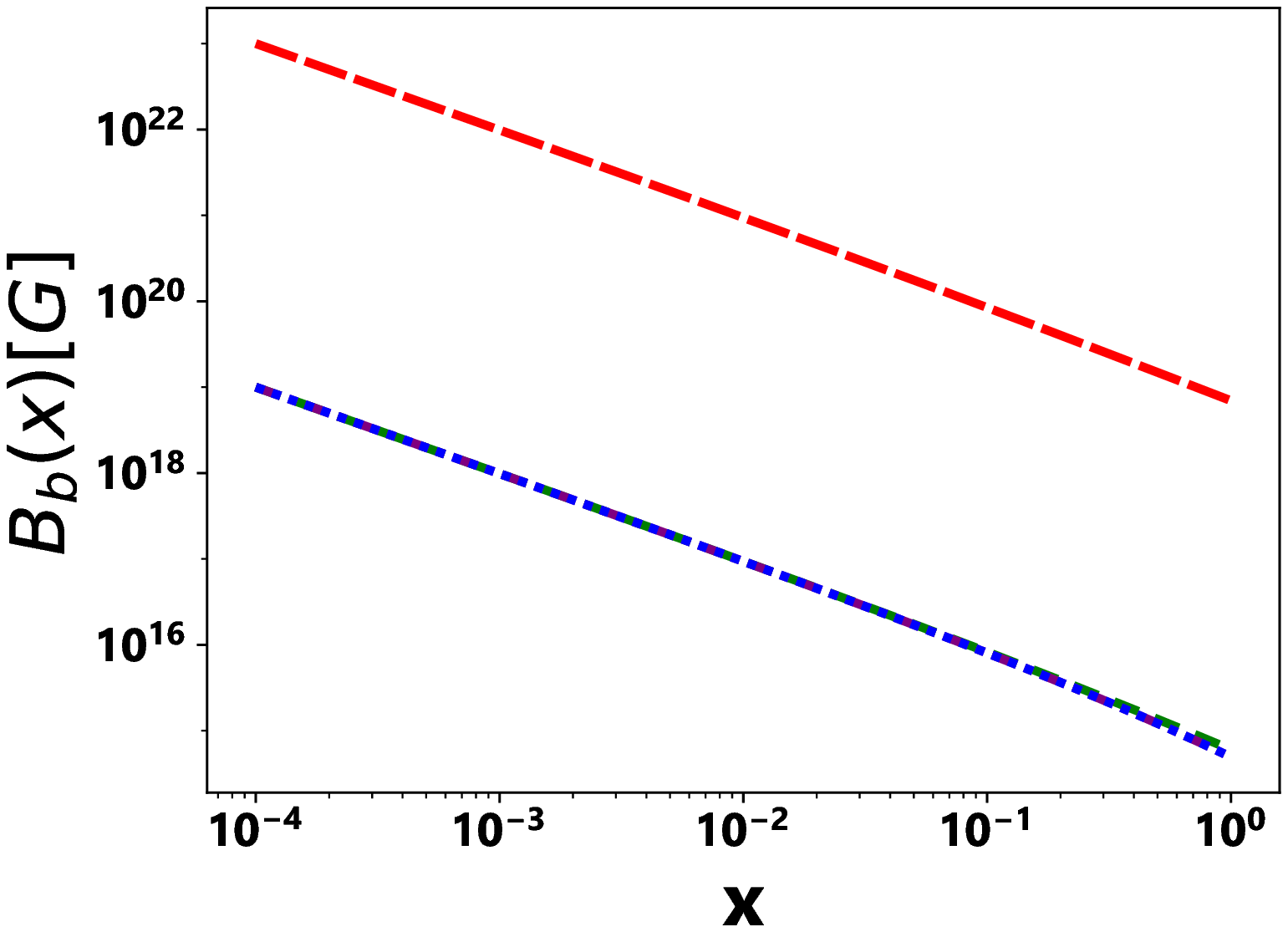}}
	\hspace{8mm}
	\subfigure[]{\label{fig:figure:in2212sa} 
		\includegraphics[width=.365\textwidth]{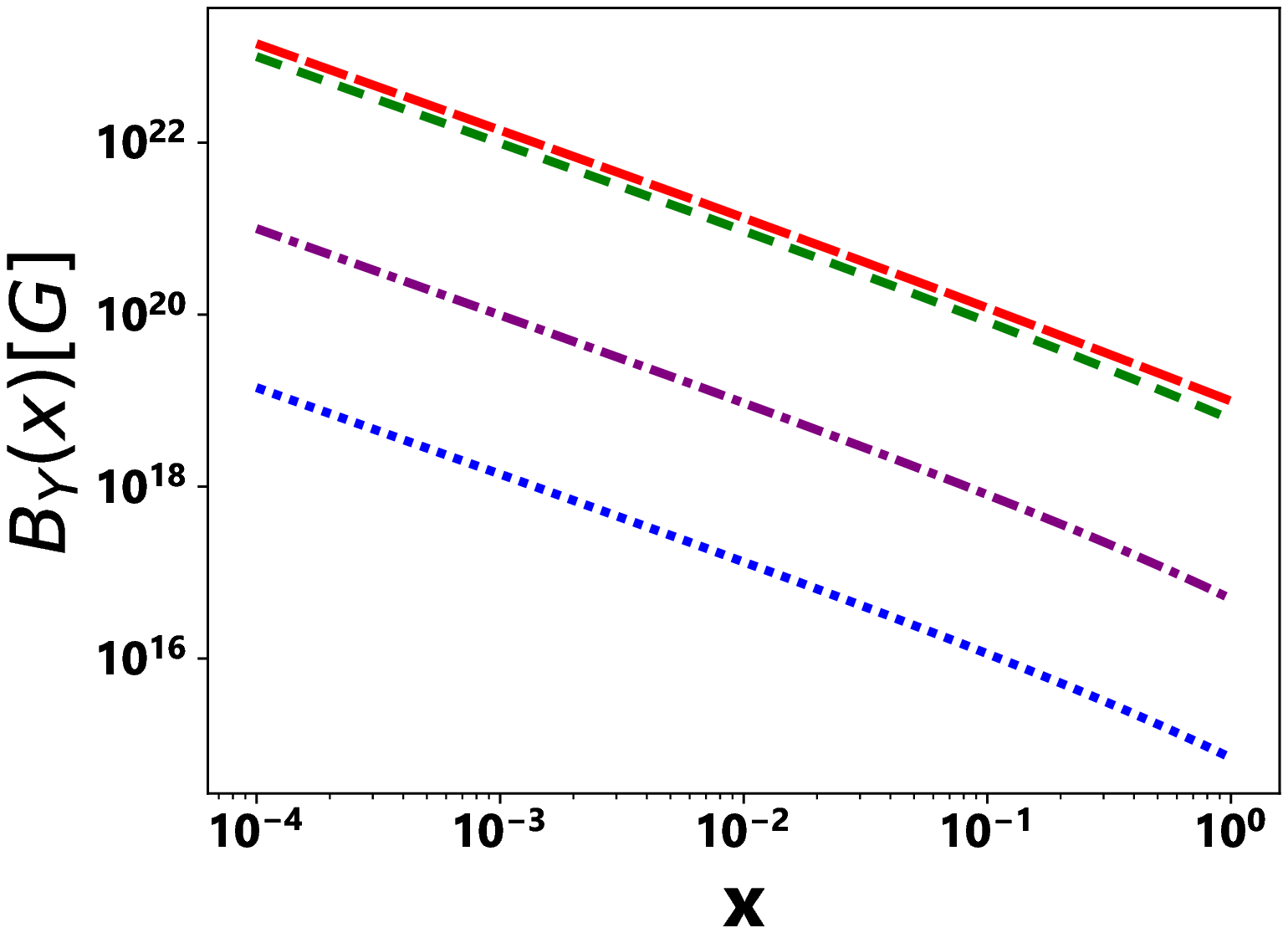}}
	\hspace{8mm}
	\subfigure[]{\label{fig:figure:1.1.311sa}
		\includegraphics[width=.365\textwidth]{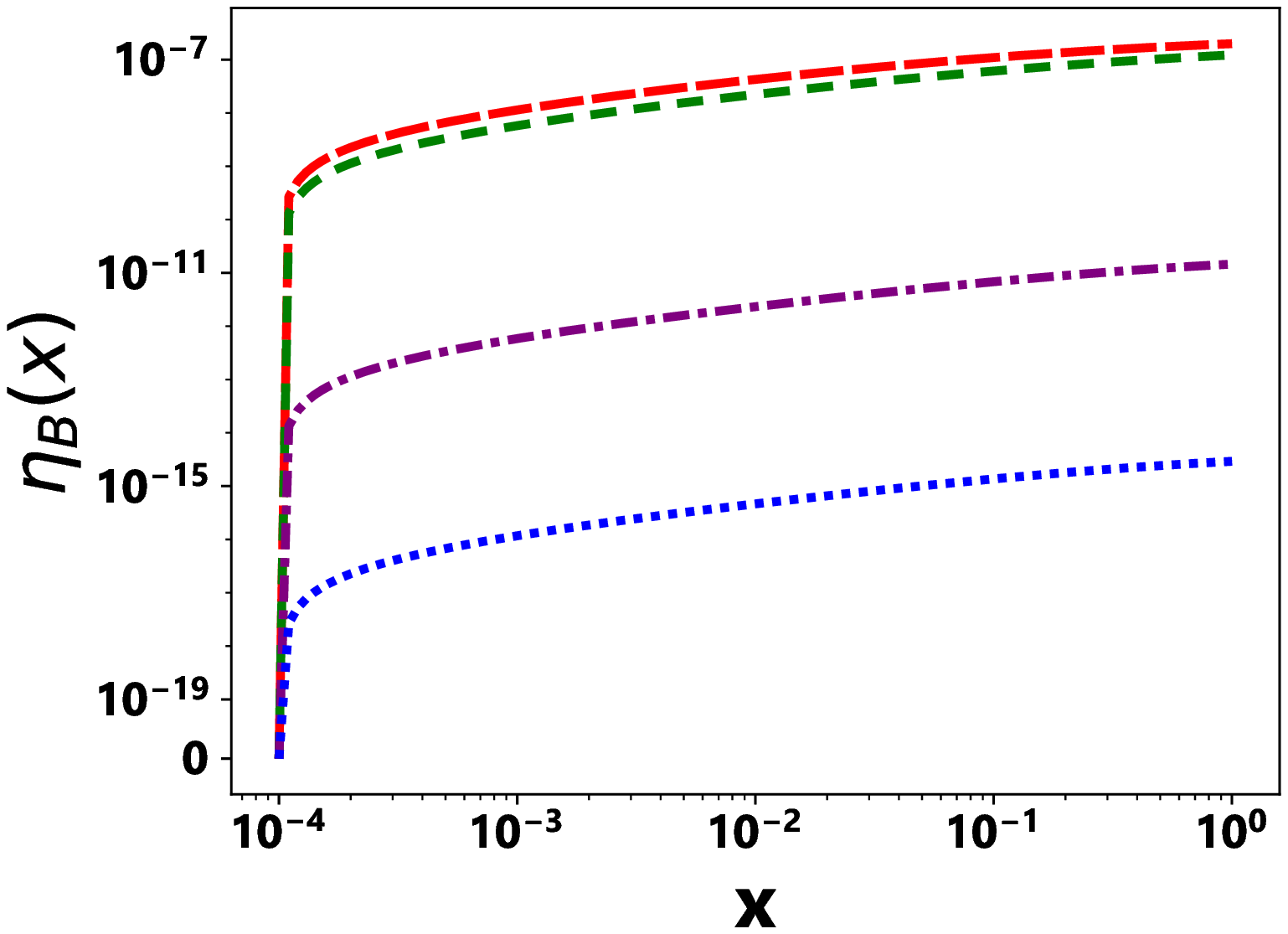}}
	\hspace{8mm}
	\subfigure[]{\label{fig:figure:1.1.211sa}
		\includegraphics[width=.365\textwidth]{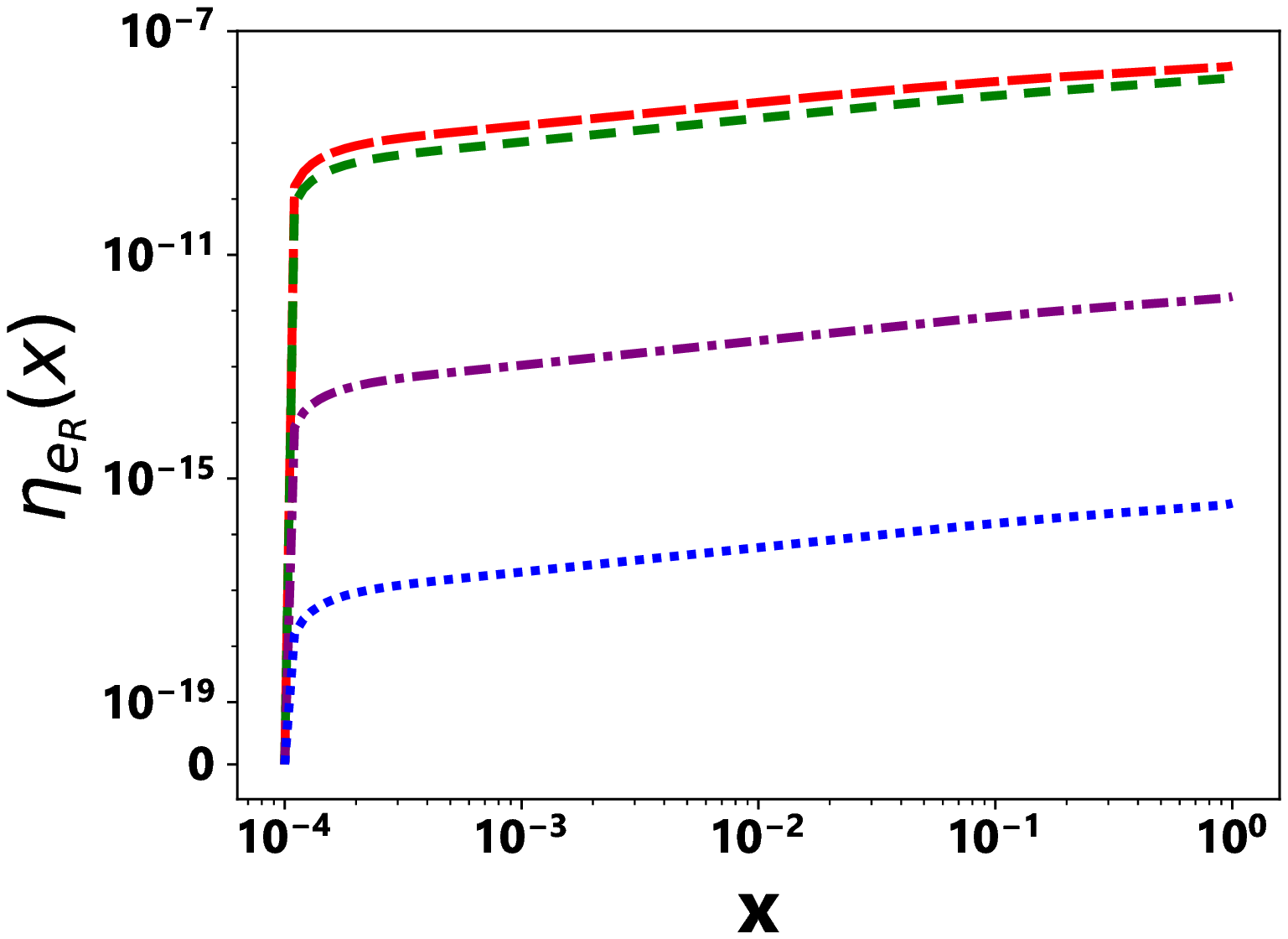}}
	\hspace{8mm}
	\subfigure[]{\label{fig:figure:1.1.311211sa}
		\includegraphics[width=.365\textwidth]{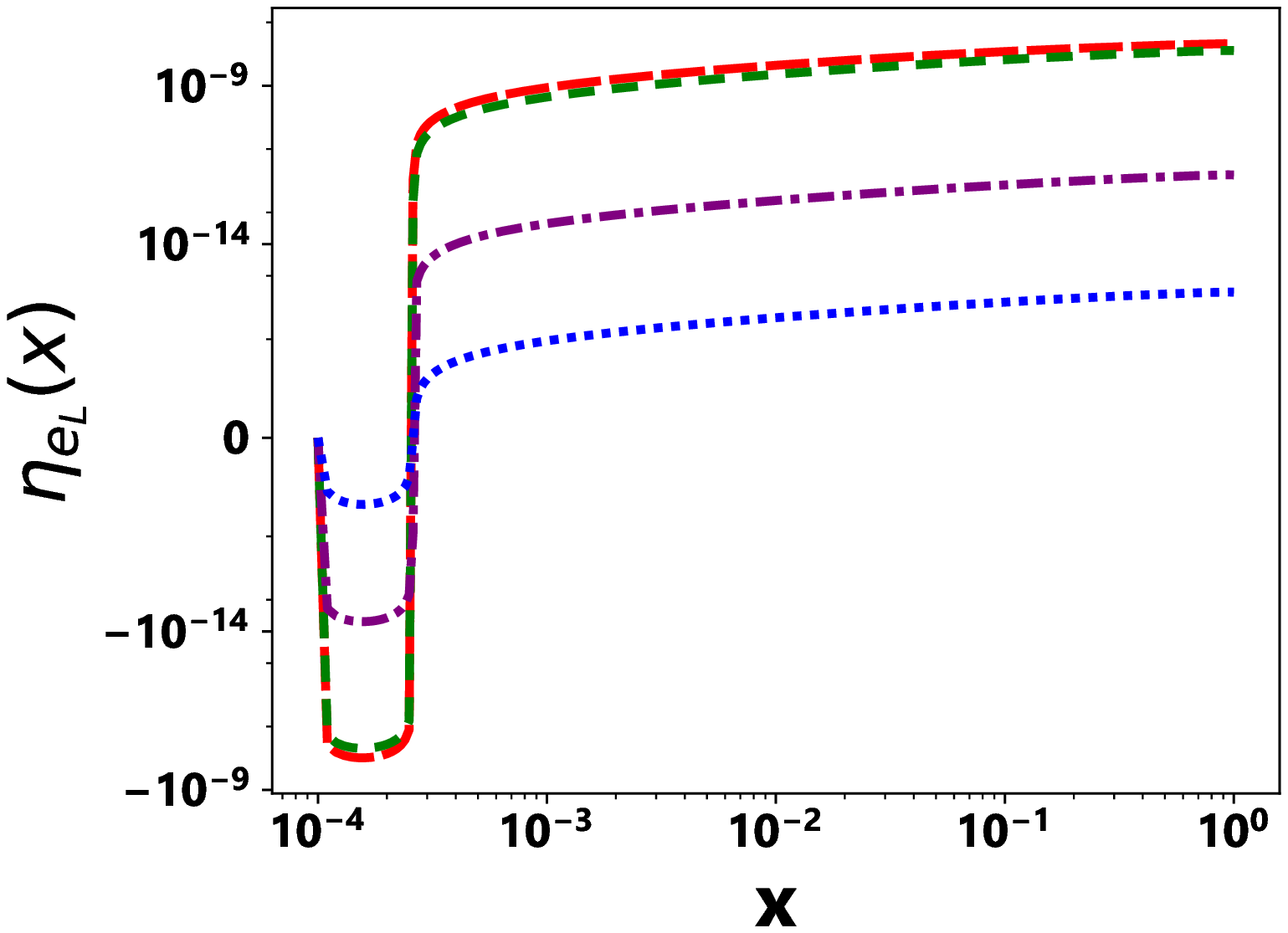}}
	\hspace{8mm}
	\subfigure[]{\label{fig:figure:1va1sa}
		\includegraphics[width=.365\textwidth]{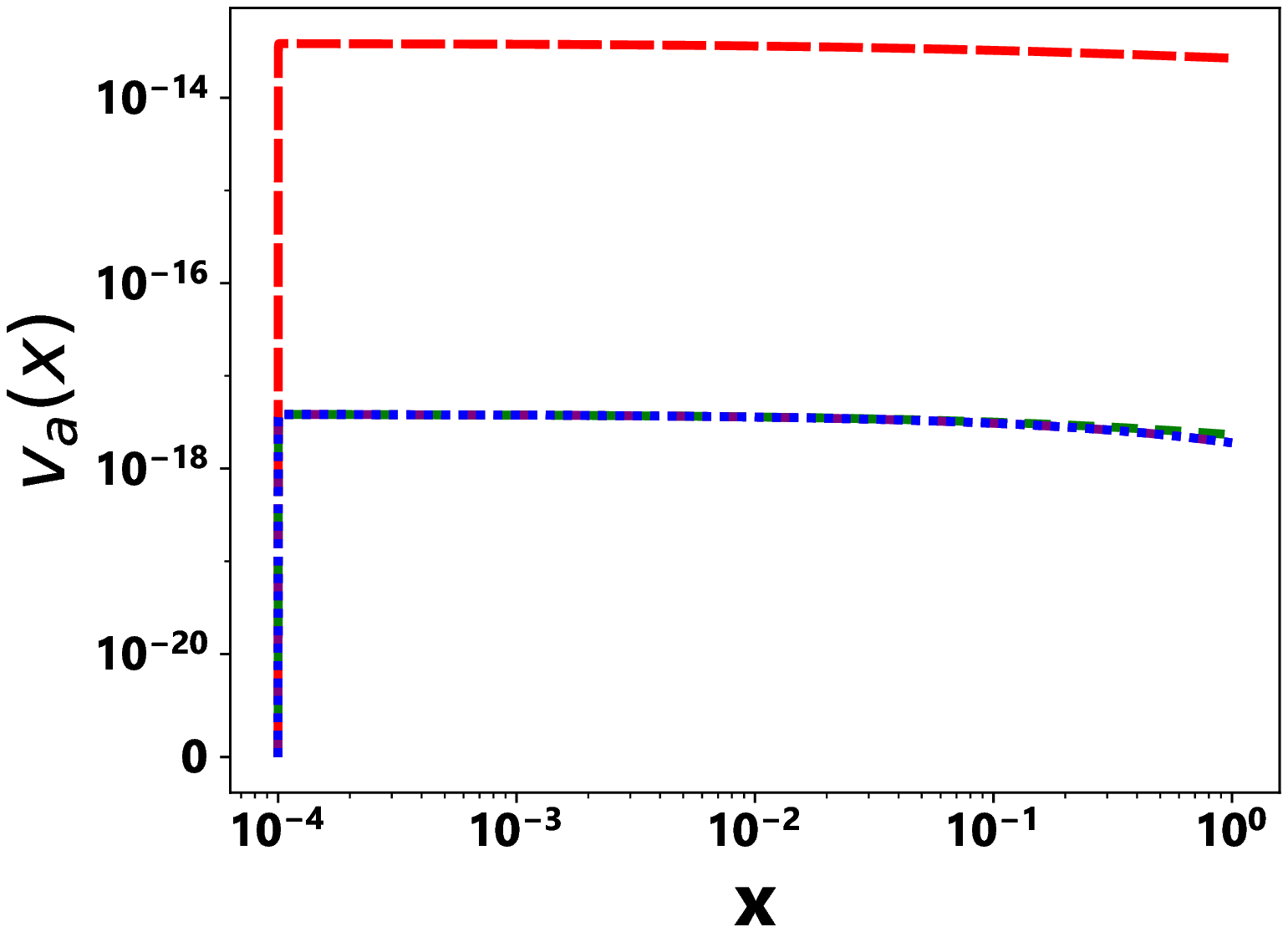}}
	\hspace{8mm}
	\subfigure[]{\label{fig:figure:1vb1sa}
		\includegraphics[width=.365\textwidth]{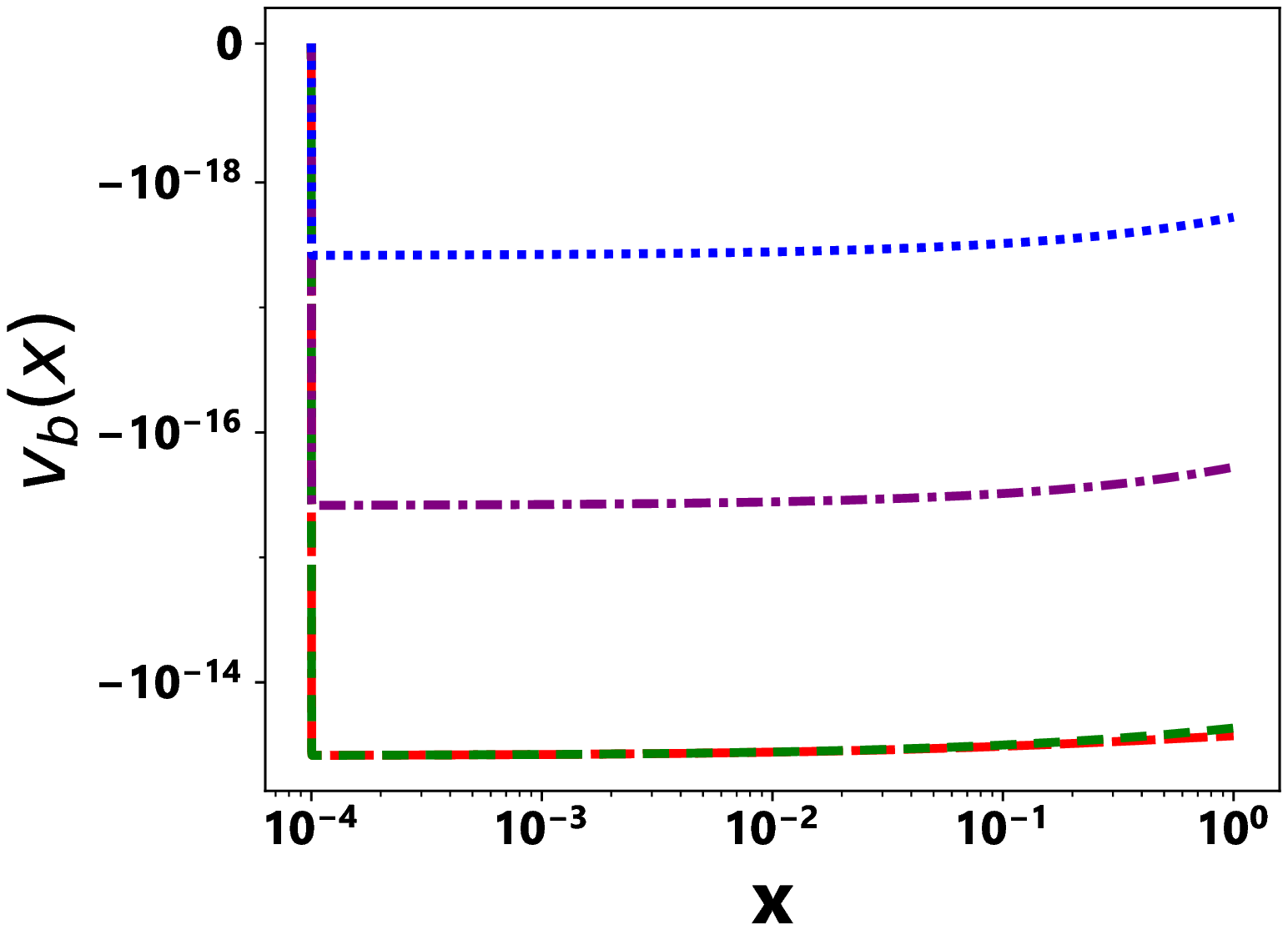}}
	\caption{\footnotesize The time plots of the helical components $B_{a}(x)$ and $ B_{b}(x)$, the hypermagnetic field amplitude $B_{Y}(x)$, the baryon asymmetry $\eta_{B}(x)$, the right-handed electron asymmetry $\eta_{e_R}(x)$, the left-handed electron asymmetry $\eta_{e_L}(x)$, and $v_{a}(x)$ and $v_{b}(x)$ with the initial conditions $B_{z}^{(0)}=10^{17} \mbox{G}$, and $\eta_{e_R}^{(0)}=\eta_{e_L}^{(0)}=\eta_{B}^{(0)}=v_{a}^{(0)}=v_{b}^{(0)}=0$. Large (red) dashed line is for $B_{a}^{(0)}=B_{b}^{(0)}=10^{23}\mbox{G}$, dashed (green) line for $B_{a}^{(0)}=10^{23}\mbox{G}$, $B_{b}^{(0)}=10^{19}\mbox{G}$, dotted-dashed (violet) line for $B_{a}^{(0)}=10^{21}\mbox{G}$, $B_{b}^{(0)}=10^{19}\mbox{G}$, and dotted (blue) line for $B_{a}^{(0)}=B_{b}^{(0)}=10^{19}\mbox{G}$.
		%The time plots of the helical components $B_{a}(x)$ and $ B_{b}(x)$, the hypermagnetic field amplitude $B_{Y}(x)$, the baryon asymmetry $\eta_{B}(x)$, the ratio of the helical part to the non-helical part $\sqrt{B_{a}^{2}(x)+B_{b}^{2}(x)}/B_{z}(x)$, the vorticity $\omega(x)$, and $v_{a}(x)$ and $\big|v_{b}(x)\big|$ with the initial conditions $B_{z}^{(0)}=10^{17} \mbox{G}$, and $y_{R}^{(0)}=y_{L}^{(0)}=y_{B}^{(0)}=v_{a}^{(0)}=v_{b}^{(0)}=0$. Large (red) dashed line is for $B_{a}^{(0)}=B_{b}^{(0)}=10^{23}\mbox{G}$, dashed (green) line for $B_{a}^{(0)}=10^{23}\mbox{G}$, $B_{b}^{(0)}=10^{19}\mbox{G}$, dotted-dashed (orange) line for $B_{a}^{(0)}=10^{21}\mbox{G}$, $B_{b}^{(0)}=10^{19}\mbox{G}$, and dotted (blue) line for $B_{a}^{(0)}=B_{b}^{(0)}=10^{19}\mbox{G}$.
	}\label{fig1.11sa}
\end{figure}

Let us investigate the possibility to produce the matter-antimatter asymmetries, and vorticity by strong  hypermagnetic fields, containing both helical and non-helical components, in the presence of the viscosity and in the temperature range $100 \mbox{GeV} \le T\le 10 \mbox{TeV}$. 
We solve the set of coupled differential equations with the initial conditions $k=10^{-7}$, $\eta_{e_R}^{(0)}=\eta_{e_L}^{(0)}=\eta_{B}^{(0)}=0$, $v_{a}^{(0)}=v_{b}^{(0)}=0$, $B_{z}^{(0)}=10^{17}\mbox{G}$, and four different sets of values for $B_{a}^{(0)}$ and $B_{b}^{(0)}$. The results are shown in Fig.\ \ref{fig1.11sa}.

As can be seen, the matter-antimatter asymmetry and the vorticity are generated from zero initial values in the presence of the strong hypermagnetic field.
Much of the analysis here is similar to that of the last subsection, part of which we repeat. Given the large values of $B_{a}^{(0)}$ and $B_{b}^{(0)}$, the $F_0$ terms in Eqs.\ (\ref{eq32}), (\ref{eq33}) and (\ref{eq34}) initially produce $\eta_{e_R}>0$, $\eta_{e_L}<0$ and $\eta_{B}>0$. Then the chirality flip processes, represented by the last terms of Eqs.\ (\ref{eq32}), (\ref{eq33}), equilibrate $\eta_{e_R}$ and $\eta_{e_L}$ both to positive values, due to the surplus production of the former. This causes the sudden turnarounds in the graphs for $\eta_{e_L}$. In this case, unlike the case with a large initial value of $\eta_{e_R}$ studied in the subsection \ref{x4}, the $\eta$s and hence $\Delta\eta$ keep increasing, but do not saturate for the initial conditions chosen. Meanwhile, the first terms in Eqs.\ (\ref{eq36},\ref{eq37}), in which the CME terms remain much smaller than the $k''$ terms, together with the last terms, arising from the expansion of the Universe, lead to exponential damping observed in Figs.\ \ref{fig:figure:in12sa} and \ref{fig:figure:in2111sa}.
The presence of both helical and non-helical components activate the JB terms in the evolution equations of the velocities Eqs.\ (\ref{eq39}) and (\ref{eq40}), which produce $v_{a}>0$ and $v_{b}<0$.  The velocities then reach their terminal values due to the viscosity terms. These terminal values depend on $B_{a}(x)$, $B_{b}(x)$ and $B_{z}(x)$, which decrease with time. Note that the values of matter-antimatter asymmetries and the velocities at the EWPT increase with increasing the amplitude of the hypermagnetic field.

\begin{figure}[H]
	\centering
	\subfigure[]{\label{fig:figure:in12sab} 
		\includegraphics[width=.365\textwidth]{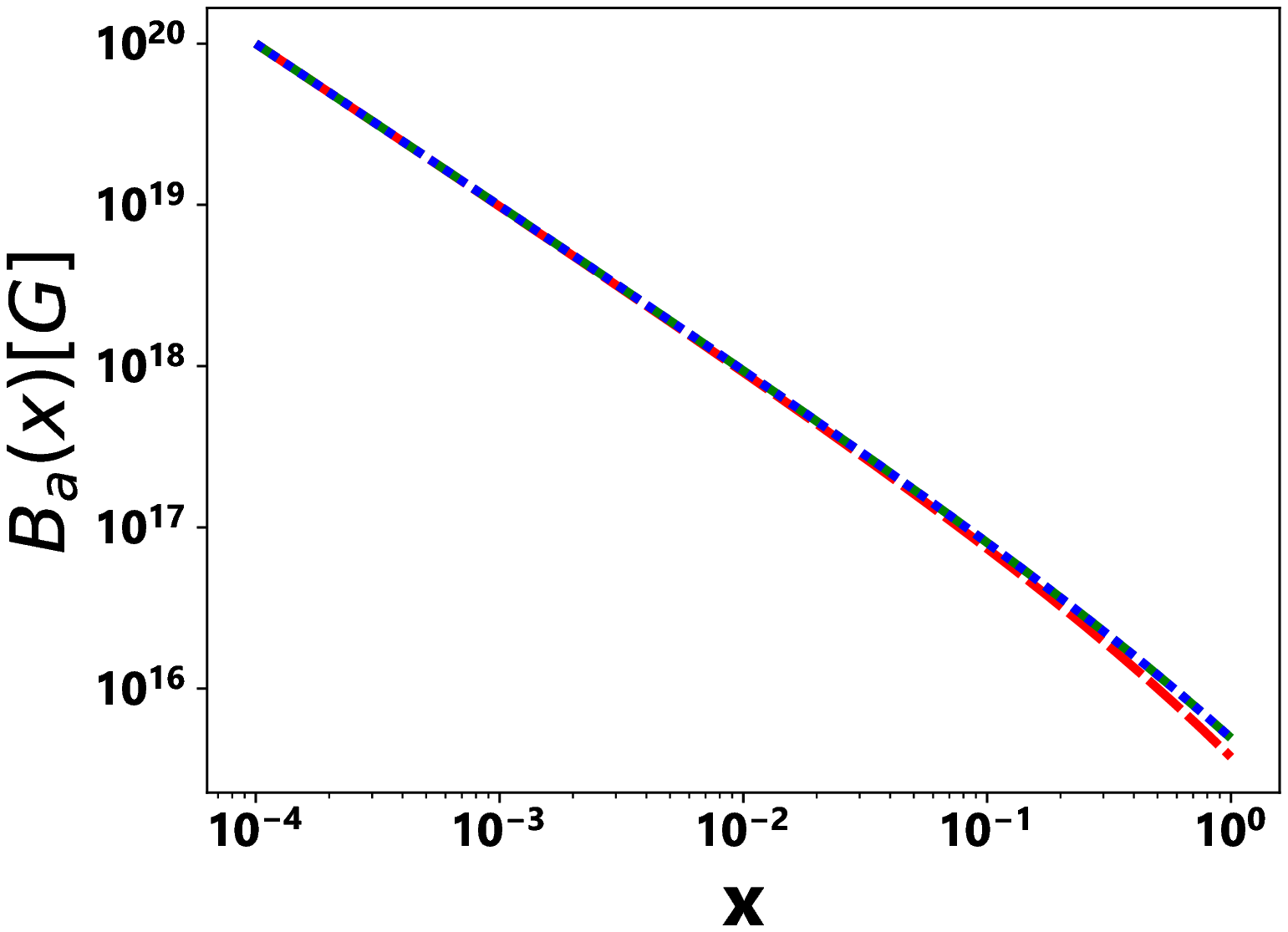}}
	\hspace{10mm}
	\subfigure[]{\label{fig:figure:in2111sab} 
		\includegraphics[width=.36665\textwidth]{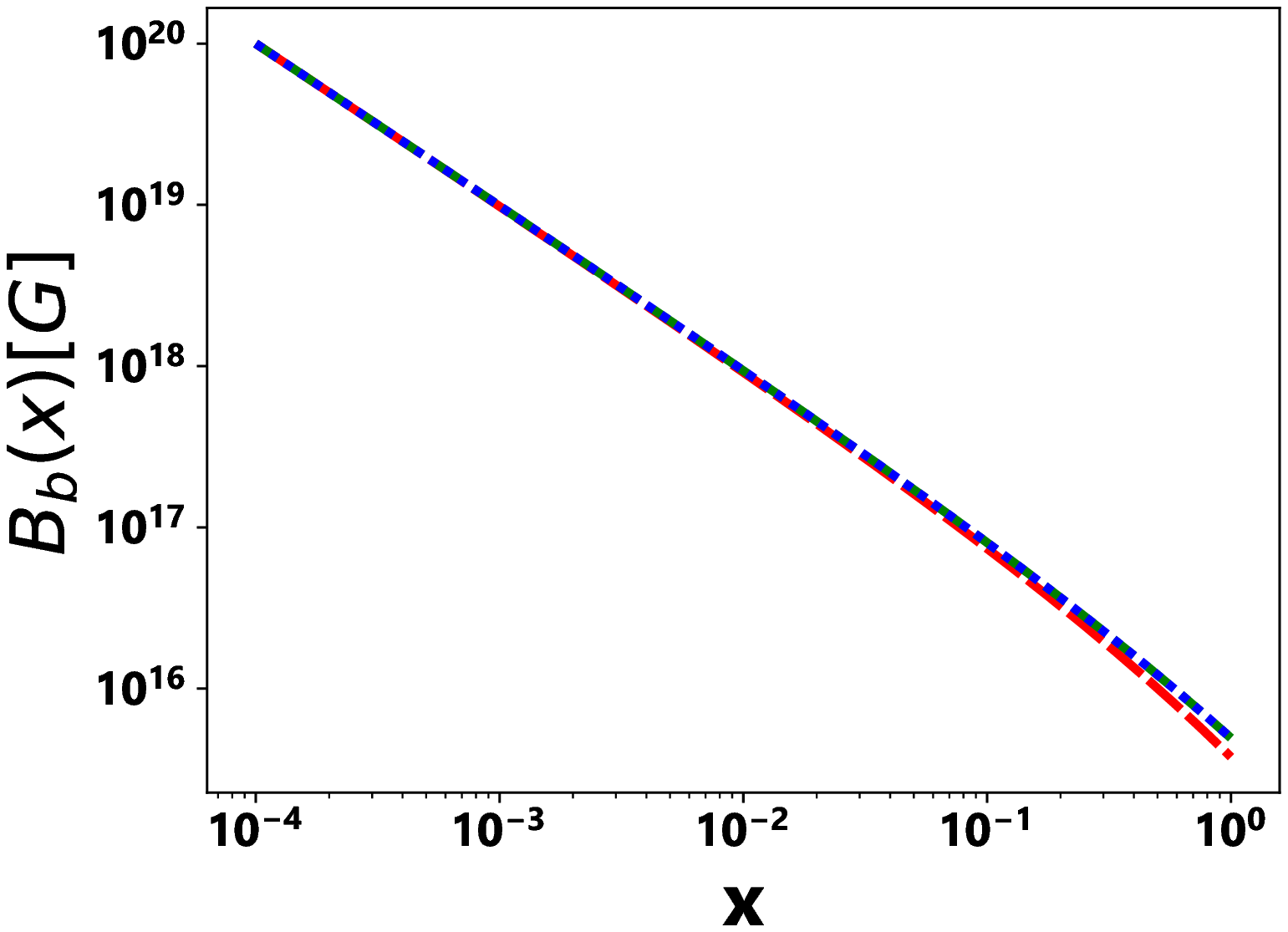}}
	\hspace{8mm}
	\subfigure[]{\label{fig:figure:in2212sab} 
		\includegraphics[width=.365\textwidth]{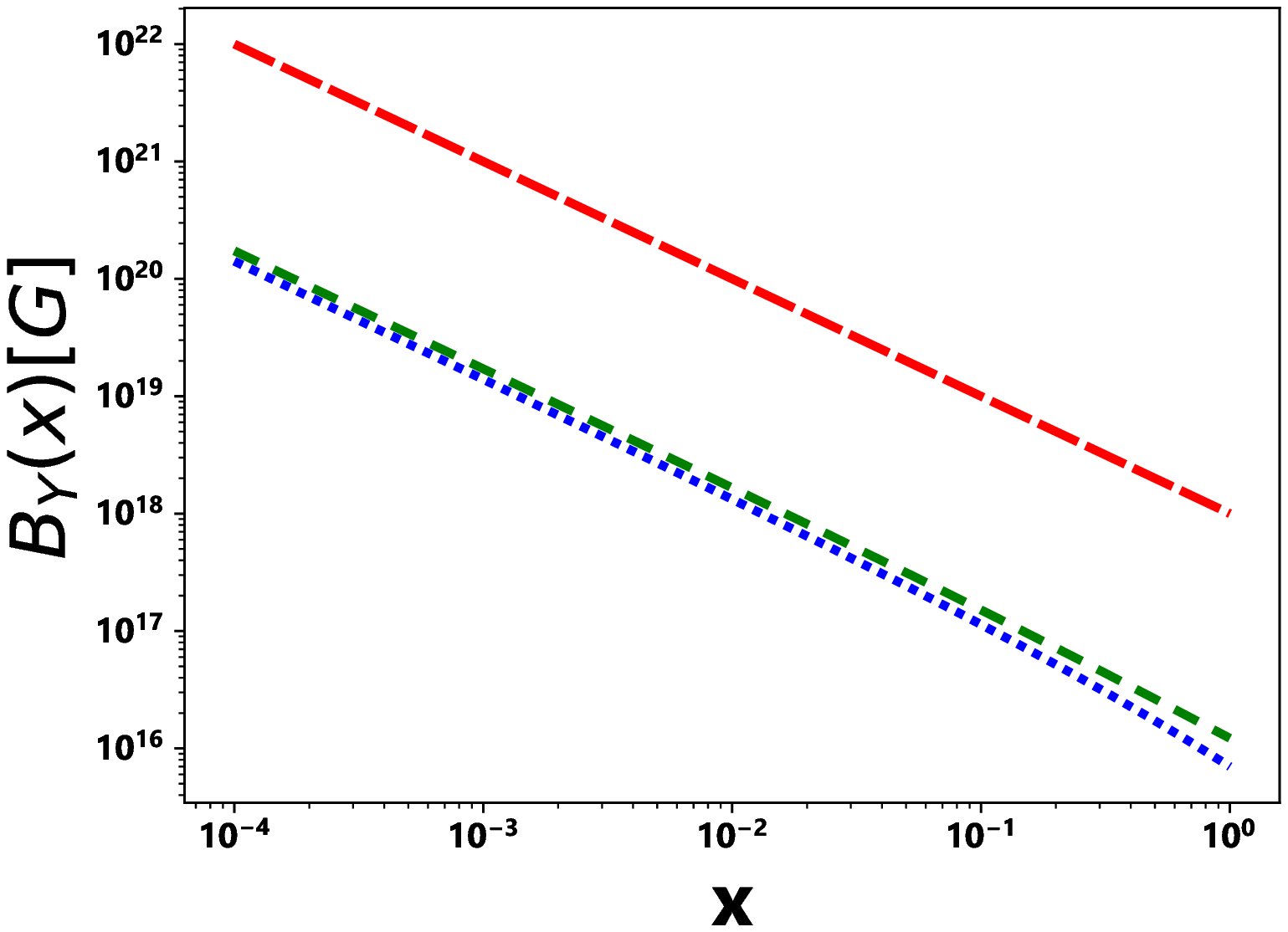}}
	\hspace{8mm}
	\subfigure[]{\label{fig:figure:1.1.311sab}
		\includegraphics[width=.365\textwidth]{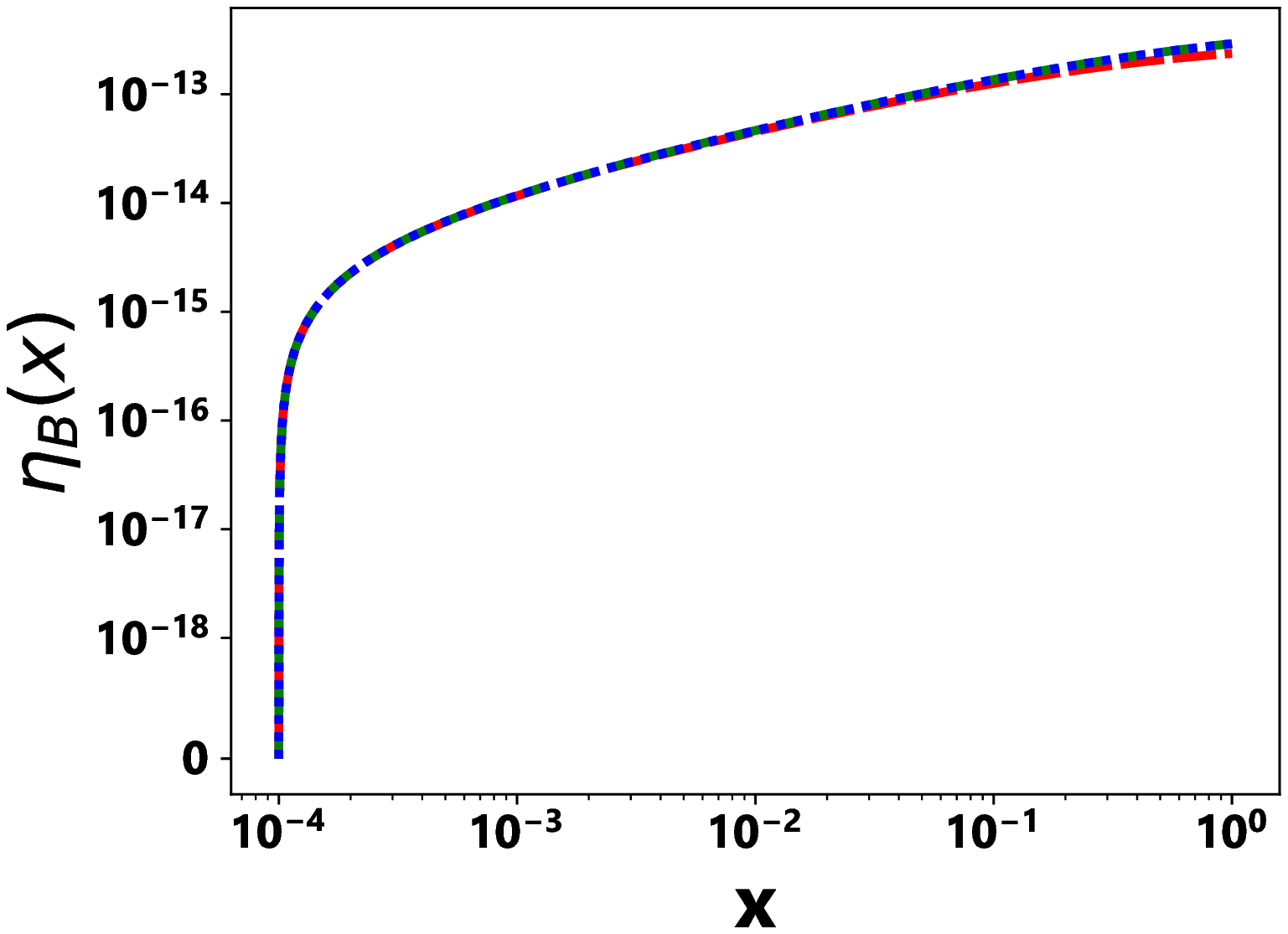}}
	\hspace{8mm}
	\subfigure[]{\label{fig:figure:1.1.211sab}
		\includegraphics[width=.365\textwidth]{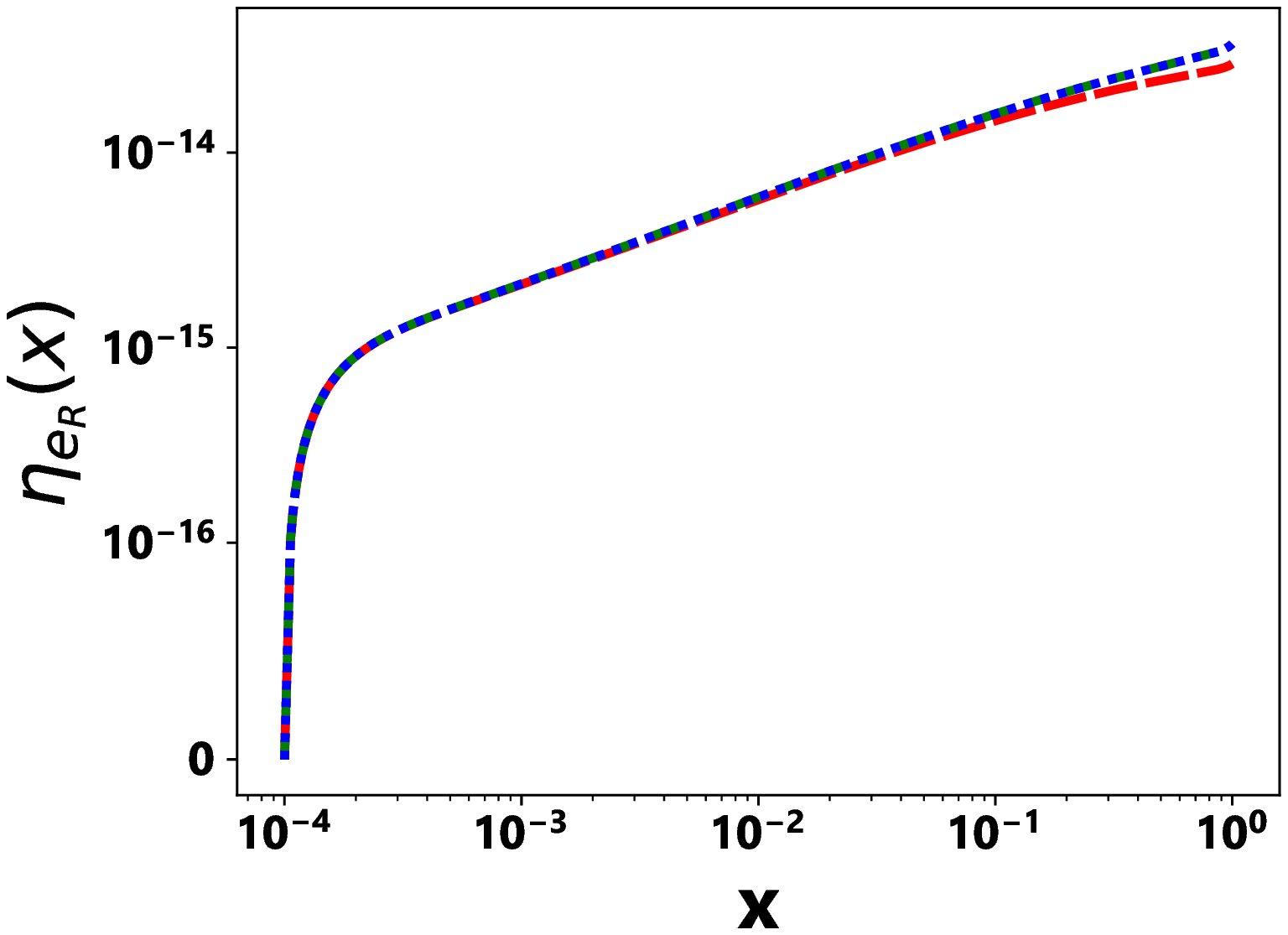}}
	\hspace{8mm}
	\subfigure[]{\label{fig:figure:1.1.311211sab}
		\includegraphics[width=.365\textwidth]{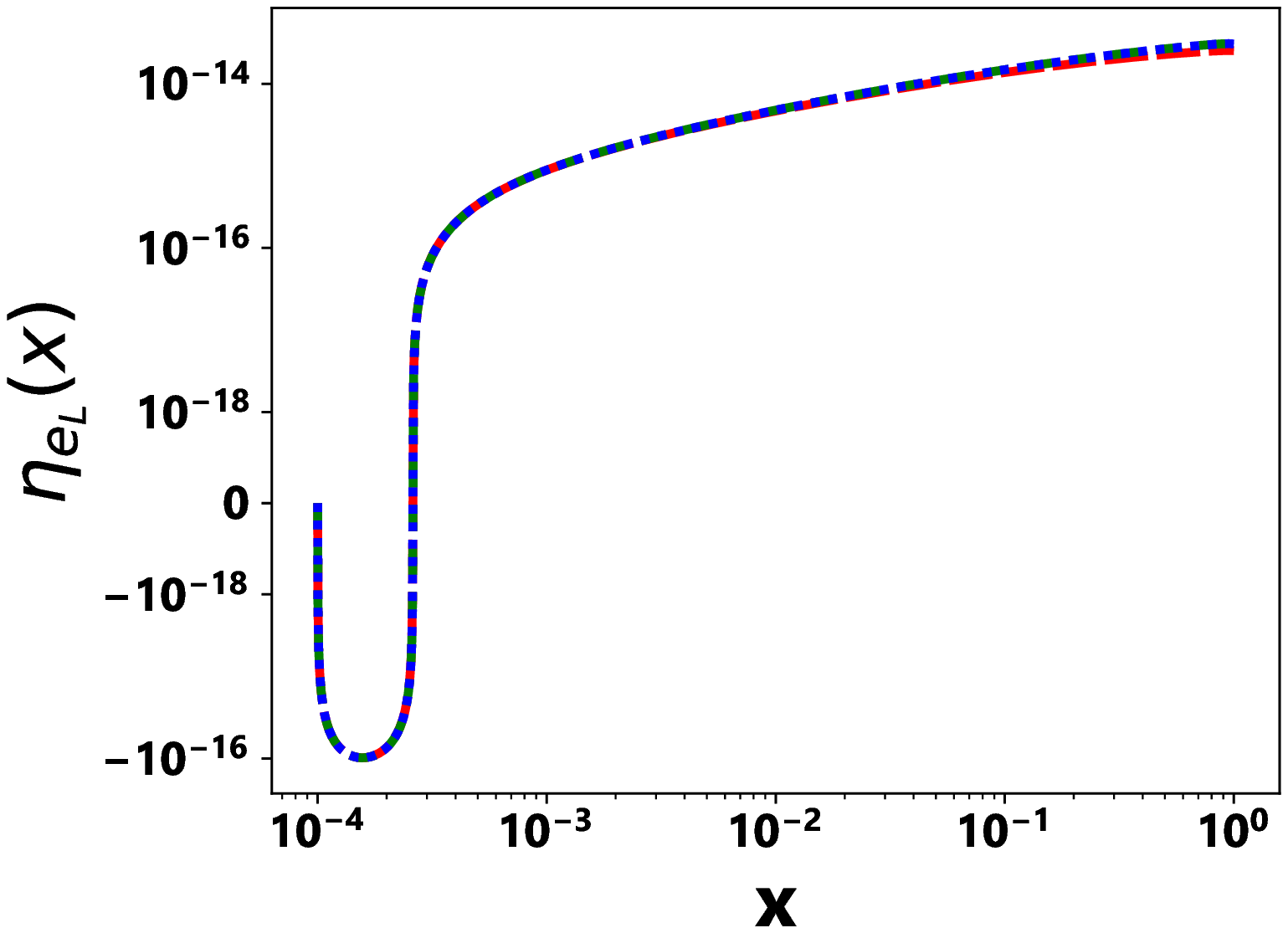}}
	\hspace{8mm}
	\subfigure[]{\label{fig:figure:1va1sab}
		\includegraphics[width=.365\textwidth]{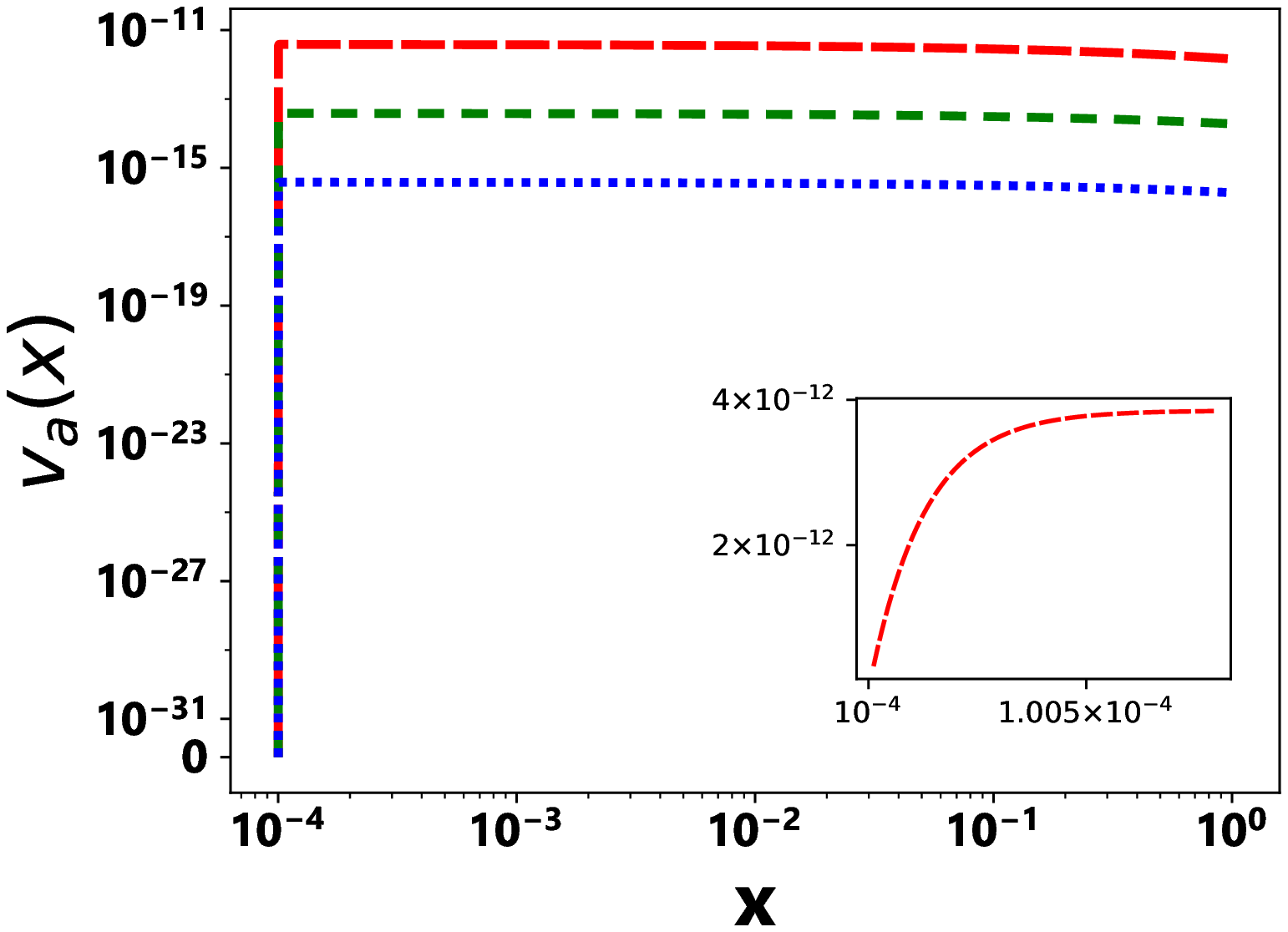}}
	\hspace{8mm}
	\subfigure[]{\label{fig:figure:1vb1sab}
		\includegraphics[width=.365\textwidth]{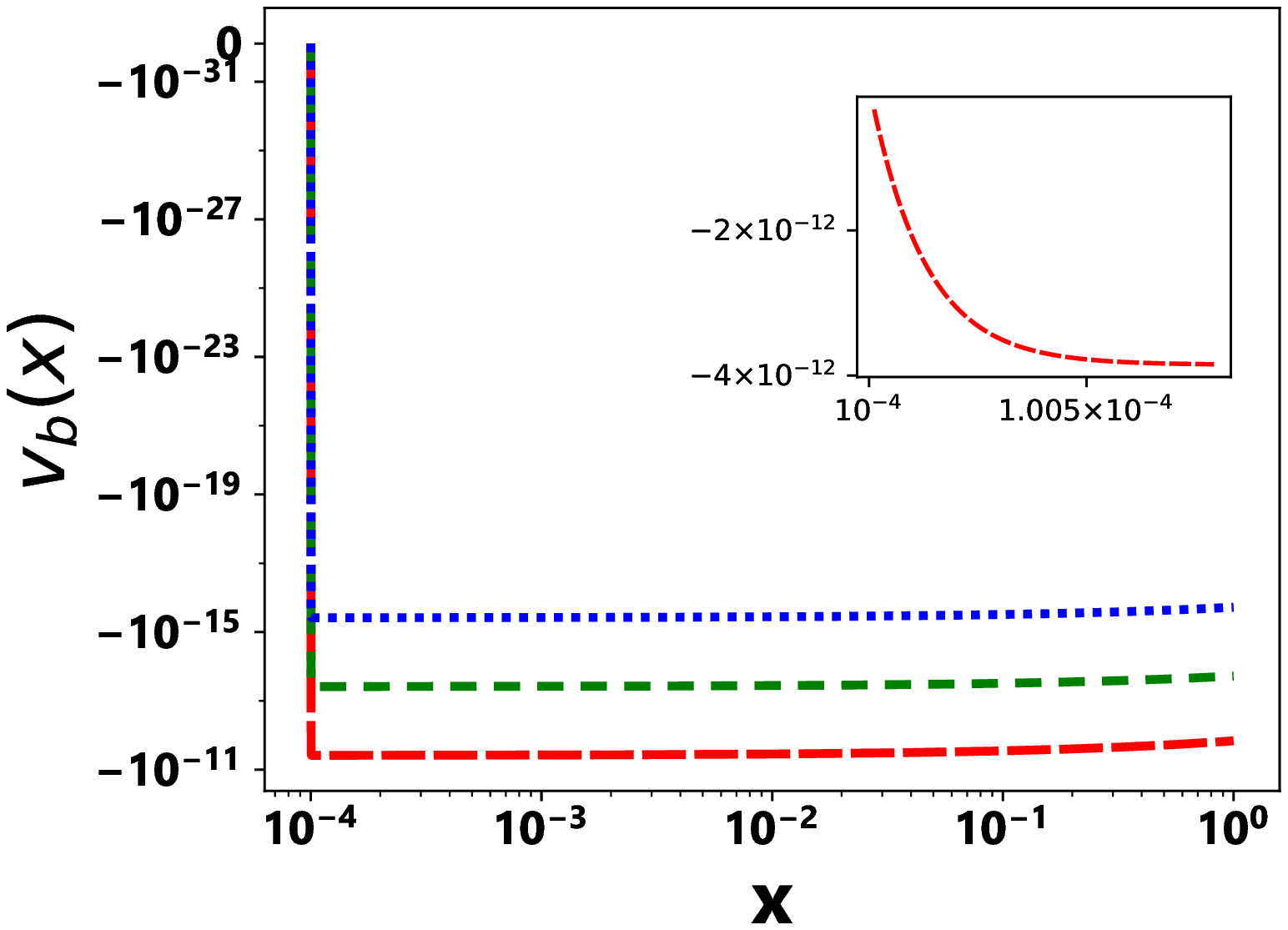}}
	\caption{\footnotesize The time plots of the helical components $B_{a}(x)$ and $ B_{b}(x)$, the hypermagnetic field amplitude $B_{Y}(x)$, the baryon asymmetry $\eta_{B}(x)$, the right-handed electron asymmetry $\eta_{e_R}(x)$, the left-handed electron asymmetry $\eta_{e_L}(x)$, and $v_{a}(x)$ and $v_{b}(x)$ with the initial conditions $B_{a}^{(0)}=B_{b}^{(0)}=10^{20}G $, $\eta_{e_R}^{(0)}=\eta_{e_L}^{(0)}=\eta_{B}^{(0)}=v_{a}^{(0)}=v_{b}^{(0)}=0$. Large (red) dashed line is for $B_{z}^{(0)}=10^{22}G$, dashed (green) line for $B_{z}^{(0)}=10^{20}G$, and doted (blue) line for $B_{z}^{(0)}=10^{18}G$.
	}\label{fig1.11sab}
\end{figure}

Let us now investigate the effect of increasing the value of $B_{z}^{(0)}$ on the evolution. We solve the set of coupled differential equations with the initial conditions $k=10^{-7}$, $\eta_{e_R}^{(0)}=\eta_{e_L}^{(0)}=\eta_{B}^{(0)}=0$, $v_{a}^{(0)}=v_{b}^{(0)}=0$, 
$B_{a}^{(0)}=B_{b}^{(0)}=10^{20}\mbox{G}$, and three different values for $B_{z}^{(0)}$, which are $B_{z}^{(0)}=10^{18}\mbox{G}$, $B_{z}^{(0)}=10^{20}\mbox{G}$ and $B_{z}^{(0)}=10^{22}\mbox{G}$. The results are shown in Fig.\ \ref{fig1.11sab}.
As can be seen from this figure, only the velocities are significantly affected by the value of $B_{z}^{(0)}$ and this is due to the JB terms. The helical components of the hypermagnetic field $B_{a}$ and $B_{b}$ are not significantly affected since the velocities generated are too low and this renders the advection terms ineffective. The matter-antimatter asymmetries are not significantly affected since when $\Delta \eta$ is below its saturation value and the components of $B_Y$ are of the same order, the CME term is at least one order of magnitude smaller than the $F_0$ term in Eqs.\ (\ref{eq32}), (\ref{eq33}) and (\ref{eq34}). Having said that, for the case $B_{z}^{(0)}=10^{22}\mbox{G}$ the effect of the  strengthened advection terms show up as slight decrease of $B_{a}$ and $B_{b}$,  which help shift the balance of power between the CME and $F_0$ terms in Eqs.\ (\ref{eq32}), (\ref{eq33}) and (\ref{eq34}) to the former and this shows up as slight decrease of the $\eta$s.

%Fig.\ \ref{fig:figure:3.31} show that the total hypermagnetic field $B_{Y}(x)$ evolved only due to the adiabatic expansion. Moreover,  Fig.\ \ref{fig4} show the  generation and evolution of the baryon asymmetry, the helical velocity\footnote{$v(x)=\sqrt{v_{a}^{2}(x)+v_{b}^{2}(x).}$} and vorticity\footnote{$\omega(x)=(kT_{EW}/\sqrt{x})\sqrt{v_{a}^{2}(x)+v_{b}^{2}(x).}$} from zero initial value in the presence of the strong hypermagnetic field. The results show that the generated baryon asymmetry does not affected, \footnote{This is true only in the limit $B_{z}\le B_{a}, B_{b}$.} by increasing the initial non-helical part $B_{z}(x)$, but the maximum value of the velocity and vorticity will be increased. Not that in contrast to the vorticity the generated velocity does not affected by the expansion.
%   Here the results show that in the presence of strong hypermagnetic field the CME is dominant in respect to the CVE, too.
\section{Conclusion  }\label{x5}

In this work we have investigated the production and evolution of	
the vorticity, matter-antimatter asymmetries and the hypermagnetic field in the symmetric phase of the early Universe, in the temperature range $100\ \mbox{GeV}\le T\le10\ \mbox{TeV}$, within the framework of AMHD. We have assumed that the hypermagnetic field includes both helical components $B_a$ and $B_b$, and non-helical component $B_z$, and have concentrated on the role of the latter in three scenarios. We have chosen similar Chern-Simons configurations for the helical components of hypermagnetic field and the fluid velocity $v_a$ and $v_b$, the latter two leading to fluid vorticity. The presence of a non-zero $B_z$ has two major and one minor effect on the evolution equations. The first major effect is to activate the  advection terms in the evolution equations of $B_a$ and $B_b$, {\it i.e.}, the second terms in Eqs.\ (\ref{eq36},\ref{eq37}), which are proportional to $B_z(x) v_b(x)$ and $-B_z(x) v_a(x)$, respectively. These terms represent the action of fluid vorticity on the helical components of the hypermagnetic field. The second major effect is to activate the JB terms in the evolution equations of $v_a$ and $v_b$, {\it i.e.}, the first terms in Eqs.\ (\ref{eq39},\ref{eq40}), which are proportional to $B_z(x) B_b(x)$ and $-B_z(x) B_a(x)$, respectively. These terms represent the back reaction of the hypermagnetic field on the fluid vorticity. The minor effect is to strengthen the CME terms in the evolution equations of the matter-antimatter asymmetries $\eta_{e_R}$, $\eta_{e_L}$, $\eta_{B}$, {\it i.e.}, the second terms in Eqs.\ (\ref{eq32},\ref{eq33},\ref{eq34}). We have shown that asymmetries of order $10^{-11}$ and helical hypermagnetic fields of order $10^{21}$G can be easily generated in the scenarios that we have investigated. An interesting observation is that the results are not very sensitive to the initial value of $B_z$, as long as it is non-zero. The specific summary and conclusions of each of the three scenarios studied are as follows. 
 
In the first scenario we have investigated how initial values of $\eta_{e_R}$, $B_z$, $v_a$ and $v_b$ can generate $\eta_{e_L}$, $\eta_{B}$, $B_a$ and $B_b$, and have obtained the time evolution of all variables up to the EWPT. In this scenario, with a large $\eta_{e_R}^{(0)}$, the CME produces $\eta_{B}(<0)$, while its effects on $\eta_{e_L}$ and  $\eta_{e_R}$ are dominated by the chirality flip processes which tend to equilibrate them. The seeds for $B_a$ and $B_b$ are produced by the advection terms or the CVE terms, the former being usually dominant with our choice of initial conditions. Subsequently, $B_a$ and $B_b$ grow mainly due to the CME. Meanwhile, when the viscosity forces the velocities to zero, the JB terms make them overshoot zero and continue to grow, as long as $B_a$ and $B_b$ do so, with the JB and the viscosity terms being almost balanced at each instant. The asymmetries continue to change, after the chirality flip processes equilibrate $\eta_{e_L}$ and $\eta_{e_R}$, until the CME terms are reduced and finally balanced by the $F_0$ terms, at which point $\Delta \eta$, which appears in the CME terms, reaches its minimum value and all $\eta$s reach their saturation values. When $\Delta \eta$, which also appears in the CME terms for $B_a$ and $B_b$, reaches its minimum value the growth factor for $B_a$ and $B_b$ is eliminated and they reach their saturation curves. These saturation curves are actually exponentially decreasing mainly due to the expansion of the Universe.

The second scenario is similar to first except $\eta_{e_R}^{(0)}$ is dispensed with. That is, we have investigated how initial values of $B_z$, $v_a$ and $v_b$ can generate $\eta_{e_R}$, $\eta_{e_L}$, $\eta_{B}$, $B_a$ and $B_b$, and have obtained the time evolution of all variables up to the EWPT. First and foremost, the advection terms produce $B_a(>0)$ and $B_b(<0)$ which have two immediate consequences. First, the combined effects of viscosity and the newly activated JB terms turn $v_a$ and $v_b$ to negative values, as explained above. Second, $\eta_{e_R}(>0)$, $\eta_{e_L}(<0)$ and $\eta_{B}(>0)$ are produced by the $F_0$ terms. The chirality flip processes equilibrate $\eta_{e_R}$ and $\eta_{e_L}$ both to positive values, due to the surplus production of the former. Hence all matter-antimatter asymmetries generated are positive in this scenario. Although the $\eta$s keep growing, they do not reach their saturation curves and the $\Delta \eta$ generated remains too low to contribute as a growth factor for $B_a$ and $B_b$ via the CME, and they reach their saturation curves very quickly, as do the velocities.

In the third scenario, we have investigated how only an initial hypermagnetic field with non-zero components $B_z$ and $B_a$ or $B_b$ can generate $\eta_{e_R}$, $\eta_{e_L}$, $\eta_{B}$, $v_a$ and $v_b$, and have obtained the time evolution of all variables up to the EWPT. First, the JB terms produce $v_a(>0)$ and $v_b(<0)$, and the $F_0$ terms produce $\eta_{e_R}(>0)$, $\eta_{e_L}(<0)$ and $\eta_{B}(>0)$. The chirality flip processes again equilibrate $\eta_{e_R}$ and $\eta_{e_L}$ both to positive values, and hence all $\eta$s generated in this scenario are positive, as well. Although the $\eta$s keep growing, they do not reach their saturation curves and the $\Delta \eta$ generated remains too low, and hence $B_a$, $B_b$, $v_a$ and $v_b$ reach their saturation curves quickly, as explained above.

 %(******************************************************************************************)
%\href{http://arxiv.org/abs/-}{[arXiv:1007.3891 [astro-ph.CO]]}
\newpage
\section{APPENDIX A}
Relativistic hydrodynamics is a powerful effective theory for describing the long-wavelength dynamics of collective phenomena in many-particle systems, such as relativistic astrophysics\cite{relativistic1a} and cosmology \cite{41}. In relativistic hydrodynamics of viscous fluids the definition of the fluid velocity is nontrivial\cite{2019jkc}. The Landau-Lifshitz (or energy) frame \cite{Landau-Lif}, and the Eckart (or conserved charge/particle) frame \cite{Eckart-1940zz} are the commonly used frames in hydrodynamics, the former being the preferred choice, particularly when the conserved charges (such as baryon asymmetry) are negligible \cite{2019jkc}. 
In the Landau-Lifshitz frame, the energy-momentum tensor $T^{\mu\nu}$ and the total electric current $J^{\mu}$ for a plasma of one component massless chiral fermions are given by \cite{saeed2,son1,Yamamoto16,Anand:2017,Landsteiner-2016} 

\begin{equation}
T^{\mu\nu}=(\rho+p)u^{\mu}u^{\nu}- p g^{\mu\nu}+\frac{1}{4}g^{\mu\nu} F^{\alpha\beta} F_{\alpha\beta}-F^{\nu\sigma}{F^{\mu}}_{\sigma}+ \tau^{\mu\nu},
	\end{equation}
	\begin{equation}
	J^{\mu}=\rho_{el} u^{\mu}+J^{\mu}_\mathrm{CME}+J^{\mu}_\mathrm{CVE}+\nu^{\mu},
	\end{equation}
	\begin{equation}
	J^{\mu}_\mathrm{CME}=(Q_\mathrm{R}\xi_{\mathrm {B,R}}+Q_\mathrm{L}\xi_{\mathrm {B,L}})B^{\mu},
	\end{equation}
	\begin{equation}
	J^{\mu}_\mathrm{CVE}=(Q_\mathrm{R}\xi_{\mathrm{v,R}}+Q_\mathrm{L}\xi_{\mathrm {v,L}})\omega^{\mu}.
	\end{equation}
	In the above equations $p$ and $\rho$ are the pressure and the energy density of the plasma, $\rho_{el}$ is the electric charge density, $F_{\alpha\beta}=\nabla_{\alpha}A_{\beta}-\nabla_{\beta}A_{\alpha}$ is the field strength tensor, $B^{\mu}=(\epsilon^{\mu\nu\rho\sigma}/2R^3)u_{\nu}F_{\rho\sigma}$ is the magnetic field four vector,  $\omega^{\mu}=(\epsilon^{\mu\nu\rho\sigma}/R^3)u_{\nu}\nabla_{\rho}u_{\sigma}$ is the vorticity four vector, with the totally
	anti-symmetric Levi-Civita tensor given by $\epsilon^{0123}=-\epsilon_{0123}=1$,  $Q_{R}$ ( $Q_{L}$) is the right-handed (left-handed) electric charge, $\nu^{\mu}=\sigma E^{\mu}-\sigma T [g^{\mu\nu}-u^{\mu}u^{\nu}]\nabla_{\nu}\big(\mu /T\big)$ is the electric diffusion current, $E^{\mu}=F^{\mu\nu}u_{\nu}$ is the electric field four vector, $\sigma$ is the electrical conductivity, 
 $\tau^{\mu\nu}$ is the viscous stress tensor, $u^{\mu}=\gamma\left(1,\vec{v}/R\right)$ is the four-velocity of the plasma normalized such that $u^{\mu}u_{\mu}=1$, and $\gamma=1/\sqrt{1-v^2}$ is the Lorentz factor. The validity of the diagonal Einstein tensor obtained from the FRW metric requires that not only the electromagnetic energy density should be small compared to the energy density of the Universe \cite{b1234ss}, but also the bulk velocity should be small, $\left |\vec{v}\right | \ll1$, which is equivalent to $\gamma\simeq1$. In this case the four vectors are given as follows \cite{saeed2}:
%
%\textcolor{red}{} \textcolor{green}{} \textcolor{green}{,} \textcolor{orange}{()} \textcolor{purple}{} %\textcolor{cyan}{} \textcolor{magenta}{}
%

\begin{equation}
	\begin{split}
	&B^{\mu}=\gamma\left(\vec{v}.\vec{B},\frac{\vec{B}-\vec{v}\times\vec{E}}{R}\right)\simeq\left(\vec{v}. \vec{B},\frac{\vec{B}}{R}\right)\\&
	\omega^{\mu}=\gamma\left(\vec{v}.\vec{\omega},\frac{\vec{\omega}-\vec{v}\times\vec{a}}{R}\right)
	\simeq	\left(\vec{v}.\vec{\omega},\frac{\vec{\omega}}{R}\right)\\&
	a^\mu=\gamma\left(\vec{v}.\vec{a},\frac{\vec{a}+\vec{v}\times\vec{\omega}}{R}\right) \simeq \left(\vec{v}.\vec{a},\frac{\vec{a}}{R}\right)\\&
	E^{\mu}=\gamma\left(\vec{v}.\vec{E},\frac{\vec{E}+\vec{v}\times\vec{B}}{R}\right) \simeq \left(\vec{v}.\vec{E},\frac{\vec{E}+\vec{v}\times\vec{B}}{R}\right),
	\end{split}
	\end{equation}
where $a^\mu=\Omega^{\mu\nu}u_\nu$ is the acceleration four vector, $\Omega_{\mu\nu}=\nabla_{\mu}u_\nu-\nabla_{\nu}u_\mu$ is the vorticity tensor, and $a^{i}=R \Omega^{0i}$ is the three vector acceleration. We have also used the assumption $\partial_{t}\sim\vec{\nabla}.\vec{v}$ in the derivative expansion of the hydrodynamics, hence $\vec{v}\times\vec{E}\simeq v^{2}\vec{B}$, and we have ignored the terms of $O(v^2)$.	
Furthermore, the CME and CVE coefficients for chiral fermions are given as \cite{saeed2,son1,Yamamoto16,Anand:2017,Landsteiner-2016,Neiman-2010zi,Landsteiner-2012kdw}
	\begin{equation}\label{a-3}
	\xi_\mathrm{B,R}=\frac{Q_\mathrm{R}\mu_\mathrm{R}}{4\pi^2}\Big[1-\frac{1}{2}\frac{(n_\mathrm{R}-\bar{n}_\mathrm{R})\mu_\mathrm{R}}{\rho+p}\Big]-\frac{1}{24}\frac{(n_\mathrm{R}-\bar{n}_\mathrm{R})T^{2}}{\rho+p}\simeq\frac{Q_\mathrm{R}\mu_\mathrm{R}}{4\pi^2},
	\end{equation}
	\begin{equation}\label{a-4}
	\xi_\mathrm{B,L}=
	-\frac{Q_\mathrm{L}\mu_\mathrm{L}}{4\pi^2}\Big[1-\frac{1}{2}\frac{(n_\mathrm{L}-\bar{n}_\mathrm{L})\mu_\mathrm{L}}{\rho+p}\Big]+\frac{1}{24}\frac{(n_\mathrm{L}-\bar{n}_\mathrm{L})T^{2}}{\rho+p}\simeq	-\frac{Q_\mathrm{L}\mu_\mathrm{L}}{4\pi^2},
	\end{equation}
	\begin{equation}\label{a-5}
	\xi_\mathrm{v,R}=\frac{\mu_\mathrm{R}^{2}}{8\pi^2}\Big[1-\frac{2}{3}\frac{(n_\mathrm{R}-\bar{n}_\mathrm{R})\mu_\mathrm{R}}{\rho+p}\Big]+\frac{1}{24}T^{2}\Big[1-\frac{2(n_\mathrm{R}-\bar{n}_\mathrm{R})\mu_\mathrm{R}}{\rho+p}\Big]\simeq\frac{\mu_\mathrm{R}^{2}}{8\pi^2}+\frac{1}{24}T^{2},
	\end{equation}
	\begin{equation}\label{a-6}
	\xi_\mathrm{v,L}=-\frac{\mu_\mathrm{L}^{2}}{8\pi^2}\Big[1-\frac{2}{3}\frac{(n_\mathrm{L}-\bar{n}_\mathrm{L})\mu_\mathrm{L}}{\rho+p}\Big]-\frac{1}{24}T^{2}\Big[1-\frac{2(n_\mathrm{L}-\bar{n}_\mathrm{L})\mu_\mathrm{L}}{\rho+p}\Big]\simeq-\frac{\mu_\mathrm{L}^{2}}{8\pi^2}-\frac{1}{24}T^{2}.
	\end{equation}
In the above equations we have used the assumption $\mu_\mathrm{R,L}/T\ll1$ in the hot plasma of the early universe, and used the relations $n_\mathrm{R,L}-\bar{n}_\mathrm{R,L}\simeq \frac{1}{6}\mu_\mathrm{R,L}T^{2}$ and $\rho=3p\simeq \frac{\pi^{2}}{30}g^{*}T^4$, where $T$ is the temperature and $n_\mathrm{R,L}$, and $(\bar{n}_\mathrm{R,L})$  are the chiral number density of the fermion, (anti-fermion), respectively.
Moreover, since $\mu /T\ll1$, we ignore the term $\sigma T [u^{\mu}u^{\nu}-g^{\mu\nu}]\nabla_{\nu}\big(\mu /T\big)$ and consider only the Ohmic effect $\nu^{\mu}=\sigma E^{\mu}$ in the dissipative current. 

In the presence of anomalous effects, the ordinary magnetohydrodynamics (MHD) is generalized to the anomalous magnetohydrodynamics (AMHD) with the following dynamical equations
	\begin{equation}
	\nabla_{\mu}T^{\mu\nu}= 0,
	\end{equation}
	\begin{equation}
	\begin{split}
	&\nabla_{\mu}F^{\mu\nu}=J^{\nu}\\&
	\nabla_{\mu}{\tilde{F}}^{\mu\nu}=0
	\end{split}
	\end{equation}
	\begin{equation}\label{eq-ab1}
	\nabla_{\mu}j^{\mu}_\mathrm{R,L}=C_\mathrm{R,L} E_{\mu}B^{\mu},
	\end{equation}
	\begin{equation}\label{eq-ab2}
	\begin{split}
	&j^{\mu}_\mathrm{R}=(n_\mathrm{R}-\bar{n}_\mathrm{R}) u^{\mu}+\xi_{\mathrm {B,R}}B^{\mu}+\xi_{\mathrm {v,R}}\omega^{\mu}+\sigma_{\mathrm{R}}E^{\mu},\\&
	j^{\mu}_\mathrm{L}=(n_\mathrm{L}-\bar{n}_\mathrm{L}) u^{\mu}+\xi_{\mathrm {B,L}}B^{\mu}+\xi_{\mathrm {v,L}}\omega^{\mu}+\sigma_{\mathrm{L}}E^{\mu},
	\end{split} 
	\end{equation} 
	where ${\tilde{F}}^{\mu\nu}=\frac{1}{2}\epsilon^{\mu\nu\rho\sigma}F_{\rho\sigma}$ is dual field tensor, $\sigma_{\mathrm{R,L}}=\big(\sigma/Q_{\mathrm{R,L}}\big)$, and $C_{R,L}$ are the corresponding right- and left-handed anomaly coefficients
	\cite{son1,Yamamoto16,Anand:2017,Landsteiner-2016}. The right- and left-handed anomalous equation (\ref{eq-ab1}) can be written as
	\begin{equation}\label{eq4aq2}
	\partial_{t}j^{0}_\mathrm{(R,L)}+ \frac{1}{R}\vec{\nabla}.\vec{j}_{(R,L)}+ 3Hj^{0}_\mathrm{(R,L)} =C_\mathrm{R,L} E_{\mu}B^{\mu},
	\end{equation} 
	where the zero and spatial component of the chiral currents are given as 
	\begin{equation}
	\begin{split}
	&j^{0}_{(R,L)}=(n_\mathrm{R,L}-\bar{n}_\mathrm{R,L}) +\xi_{\mathrm {B,(R,L)}}\vec{v}.\vec{B}+\xi_{\mathrm {v,(R,L)}}\vec{v}.\vec{\omega}+\sigma_{\mathrm{(R,L)}}\vec{v}.\vec{E},\\&
	\vec{j}_{(R,L)}=(n_\mathrm{R,L}-\bar{n}_\mathrm{R,L})\vec{v}+\xi_{\mathrm {B,(R,L)}}\vec{B}+\xi_{\mathrm{v,(R,L)}}\vec{\omega}+\sigma_{\mathrm{(R,L)}}\big(\vec{E}+\vec{v}\times\vec{B}\big).
	\end{split}
	\end{equation} 	
	After taking the spatial average of Eq.\ (\ref{eq4aq2}), the divergent terms $\frac{1}{R}\vec{\nabla}.\vec{j}_{(R,L)}$ vanish and we obtain \cite{saeed2} 
	\begin{equation}\label{eq4aq3w1}
	\begin{split}
	\partial_{t}\left(\frac{n_\mathrm{R,L}-\bar{n}_\mathrm{R,L}}{s}\right)= &-\partial_{t}\Big[\frac{\xi_\mathrm{B,(R,L)}}{s}\langle\vec{v}.\vec{B}\rangle+\frac{\xi_\mathrm{v,(R,L)}}{s}\langle\vec{v}.\vec{\omega}\rangle+\frac{\sigma}{Q_\mathrm{R,L}s}\langle\vec{v}.\vec{E}\rangle\Big]\\&-\frac{C_\mathrm{R,L}}{s} \langle\vec{E}.\vec{B}\rangle,
	\end{split}
	\end{equation}
	where $s$ is the entropy density and we have used the relation $\dot{s}/s=-3H$\footnote{Note that $E_{\mu}B^{\mu}=(\vec{v}.\vec{E})(\vec{v}.\vec{B})-(\vec{E}+\vec{v}\times\vec{B}).(\vec{B}-\vec{v}\times\vec{E})= -\vec{E}.\vec{B}+O(v^2)$, and we have ignored the terms of $O(v^2)$.}. Using Eqs.\ (\ref{a-3}-\ref{a-6}) and $x=t/t_\mathrm{EW}=(T_\mathrm{EW}/T)^2$  we obtain 	
	\begin{equation}
	\begin{split}
	\frac{d\eta_\mathrm{R}}{dx}=&\frac{1}{\Big[1+\lambda_{R}(x)\Big]}\Bigg[\Lambda_{R}(x)-\frac{t_\mathrm{EW}C_\mathrm{R}}{s}\langle\vec{E}.\vec{B}\rangle\Bigg],
	\end{split}
	\end{equation}  
	\begin{equation}
	\begin{split}
	\frac{d\eta_\mathrm{L}}{dx}=&\frac{1}{\Big[1+\lambda_{L}(x)\Big]}\Bigg[\Lambda_{L}(x)-\frac{t_\mathrm{EW}C_\mathrm{L}}{s}\langle\vec{E}.\vec{B}\rangle\Bigg],
	\end{split}
	\end{equation} 
	where $M=2\pi^{2}g^{*}/45$, and $\lambda_{R,L}(x)$ and $\Lambda_{R,L}(x)$ are given as follow
	\begin{equation}
	\begin{split}
	&\lambda_{R}(x)=\frac{6Q_\mathrm{R}}{4\pi^{2}}\frac{\langle\vec{v}.\vec{B}\rangle}{10^{20}G}\frac{x}{5000}-\frac{36M}{4\pi^{2}}\eta_\mathrm{R}kv^{2},\\
	&\lambda_{L}(x)=-\frac{6Q_\mathrm{L}}{4\pi^{2}}\frac{\langle\vec{v}.\vec{B}\rangle}{10^{20}G}\frac{x}{5000}+\frac{36M}{4\pi^{2}}\eta_\mathrm{L}kv^{2},\\
	\Lambda_{R}(x)=&-\frac{6Q_\mathrm{R}}{4\pi^{2}}\frac{x}{5000}\frac{\eta_\mathrm{R}}{10^{20}G}\Big[\langle\vec{v}.\partial_{x}\vec{B}\rangle+\langle\frac{\vec{v}.\vec{B}}{x}\rangle+\langle\vec{B}.\partial_{x}\vec{v}\rangle\Big]\\&+\Big[\frac{36M}{4\pi^{2}}\eta_\mathrm{R}^{2}+\frac{1}{12M}\Big]k\vec{v}.\partial_{x}\vec{v}-\frac{x}{50MQ_\mathrm{R}10^{20}  G}\Big[\langle\vec{v}.\partial_{x}\vec{E}\rangle+\langle\frac{\vec{v}.\vec{E}}{x}\rangle+\langle\vec{E}.\partial_{x}\vec{v}\rangle\Big],\\
	\Lambda_{L}(x)=&\frac{6Q_\mathrm{L}}{4\pi^{2}}\frac{x}{5000}\frac{\eta_\mathrm{L}}{10^{20}G}\Big[\langle\vec{v}.\partial_{x}\vec{B}\rangle+\langle\frac{\vec{v}.\vec{B}}{x}\rangle+\langle\vec{B}.\partial_{x}\vec{v}\rangle\Big]\\&-\Big[\frac{36M}{4\pi^{2}}\eta_\mathrm{L}^{2}+\frac{1}{12M}\Big]k\vec{v}.\partial_{x}\vec{v}-\frac{x}{50MQ_\mathrm{L}10^{20}  G}\Big[\langle\vec{v}.\partial_{x}\vec{E}\rangle+\langle\frac{\vec{v}.\vec{E}}{x}\rangle+\langle\vec{E}.\partial_{x}\vec{v}\rangle\Big].
	\end{split}
	\end{equation}
	 Here we assume that not only $\big|\lambda_{R,L}\big|\ll1$, but also $\big|\Lambda_{R,L}(x)\big|\ll\big|(t_\mathrm{EW}C_\mathrm{R,L}/s)\langle\vec{E}.\vec{B}\rangle\big|$.\footnote{This is due to the fact that max $(\mu_\mathrm{R,L}/T,v)\ll1$.}  Therefore, we can write $j^{0}_{(R,L)}\simeq n_\mathrm{(R,L)}-\bar{n}_\mathrm{(R,L)}$ and the anomaly equations reduce to the form \cite{saeed2,Anand:2017}
\begin{equation}
	 \partial_{t}\left(\frac{n_{R,L}-\bar{n}_\mathrm{(R,L)}}{s}\right)=-\frac{C_\mathrm{R,L}}{s} \langle\vec{E}.\vec{B}\rangle.
	 	\end{equation}
	Upon using the relation $(n_{f}-\bar{n}_{f})/s=\mu_{f}T^{2}/6s=\eta_{f}$ and considering the chirality flip processes we obtain the Eqs.\ (\ref{eq32}-\ref{eq33}) \cite{saeed2}. Here we have considered only the spatial components of CME and CVE currents in anomaly equations. We have shown that the temporal components make negligible contributions to the evolution of physical quantities \cite{saeed2}.  
	The chiral vorticity and magnetic coefficients $c_{\mathrm{v}}$ and $c_{\mathrm{B}}$ in the early Universe plasma, which consists of all three generations of leptons and quarks, are given as follows \cite{saeed2,saeed,31,34,shiva3}: %\textcolor{orange}{(Reference? Is this the first time we are presenting this?)}
	\begin{equation}\label{eq22wq}
	\begin{split}
	c_{\mathrm{v}}(t)=&\sum_{i=1}^{n_{G}}\Big[\frac{g'}{48}\Big(-Y_{R}T^{2}+Y_{L}T^{2}N_{w}-Y_{d_{R}}T^{2}N_{c}-Y_{u_{R}}T^{2}N_{c}+Y_{Q}T^{2}N_{c}N_{w}\Big)\\&+\frac{{g'}}{16\pi^{2}}\Big(-Y_{R}\mu_{R_{i}}^{2}+Y_{L}\mu_{L_{i}}^{2}N_{w}-Y_{d_{R}}\mu_{d_{R_{i}}}^{2}N_{c}-Y_{u_{R}}\mu_{u_{R_{i}}}^{2}N_{c}+Y_{Q}\mu_{Q_{i}}^{2}N_{c}N_{w}\Big)\Big],	
	\end{split}
	\end{equation}
	\begin{equation}\label{eqc_Bwq} 
	\begin{split}
	c_{\mathrm{B}}(t)=&
	-\frac{g'^{2}}{8\pi^{2}}\sum_{i=1}^{n_{G}}\Big[-\Big(\frac{1}{2}\Big)Y_{R}^{2}\mu_{R_{i}}-\Big(\frac{-1}{2}\Big)Y_{L}^{2}\mu_{L_{i}}N_{w}-\Big(\frac{1}{2}\Big)Y_{d_{R}}^{2}\mu_{d_{R_{i}}}N_{c}\\&-\Big(\frac{1}{2}\Big)Y_{u_{R}}^{2}\mu_{u_{R_{i}}}N_{c}-\Big(\frac{-1}{2}\Big)Y_{Q}^{2}\mu_{Q_{i}}N_{c}N_{w}\Big], 
	\end{split}
	\end{equation}
%
%\textcolor{red}{} \textcolor{green}{} \textcolor{green}{,} \textcolor{orange}{()} \textcolor{purple}{} %\textcolor{cyan}{} \textcolor{magenta}{}
%
	where $n_{G}$ is the number of generations, $N_{c}=3$ and $N_{w}=2$ are the ranks of the non-Abelian SU$(3)$ and SU$(2)$ gauge groups, $\mu_{L_i}$($\mu_{R_i}$), $\mu_{Q_i}$, and $\mu_{{u_R}_i}$ ($\mu_{{d_R}_i}$) are the common chemical potentials of left-handed (right-handed) leptons, left-handed quarks with different colors, and up (down) right-handed quarks with different colors, respectively.
	After substituting the relevant  hypercharges in Eqs.\ (\ref{eq22wq}) and (\ref{eqc_Bwq}), we obtain
	\begin{equation}\label{eq24wq}
	\begin{split}
	c_{\mathrm{v}}(t)=&\sum_{i=1}^{n_{G}}\Big[\frac{{g'}}{8\pi^{2}}\left(\mu_{R_{i}}^{2}-\mu_{L_{i}}^{2}+\mu_{d_{R_{i}}}^{2}-2\mu_{u_{R_{i}}}^{2}+\mu_{Q_{i}}^{2}\right)\Big],
	\end{split}	
	\end{equation}
	\begin{equation}\label{eq26wq}
	\begin{split} 
	&c_{\mathrm{B}}(t)=\frac{-g'^{2}}{8\pi^{2}} \sum_{i=1}^{n_{G}}\left[-2\mu_{R_{i}}+\mu_{L_{i}}-\frac{2}{3}\mu_{d_{R_{i}}}-\frac{8}{3}\mu_{u_{R_{i}}}+\frac{1}{3}\mu_{Q_{i}}\right]. 
	\end{split}
	\end{equation}
	which are the simplified coefficients obtained in Ref \cite{saeed}.
%\textcolor{blue}{ We assume that all quarks-Yukawa interactions are in equilibrium and the Higgs asymmetry is zero and obtain $\mu_{d_{R}}=\mu_{u_{R}}=\mu_{Q}$ for all generations of the quarks \cite{34,saeed}.	Moreover, we assume that only the contributions of the baryonic and the first-generation leptonic chemical potentials to the chiral vorticity and magnetic coefficients $c_{\mathrm{v}}$ and $c_{\mathrm{B}}$ are nonnegligible. Upon using these simplification we obtain the Eqs.\ (\ref{eq12}) and (\ref{eq13}).}

\end{document}